# Extended p-median problems for balancing service efficiency and equality

Yunfeng Kong[1,2*], Chenchen Lian[1], Guangli Zhang[1], Shiyan Zhai[1,2*]

1 College of Geography and Environmental Science, Henan University, Kaifang, China; 2 Key Laboratory of Geospatial Technology for the Middle and Lower Yellow River Regions, Henan University, Kaifang, China
*Correspondence: yfkong@henu.edu.cn; zsycenu@hotmail.com.

**Abstract**: This article deals with the location problem for balancing the service efficiency and equality. In public service systems, some individuals may experience envy if they have to travel longer distances to access services compared to others. This envy can be simplified by comparing an individual's travel distance to a service facility against a threshold distance. Four extended p-median problems are proposed, utilizing the total travel distance and total envy to balance service efficiency and spatial equality. The new objective function is designed to be inequity-averse and exhibits several analytical properties that pertain to both service efficiency and equality. The extended problems were extensively tested on two sets of benchmark instances and one set of geographical instances. The experimentation shows that the equality measures, such as the standard deviation, mean absolute deviation, and Gini coefficient between travel distances, can be substantially improved by slightly increasing the travel distance. Additionally, the advantages of the proposed problems were validated through Pareto optimality analysis and comparisons with other location problems.

**Keywords:** location problem; spatial equality; spatial envy; analytical property; case analysis.

# 1 Introduction

Location theory is one of the fundamental research topics in regional science and has been widely used in public facility planning. Mainstream location problems are usually designed to improve the service efficiency, such as the facility cost, the travel cost, and the amount of customers to be served. Nevertheless, delivering quality services in an efficient and equitable manner is critical to general public. Some location models aim to optimize one of the spatial equality measures, such as variance, standard deviation, and the Gini coefficient between the travel distances (e.g. Maimon, 1986; Berman, 1990; Mulligan, 1991; Ogryczak, 2000; Barbati & Bruno, 2018). Definitions of equity in public services are systematically surveyed by Cepiku D. and Mastrodascio (2021). Among the multiple understandings of equity, spatial equality ensures that public services such as healthcare, education, and transportation are evenly distributed, allowing all individuals, regardless of their location, to access essential services. However, facility locations and their service areas could be seriously distorted by minimizing inequality indicators. Consequently, only modest progress has been made in research on location models addressing equality issues (Wang, 2022).

This article aims to extend the classical p-median problem (PMP) with efficiency and equality considerations. In public service systems, it is assumed that some individuals may experience envy if they travel longer distances to access services than others. This envy can be simplified by comparing one's travel distance to a service facility against a threshold distance (Bolton and Ockenfels, 2000). Based on the idea of spatial envy, a minimum envy location problem (MELP) is formulated by replacing the objective function of the PMP with the spatial envy function. More generally, a minimum distance and envy location problem (MDELP) is formulated by adding the spatial envy function to the objective function of the PMP. The total travel distance, as an efficient indicator, and the total spatial envy, as an equality indicator, are simultaneously minimized by the weighted objective function. In addition, the capacitated MELP (CMELP) and the capacitated MDELP (CMDELP) can be similarly formulated. The performance of the newly proposed problems is intensively tested on three sets of benchmark instances. Experiments show that all the instances can be solved with highly satisfactory solutions, in which the equality measures such as standard deviation, mean absolute deviation, and the Gini coefficient between travel distances can be substantially improved.

This research has made two contributions to the location science. First, four new location problems are formulated as mixed integer linear programming (MILP) models for trade-off between service efficiency and spatial equality. The problem models are simple, effective, and relatively easy to solve. Second, the effectiveness of the proposed problems is verified by benchmark test, Pareto optimality analysis and comparison with other approaches.

The remainder of the article is organized as follows. Section 2 provides a brief review of related works. Section 3 introduces the newly proposed problems and their properties. In Section 4, the problems are tested and analyzed using three sets of instances. Section 5 discusses the advantages, potential real-world applications, and possible limitations of the proposed problems. Finally, Section 6 summarizes several concluding remarks.

# 2 Related works

Various location problems have been investigated since 1950s (Laporte et al., 2019). Among the family of location problems, quite a few have been proposed to search for fair solutions. A straightforward way is to optimize one of the equality measures. The simplest and most common method to improve service equitability is to minimize the worst-case scenario of service. The p-center problem (PCP) (Hakimi, 1964) and the capacitated PCP (CPCP) (Barilan et al., 1993), which minimize the worst-off travel cost, has been widely considered in public sector applications. There are many equality measures and choosing one implies certain assumptions about the decision maker's attitude to service equality. Some of these measures have been widely utilized to propose location models that consider service equality, such as the variance of the of travel distances (Maimon, 1986; Berman, 1990), the Gini coefficient (Maimon, 1988; Mulligan, 1991; Drezner et al., 2009; Barbati & Bruno, 2018), the sum of absolute deviations (Ogryczak, 2000; Smith et al., 2013), and the sum of absolute difference (Ohsawa et al., 2008; Lejeune & Prasad, 2013). In these location models, the equality measure is employed in objective function or is occasionally used in a constraint. An equality measure can also be defined as an objective alongside other efficiency objectives, resulting in a multi-objective optimization model. The equality measures in location theory have been surveyed in several articles (Erkut, 1993; Truelove, 1993; Marsh & Schilling, 1994; Ogryczak 2000; Ogryczak 2009; Karsu & Morton, 2015; Barbati & Piccolo, 2016).

The second approaches to service equality are based on the ordered median function (Nickel & Puerto, 2005; Puerto & Rodríguez-Chía 2019). Its objective function is a weighted total cost function, where the weights are rank-dependent. This approach provides a common framework for several classical location problems such as the median, the center, the k-centrum, and the trimmed-mean problems. The trade-off between efficiency and equality can be implementing by assigning appropriate weights to the ordered costs (Ogryczak & Zawadzki, 2002; Ogryczak & Olender 2016; Filippi et al., 2021a; Filippi et al., 2021b). López-de-los Mozos et al. (2008) have extended the concept of ordered median to define the ordered weighted average of the absolute deviations; and thus translated the k-centrum problem (Tamir, 2001) to the k-sum deviation problem. Marín et al. (2010) also show that both the range and sum of pair-wise absolute differences functions can be similarly modeled. Chen and Hooker (2022) propose a set of objective functions that balance Rawlsian leximax fairness and efficiency. Their approach aims to address some of the limitations of previous methods. The ordered median approaches are very inequality averse and considered by some researchers as the "most equitable" solution (Karsu & Morton, 2015).

The third approaches are based on the concept of spatial accessibility. The service efficiency and equality are balanced by a two-step optimization (2SO4SAI) method: select best facility locations by solving a classical location problem such as the PMP, the MCLP, or the SCLP; and then allocate resources to facility locations by minimizing the variance of spatial accessibility (Wang & Tang, 2013; Tao et al., 2021; Li et al., 2017; Li et al., 2022; Wang, 2022). Nevertheless, the solution result from the 2SO4SAI method is sensitive to the threshold distance and the distance decay function in the second step of resource allocation.

Various equality measures and mathematical models for equality-efficiency trade-off in location science have been explored theoretically and somewhat practically. It is agreed that the spatial equality of public service can be achieved by minimizing an inequality measure, an ordered weighed function, or the variance of spatial accessibility values. However, quite a few issues have remained uninvestigated. First, it is difficult to select a location problem for real-world applications with efficiency and equality considerations. Usually, the computational complexity is quite high for minimizing inequality objective functions such as the Gini coefficient, the coefficient of variation, and the Schutz index (Barbati & Bruno, 2018). The solutions from the ordered median problems are very sensitive to the weight vector (Filippi et al., 2021a). Sensitive analysis shows that the parameter $k$ is critical to the $k$-centrum problem or the $k$-sum deviation problem. Second, in most cases, equality and efficiency may seriously conflict: increasing equality concerns the results in a decrease in efficiency. There are some initial attempts to numerically analyze the trade-off between the efficiency and equality (e.g. Mulligan, 1991; Lopez-de-losmozos & Mesa, 2003; Filippi et al., 2021a; Filippi et al., 2021b). It is essential to know how much you may sacrifice in efficiency when a fair solution is achieved. Nevertheless, the relationships between service cost, spatial access and spatial equality still remain unclear. In this article, four extended p-median problems are proposed. The new models are structurally simple, relatively easy to solve, but effective to balance service efficiency and equality.

# 3 Extended p-median problems

## 3.1 Model formulation

Let $I = \{1,2 \cdots n\}$ be a set of $n$ candidate facility locations, each location $i$ $(i \in I)$ has a maximum service capacity $s_i$. Let $J = \{1,2 \cdots m\}$ be a set of $m$ customer locations, each location $j$ $(j \in J)$ has service demand $w_j$. Variable $d_{ij}$ is the distance between locations $i$ and $j$. Let $y_i$ $(i \in I)$ be a binary variable indicating whether a facility is opened at location $i$. Let $x_{ij}$ $(i \in I, j \in J)$ be a binary variable denoting whether customer $j$ is served by a facility at location $i$. The well-known PMP, first introduced by Hakimi (1964), can be formulated as follows.

$$\text{Minimize:} \sum_{i\in I}\sum_{j\in J} w_j d_{ij} x_{ij} \tag{1}$$

$$\text{Subject to:} \ \sum_{i\in I} x_{ij} = 1, \forall j \in J \tag{2}$$

$$x_{ij} \le y_i, \forall i \in I, j \in J \tag{3}$$

$$\sum_{i\in I} y_i = P \tag{4}$$

$$x_{ij}, y_i \in \{0,1\}, \forall i \in I, j \in J \tag{5}$$

The objective function (1) minimizes the total travel distance. The constraints (2) ensure that customers at each location must be serviced by a facility. Constraints (3) guarantee that customers at each location must be served by an operational facility. Constrain (4) limits the number of facilities.

The capacitated version of the PMP (CPMP) with single-source allocation (Osman and Christofides, 1994) can be expressed as follows. Unlike the PMP, the constrains (8) guarantee that the total customers serviced by an operational facility does not exceed its capacity.

$$\text{Minimize:} \sum_{i\in I}\sum_{j\in J} w_j d_{ij} x_{ij} \tag{6}$$

$$\text{Subject to:} \ \sum_{i\in I} x_{ij} = 1, \forall j \in J \tag{7}$$

$$\sum_{j\in J} w_j x_{ij} \le s_i y_i, \forall i \in I \tag{8}$$

$$\sum_{i\in I} y_i = P \tag{9}$$

$$x_{ij}, y_i \in \{0,1\}, \forall i \in I, j \in J \tag{10}$$

In order to improve the spatial equality of service, we extend the PMP and the CPMP by introducing the concept of spatial envy. The minimization of an envy function was first introduced in location problems by Espejo et al. (2009), and then redefined by Chanta et al. (2011) and Chanta et al. (2014). We simplify the envy function by comparing an individual's travel distance to a service facility against a threshold distance, such as the mean travel distance of a preferred distance. This approach is based on the theory of equity, reciprocity, and competition (Bolton and Ockenfels, 2000), where inequality is defined as the deviation of an individual from the group average. Accordingly, the envy of an individual at location $j$, $v_j$, is expressed as follow:

$$v_j = \begin{cases} 0, \forall d_j \le d^* \\ \left(d_j - d^*\right)^2, \forall d_j > d^* \end{cases} \tag{11}$$

In formula (11), $d_j$ is the travel distance from location $j$ to its facility, and $d^*$ is a threshold travel distance. The sum of spatial envies can serve as an optimization objective (12), where $(d_{ij} - d^*)_+ = \max(0, d_{ij} - d^*)$. Note that one of the inequality measures, the standard upper semi-deviation discussed in Ogryczak (2009) is a special cases of the function (12) when distance $d^*$ is set to the mean travel distance.

$$\text{Minimize} \ \sum_{j\in J} w_j v_j = \sum_{i\in I}\sum_{j\in J} w_j \left(d_{ij} - d^*\right)_+^2 x_{ij} \tag{12}$$

The objective functions (1) and (2) can be replaced by the inequity-averse function (12). As a result, a minimum envy location problem (MELP) can be defined by the objective function (12) and the constraints (2), (3), (4) and (5). Similarly, the capacitated MELP (CMELP) can be defined by the objective function (12) and the constraints (7), (8), (9) and (10).

The envy objective can also be combined with the distance objective, shown as (13):

$$\text{Minimize:} \sum_{i\in I}\sum_{j\in J} w_j d_{ij} x_{ij} + \beta \sum_{i\in I}\sum_{j\in J} w_j \left(d_{ij} - d^*\right)_+^2 x_{ij} \tag{13}$$

The objective (13) aims to balance the total travel distance and the total service envy by the weight $\beta(\beta \ge 0)$. Using this objective, a minimum distance and envy location problem for trade-off between efficiency and equality (MDELP) can be formulated by the objective function (13) and the constraints (2), (3), (4) and (5). Similarly, the capacitated version (CMDELP) can be formulated by the objective function (13) and the constraints (7), (8), (9) and (10). It is obvious that the PMP (CPMP) is a special case of MDELP (CMDELP) when $\beta = 0$; and the MELP (CMELP) is equivalent to the MDELP (CMDELP) when $\beta \to \infty$.

The weight $\beta$ is sensitive to model performance. If its value is too small, the model solution will be dominated by the total travel distance. On the other side, if its value is large enough, the model solution will be controlled by the total envy. In case that the values of $\beta$ and $d^*$ can be estimated satisfying the equation $\sum_{i\in I}\sum_{j\in J} w_j d_{ij} x_{ij} \approx \beta \sum_{i\in I}\sum_{j\in J} w_j \left(d_{ij} - d^*\right)_+^2 x_{ij}$, the total distance and the total envy will be optimized in a balanced manner.

Let $d^* = \bar{d}$ be the mean travel distance, then $\sum_{i\in I}\sum_{j\in J} w_j d_{ij} x_{ij} = \sum_{j\in J} w_j\, d^*$. Let $\sigma$ be the standard deviation of travel distance, and assume that $\sum_{i\in I}\sum_{j\in J} w_j \left(d_{ij} - d^*\right)_+^2 x_{ij} \approx 0.5 \sum_{i\in I}\sum_{j\in J} w_j (d_{ij} - d^*)^2 x_{ij} = 0.5 \sum_{j\in J} w_j\, \sigma^2$, then the objective function (13) is approximately equal to $\sum_{j\in J} w_j (\, d^* + 0.5\sigma^2)$. If $d^* \approx \bar{d}$ and $\beta \approx 2d^*/\sigma^2$, the two parts of the objective function will be optimized in a nearly equivalent manner.

Using the recommended parameter values, the following approximate equations hold: $\beta \sum_{i\in I}\sum_{j\in J} w_j \left(d_{ij} - d^*\right)_+^2 x_{ij} \approx \beta \sum_{i\in I}\sum_{j\in J} w_j \left(d_{ij} - \overline{d}\right)_+^2 x_{ij} \approx 0.5\beta \sum_{j\in J} w_j\, \sigma^2$. Given $\beta \approx 2d^*/\sigma^2$, $0.5\beta \sum_{j\in J} w_j\, \sigma^2 \approx 2d^*/\sigma^2 * 0.5 \sum_{j\in J} w_j\, \sigma^2 \approx \sum_{j\in J} w_j\, \overline{d}$. This means that the total envy value can be adjusted by the weight $\beta$ to compare with the total travel distance. As a result, the total travel distance and the total envy can be theoretically combined as a linear function.

Compared to commonly used equality measures such as variance, mean absolute deviation, sum of absolute differences, and the Gini coefficient, the envy measure has two advantages. First, new location models, such as MELP, MDELP, CMELP, and CMDELP, can be straightforwardly formulated using the spatial envy function. This allows for balancing the efficiency measure (total travel distance) and the equality measure (total spatial envy) by optimizing the weighted objective (13). Second, the envy objective is defined as a linear function without introducing additional decision variables and related constraints, making it relatively easier to solve than non-linear inequality objectives or linear inequality objectives subject to additional variables and constraints.

## 3.2 Model properties

One way to achieve an equity-efficiency trade-off is to use an inequity-averse aggregation function (Karsu & Morton, 2015). The monotonicity, anonymity, and Pigou-Dalton principle of transfers are the essential standards to formulate an equitable aggregation function (Kostreva et al. 2004; Karsu & Morton, 2015). The objective (13) is an anonymous and monotonic increasing function, and weakly satisfies the Pigou-Dalton principle of transfers. These properties are essential for ensuring fairness and efficiency in location problems. The inequity-averse objective function also benefits in solving the MELP and MDELP problems by adapting existing heuristic algorithms for the PMP. The proof of the inequity-averse property of the function (13) can be found in Appendix 1 in the Supplementary Material.

There are four analytical properties to understand the relationships between parameter $\beta$ and model performance.

Let $s_m$ be the optimal solution of the PMP. Let $\mu_m$ and $\gamma_m$ be the total travel distance and the total spatial envy of solution $s_m$, respectively.

Let $s_e$ be the optimal solution of the MELP. Let $\mu_e$ and $\gamma_e$ be the total travel distance and the total spatial envy of solution $s_e$, respectively. the total spatial envy.

Let $s_1$ be the optimal solution of the MDELP with parameters $\beta_1$ and $d_1^*$. Let $f_1 = \mu_1 + \beta_1\gamma_1$ be the optimal objective value, where $\mu_1$ is total travel distance, and $\gamma_1$ is the total spatial envy.

Let $s_2$ be the optimal solution of the MDELP with parameters $\beta_2$ and $d_2^*$. Let $f_2 = \mu_2 + \beta_2\gamma_2$ be the optimal objective value, where $\mu_2$ is total travel distance, and $\gamma_2$ is the total spatial envy.

***Property 1: If*** $\boldsymbol{\beta_1 \geq 0}$ ***and*** $\boldsymbol{d_1^* \geq 0}$***, then*** $\boldsymbol{\mu_1 \geq \mu_m}$ ***and*** $\boldsymbol{\gamma_1 \geq \gamma_e}$***.***

*Property* 1 indicates that, for the optimal solution of MDELP with any parameters $\beta$ and $d^*$, $\mu_m$ is the lower bound of total travel distance, and $\gamma_e$ is the lower bound of total spatial envy.

***Property 2: If*** $\boldsymbol{0 \leq \beta_1 < \beta_2}$ ***and*** $\boldsymbol{d_1^* = d_2^*}$***, then*** $\boldsymbol{f_1 \leq f_2}$***.***

*Property* 2 indicates that keeping the value of $d^*$ the same, and increasing the weight of spatial envy, $\beta$, the optimal objective of the MDELP will be increased or remained the same.

***Property 3: If*** $\boldsymbol{0 \leq \beta_1 < \beta_2}$ ***and*** $\boldsymbol{d_1^* = d_2^*}$***, then*** $\boldsymbol{\mu_1 \leq \mu_2}$ ***and*** $\boldsymbol{\gamma_1 \geq \gamma_2}$***.***

*Property* 3 shows that keeping the value of parameter $d^*$ the same, and increasing the weight of spatial envy, $\beta$, the optimal travel distance of the MDELP will be increased or remained the same, and the optimal spatial envy of the MDELP will be reduced or remained the same.

***Property 4: If*** $\boldsymbol{\beta_1 \geq 0}$ ***and*** $\boldsymbol{d_1^* \geq 0}$***, then*** $\boldsymbol{\mu_1 \leq \mu_e}$ ***and*** $\boldsymbol{\gamma_1 \leq \gamma_m}$***.***

*Properties* 1, 3 and 4 show that $\mu_m \leq \mu_1 \leq \mu_e$ and $\gamma_e \leq \gamma_1 \leq \gamma_m$. This indicates that the optimal travel distance of the MDELP located in the interval $[\mu_m, \mu_e]$, and the optimal spatial envy of the MDELP located in the interval $[\gamma_e, \gamma_m]$.

The proof of the analytical properties of the MDELP can be found in Appendix 2 in the Supplementary Material.

Similarly, the CMDELP also exhibits the aforementioned analytical properties.

# 4 Experiments

## 4.1 Instances and algorithms

The proposed models, MDELP and CMDELP, were tested on three sets of benchmark instances with 100-1000 facilities and 300-6752 customers. The dataset name, source, instance sizes and the number of facilities for each instance are shown in Table 1. Datasets ORlib (Ahuja et al., 2004), and Tbed1 (Avella & Boccia, 2009) can be downloaded from webpage https://or-brescia.unibs.it/instances/instances_sscflp. The datasets ORlib and Tbed1 were randomly generated for the facility location problem. The third dataset Geo (https://github.com/yfkong/PMPequality) was designed by the authors using geographic data such as the administrative boundaries, residential points, population density, socio-economic statistics, schools, and healthcare centers in Henan Province, China. One rural region (GY) and three urban regions (ZY, KF and ZZ) were selected as real-world instances. The instances GY and ZY were originally designed for site-selection of compulsory schools; the instances KF and ZZ were originally prepared for facility planning of community health centers. For each MDELP (CMDELP) instance, five P values were subjectively selected.

The parameters $d^*$ and $\beta$ for each MDELP (CMDELP) instance were estimated according to its PMP (CPMP) solution. Let $\bar{d}$ be the mean distance, and $\sigma$ be the standard deviation of the travel distances in the PMP (CPMP) solution, we set the model parameters $d^* \approx \bar{d}$ and $\beta \approx 2\bar{d}/\sigma^2$.

The problem instances were solved by the Gurobi optimizer (https://www.gurobi. com). In case that the instance cannot be optimally or near-optimally solved in 2 hours, heuristic algorithms were used. Given the inequity-averse nature of function (13), the interchange algorithms for the PMP can be readily adapted to address the MDELP and MELP, and certain heuristics for the CPMP can be modified to solve the CMDELP and CMELP. Almost all PMP, MELP, and MDELP instances were solved optimally by the Gurobi 9.1.1. The uncapacitated instances of ZZ were solved by an iterative local search (ILS) algorithm based on the fast interchange procedure (Whitaker, 1983). Most CPMP, CMELP, and CMDELP instances can be solved optimally or near-optimally by the Gurobi 9.1.1. The capacitated instances of ZZ were solved by a modified version of the matheuristic algorithm (Kong, 2021). The two heuristic methods were implemented using the Python programming and run in PyPy 7.0 (http://pypy.org), a fast and compliant implementation of the Python language. All the instances were solved on a desktop workstation: Dell Precision Tower 7910 with dual CPU (Intel Xeon E5-2630 v4, 2.20GHz), 32GB RAM and Windows 10 operating system.

Table 1: Benchmark instances

| Dataset | Instance | $\lvert I \rvert$ | $\lvert J \rvert$ | P for MDELP | P for CMDELP |
|---|---|---|---|---|---|
| ORlib | Capa1 | 100 | 1000 | 10,11,12,13,14 | 10,11,12,13,14 |
| ORlib | Capb1 | 100 | 1000 | 12,13,14,15,16 | 12,13,14,15,16 |
| ORlib | Capc1 | 100 | 1000 | 12,13,14,15,16 | 12,13,14,15,16 |
| Tbed1 | i300_1 | 300 | 300 | 10,20,30,40,50 | 36,38,40,42,44 |
| Tbed1 | i300_6 | 300 | 300 | 10,20,30,40,50 | 22,24,26,28,30 |
| Tbed1 | i3001500_1 | 300 | 1500 | 10,20,30,40,50 | 105,110,115,120,125 |
| Tbed1 | i3001500_6 | 300 | 1500 | 10,20,30,40,50 | 50,53,56,59,62 |
| Tbed1 | i500_1 | 500 | 500 | 10,20,30,40,50 | 70,73,76,79,82 |
| Tbed1 | i500_6 | 500 | 500 | 10,20,30,40,50 | 36,38,40,42,44 |
| Tbed1 | i700_1 | 700 | 700 | 10,20,30,40,50 | 95,100,105,110,115 |
| Tbed1 | i700_6 | 700 | 700 | 10,20,30,40,50 | 50,53,56,59,62 |
| Tbed1 | i1000_1 | 1000 | 1000 | 10,20,30,40,50 | 140,145,150,155,160 |
| Tbed1 | i1000_6 | 1000 | 1000 | 10,20,30,40,50 | 72,76,80,84,88 |
| Geo | ZY | 105 | 324 | 10,11,12,13,14 | 10,11,12,13,14 |
| Geo | GY | 135 | 1276 | 22,24,26,28,30 | 22,24,26,28,30 |
| Geo | KF | 146 | 2999 | 18,20,22,24,26 | 18,20,22,24,26 |
| Geo | ZZ | 320 | 6752 | 48,52,56,60,64 | 48,52,56,60,64 |

The efficiency and equality of the solutions from the benchmark instances are evaluated in terms of mean, standard deviation (SD), mean absolute deviation (MAD), and the Gini coefficient (GC) between the travel distances. Let $d_j = \sum_{i\in I} d_{ij}x_{ij}$ be the distance from location $j$ to its facility, and $\overline{d} = \sum_{i\in I}\sum_{j\in J} w_j d_{ij} x_{ij} / \sum_{j\in J} w_j$ be the mean travel distance, the SD, MAD, and GC can be calculated as follows. Note that the demand weights are considered in the three formulas.

$$SD = \sqrt{\sum_{j\in J} w_j\left(d_j - \overline{d}\right)^2 / \sum_{j\in J} w_j} \tag{14}$$

$$MAD = \sum_{j\in J} w_j \left|d_i - \overline{d}\right| / \sum_{j\in J} w_j \tag{15}$$

$$GC = \sum_{i\in J}\sum_{j\in J} w_i w_j \left|d_i - d_j\right| / (2\sum_{j\in J} w_j * \sum_{j\in J} w_j d_j) \tag{16}$$

## 4.2 Computational results

Six problems, the PMP, MELP, MDELP, CPMP, CMELP, and CMDELP, were solved for each instance with each parameter P, the number of facilities. The detailed solutions from the 17 benchmark instances are shown in Table S1 ~ Table S17 in the Supplementary Material (https://github.com/yfkong /PMPequality/equalPMP_supplementary_material.docx). The objective values, optimal gaps, and computational times (in seconds) are shown in columns ‘Objective’, ‘Gap’ and ‘Time’, respectively. The ‘opt’ in column ‘Gap’ indicates the corresponding instance was solved optimally by the Gurobi optimizer. The mean, standard deviation, mean absolute deviation, and the Gini coefficient between the travel distances are shown in columns ‘Mean’, ‘SD’, ‘MAD’ and ‘Gini’, respectively.

There are some findings from the solutions. First, according to solution times, the uncapacitated problems (PMP, MDELP and MELP) are relatively easier to solve than the capacitated problems (CPMP, CMDELP and CMELP); and it is harder to optimize the envy objective than the travel distance objective. Second, statistics on the travel distances from customers to their facilities show that the equality measures, especially the standard deviation and mean absolute deviation, are significantly reduced by the MDELP (CMDELP) and the MELP (CMELP), while the spatial accessibility in terms of mean distance is increased in an acceptable manner. Third, according to the model properties shown in Section 3.2, the lower and upper bounds on the mean distance and standard deviation for each instance can be obtained by solving the PMP (CPMP) and the MELP (CMELP).

Table 2: Statistics on the distance increase and the deviation decrease

| Instance | MDELP | | CMDELP | |
|---|---|---|---|---|
| | dM (dMmax) | dSD (dSMmax) | dM (dMmax) | dSD (dSMmax) |
| Capa1 | 0.90% (1.24%) | -6.22% (-6.40%) | 0.90% (1.24%) | -6.22% (-6.40%) |
| Capb1 | 1.08% (2.61%) | -6.62% (-8.46%) | 1.22% (2.72%) | -7.04% (-8.61%) |
| Capc1 | 1.31% (1.58%) | -7.90% (-8.19%) | 1.24% (1.55%) | -7.93% (-8.30%) |
| i300_1 | 2.94% (4.66%) | -17.07% (-20.05%) | 2.90% (4.46%) | -13.10% (-15.27%) |
| i300_6 | 3.49% (5.09%) | -14.24% (-16.77%) | 2.01% (3.93%) | -9.67% (-12.59%) |
| i3001500_1 | 1.56% (2.10%) | -8.98% (-9.78%) | 1.55% (2.37%) | -7.04% (-8.45%) |
| i3001500_6 | 1.36% (1.91%) | -5.99% (-7.30%) | 1.91% (2.42%) | -7.82% (-8.57%) |
| i500_1 | 3.24% (4.34%) | -15.07% (-16.68%) | 4.63% (5.88%) | -17.43% (-19.01%) |
| i500_6 | 3.05% (3.81%) | -16.50% (-18.07%) | 3.50% (4.16%) | -16.19% (-17.15%) |
| i700_1 | 3.11% (4.15%) | -16.72% (-18.43%) | 4.26% (5.22%) | -13.76% (-15.90%) |
| i700_6 | 2.54% (3.18%) | -14.12% (-14.90%) | 3.31% (4.24%) | -12.88% (-14.18%) |
| i1000_1 | 2.86% (3.11%) | -14.04% (-14.56%) | 3.30% (5.32%) | -17.75% (-20.81%) |
| i1000_6 | 2.00% (2.96%) | -12.90% (-14.73%) | 2.94% (4.94%) | -12.84% (-15.17%) |
| GY | 5.42% (8.73%) | -14.52% (-18.28%) | 4.30% (13.43%) | -11.61% (-21.58%) |
| ZY | 3.10% (3.26%) | -14.10% (-14.30%) | 3.42% (3.69%) | -13.76% (-13.87%) |
| KF | 4.01% (5.80%) | -12.48% (-14.17%) | 3.70% (5.93%) | -12.09% (-14.08%) |
| ZZ | 3.25% (4.12%) | -14.06% (-15.54%) | 1.75% (3.66%) | -8.28% (-10.71%) |
| Average | 2.66% (3.69%) | -12.44% (-13.92%) | 2.76% (4.42%) | -11.49% (-13.57%) |

The spatial accessibility and the spatial equality are effectively balanced by the weighted objective function. Compared with the PMP (CPMP) solutions, the standard deviation of travel distances in the MDELP (CMDELP) solutions is decreased significantly, while the mean distance is increased slightly. Table 2 shows the statistics on the increase of the mean distance and the decrease of standard deviation. The column ‘dM’ presents the percentage of the increase of the mean distance; and the column ‘dSD’ presents the percentage of the decrease of the standard deviation. Note that $dM = (M_{MDELP} - M_{PMP})/M_{PMP} * 100\%$ and $dSD = (SD_{MDELP} - SD_{PMP})/SD_{PMP} * 100\%$, where $M$ refers to the mean distance and $SD$ refers to the standard deviation. The bound values (dMmax and dSMmax) are also shown in their following parentheses.

In table 2, since the instances vary considerably in terms of demand, supply, number of facilities,

and the data generation method, the changes in travel distance and its standard deviation vary considerably. Nevertheless, the service equality can be improved in terms of the standard deviation of travel distance for all the instances. It is also observed that, for the instances capa1, capb1, capc1, i3001500_1 and i3001500_6, the dMmax and dSDmax values are relatively small. However, the dMmax and dSDmax values for the instance GY, a rural area with clustered distribution of service demand, are relatively large.

The solutions from the geographical instances can be visualized in maps. Figure 1 shows the PMP solution (left) and the MDELP solution (right) from the instance KF (P=20). The black crosses indicate the facility locations; and the colored polygons show the service areas of the facilities. It can be seen that the two solutions are different in facility locations and their service areas. For the PMP solution, the mean distance, standard deviation, mean absolute deviation, and Gini coefficient are 0.787 km, 0.434 km, 0.338 km and 0.303, respectively. Nevertheless, for the MDELP solution, the four measures are 0.821 km, 0.381 km, 0.304 km, and 0.260, respectively. Compared with the PMP solution, the MDELP increases the mean distance by 4.3%, and reduces the standard deviation, mean absolute deviation, and Gini coefficient by 12.2%, 10.1%, and 14.2%, respectively.

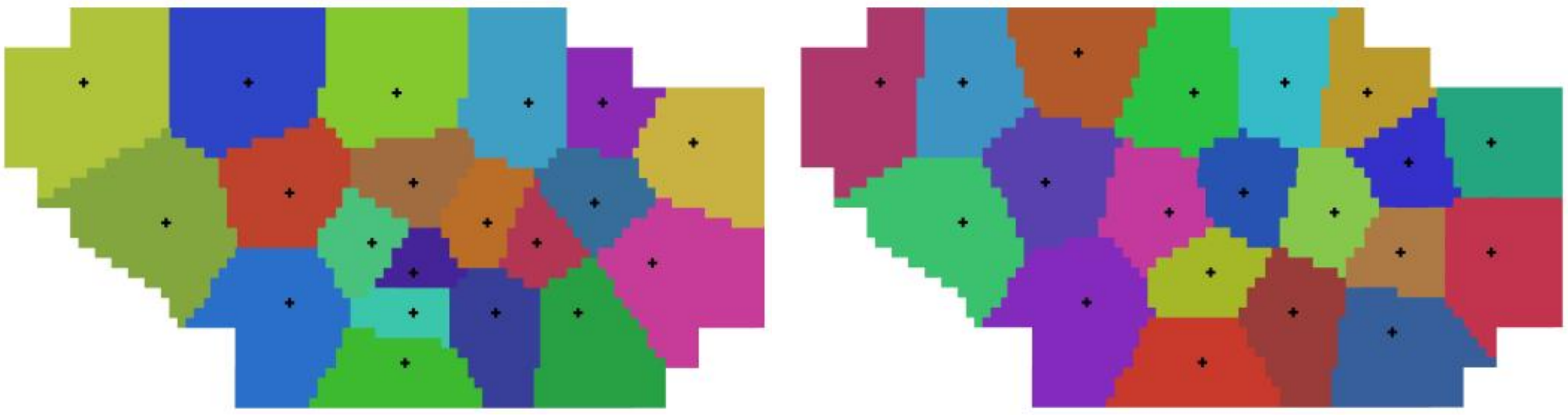

Figure 1: The PMP (left) solution and the MDELP (right) solution from the instance KF (P=20)

Figure 2 shows the CPMP solution and the CMDELP solution from the instance ZZ (P=52). It can also be observed that the two solutions are different in facility locations and their service areas. For the CPMP solution, the mean distance, standard deviation, mean absolute deviation, and the Gini coefficient are 0.815 km, 0.452 km, 0.355 km, and 0.305, respectively. Nevertheless, for the CMDELP solution, the four measures are 0.828 km, 0.415 km, 0.329 km, and 0.279, respectively. Compared with the CPMP solution, the CMDELP is able to increase the mean distance by 1.6%, and reduce the standard deviation, mean absolute deviation, and Gini coefficient by 8.2%, 7.3%, and 8.5%, respectively.

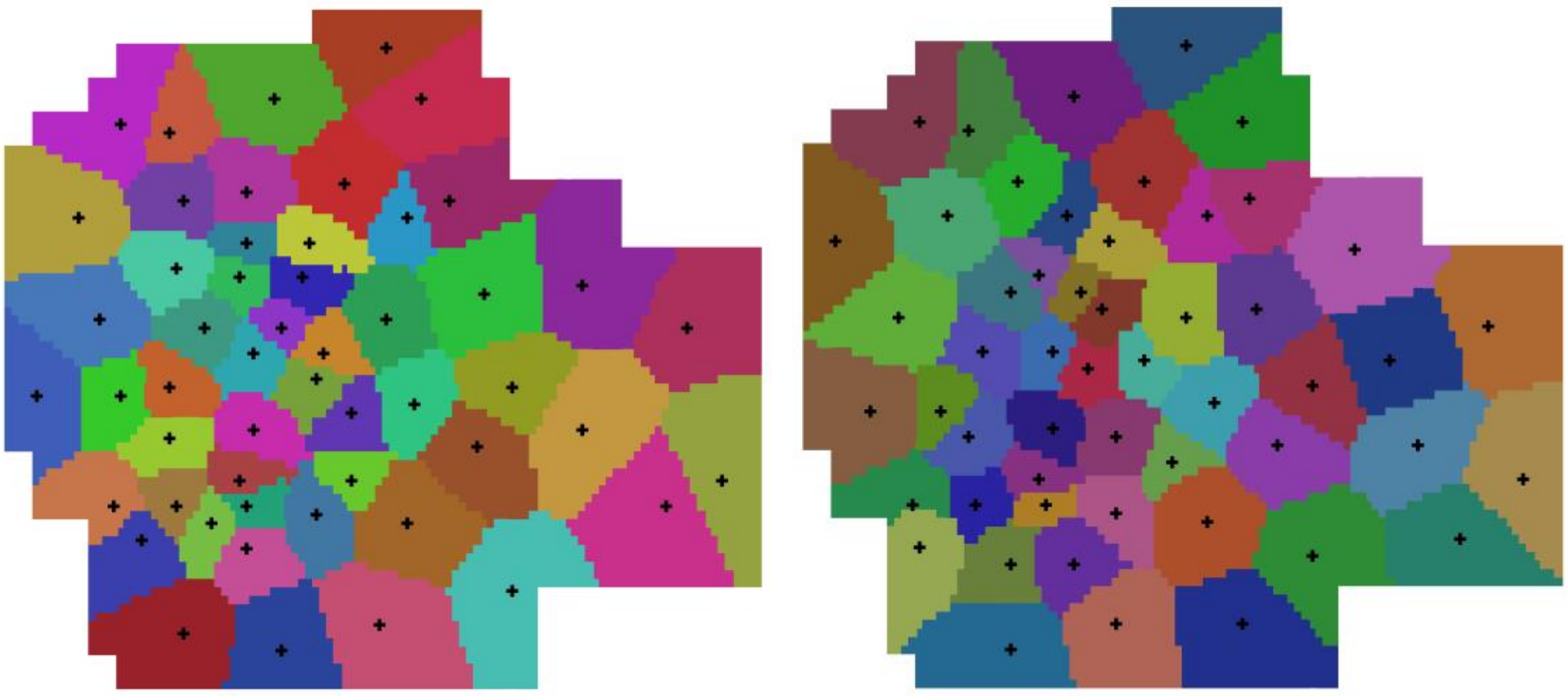

Figure 2: The CPMP solution (left) and the CMDELP solution (right) from the instance ZZ (P=52)

## 4.3 Sensitive analysis of the model parameters

The effectiveness of the MDELP (CMDELP) depends on the model parameters $d^*$ and $\beta$. If $d^*$ is large enough, the MDELP (CMDELP) solution will be the same as the PMP (CPMP) solution. If $\beta$ is large enough, the MDELP (CMDELP) will be equivalent to the MELP (CMELP). When the equality of service is preferred, the MELP (CMELP) could be the best choice to maximize the spatial equality.

In order to test the sensitivity of parameters $d^*$ and $\beta$, we selected five instances: GY (P=24), ZY (P=12), KF (P=20), i300_6 (P=30), and i500_6 (P=40). For each instance, the baseline parameters were set according to its PMP (CPMP) solution: $d_0 \approx \bar{d}$, and $\beta_0 \approx 2\bar{d}/\delta^2$, where $\bar{d}$ is the mean distance and $\delta$ is the standard deviation. The sensitivity of parameters $\beta$ was tested by setting $d^* \approx d_0$ and $\beta \approx 0.25\beta_0$, $0.5\beta_0$, $\beta_0$, $1.5\beta_0$, and $2.0\beta_0$. The effect of parameter $\beta$ was evaluated by comparing the means and the standard deviations. Similarly, the sensitivity of parameters $d^*$ was tested by setting $\beta = \beta_0$ and $d^* \approx 0.8d_0, 0.9d_0, d_0, 1.1d_0$, and $1.2d_0$.

Table 3: Solutions from selected instances with different values of the parameter $\beta$

| | | MDELP | | | | | | CMDELP | | | | | |
|---|---|---|---|---|---|---|---|---|---|---|---|---|---|
| Inst | P | $d^*$ | $\beta$ | Mean | SD | dM | dSD | $d^*$ | $\beta$ | Mean | SD | dM | dSD |
| GY | 24 | 1.82 | 0.5 | 1.859 | 1.262 | 2.03% | -6.88% | 1.87 | 0.5 | 1.924 | 1.357 | 2.59% | -8.33% |
| GY | 24 | 1.82 | 1.0 | 1.893 | 1.179 | 3.91% | -12.98% | 1.87 | 1.0 | 1.940 | 1.343 | 3.49% | -9.23% |
| GY | 24 | 1.82 | 2.0 | 1.922 | 1.159 | 5.48% | -14.44% | 1.87 | 2.0 | 1.973 | 1.317 | 5.24% | -11.04% |
| GY | 24 | 1.82 | 3.0 | 1.922 | 1.159 | 5.48% | -14.44% | 1.87 | 3.0 | 1.991 | 1.300 | 6.17% | -12.19% |
| GY | 24 | 1.82 | 4.0 | 1.938 | 1.150 | 6.37% | -15.10% | 1.87 | 4.0 | 2.009 | 1.271 | 7.13% | -14.12% |
| ZY | 12 | 0.39 | 5.0 | 0.402 | 0.191 | 0.93% | -7.56% | 0.39 | 5.0 | 0.403 | 0.193 | 1.20% | -6.99% |
| ZY | 12 | 0.39 | 10.0 | 0.413 | 0.174 | 3.77% | -15.83% | 0.39 | 10.0 | 0.414 | 0.175 | 4.00% | -15.70% |
| ZY | 12 | 0.39 | 20.0 | 0.413 | 0.174 | 3.77% | -15.83% | 0.39 | 20.0 | 0.414 | 0.175 | 4.00% | -15.70% |
| ZY | 12 | 0.39 | 30.0 | 0.413 | 0.174 | 3.77% | -15.83% | 0.39 | 30.0 | 0.414 | 0.175 | 4.00% | -15.70% |
| ZY | 12 | 0.39 | 40.0 | 0.413 | 0.174 | 3.77% | -15.83% | 0.39 | 40.0 | 0.414 | 0.175 | 4.00% | -15.70% |
| KF | 20 | 0.79 | 2.5 | 0.803 | 0.395 | 2.09% | -9.07% | 0.80 | 2.5 | 0.804 | 0.395 | 2.10% | -9.06% |
| KF | 20 | 0.79 | 5.0 | 0.808 | 0.391 | 2.71% | -10.02% | 0.80 | 5.0 | 0.808 | 0.391 | 2.73% | -10.02% |
| KF | 20 | 0.79 | 10.0 | 0.821 | 0.381 | 4.37% | -12.21% | 0.80 | 10.0 | 0.822 | 0.381 | 4.39% | -12.21% |
| KF | 20 | 0.79 | 15.0 | 0.821 | 0.381 | 4.37% | -12.21% | 0.80 | 15.0 | 0.822 | 0.381 | 4.39% | -12.21% |
| KF | 20 | 0.79 | 20.0 | 0.821 | 0.381 | 4.37% | -12.21% | 0.80 | 20.0 | 0.822 | 0.381 | 4.39% | -12.21% |
| i300_6 | 30 | 0.57 | 3.2 | 0.585 | 0.260 | 1.07% | -10.86% | 0.60 | 3.2 | 0.616 | 0.268 | 1.31% | -12.58% |
| i300_6 | 30 | 0.57 | 6.5 | 0.587 | 0.258 | 1.46% | -11.67% | 0.60 | 6.5 | 0.618 | 0.267 | 1.64% | -13.03% |
| i300_6 | 30 | 0.57 | 13.0 | 0.601 | 0.243 | 3.80% | -16.63% | 0.60 | 13.0 | 0.618 | 0.267 | 1.64% | -13.03% |
| i300_6 | 30 | 0.57 | 20.0 | 0.601 | 0.243 | 3.80% | -16.63% | 0.60 | 20.0 | 0.618 | 0.267 | 1.64% | -13.03% |
| i300_6 | 30 | 0.57 | 26.0 | 0.601 | 0.243 | 3.80% | -16.63% | 0.60 | 26.0 | 0.629 | 0.258 | 3.40% | -15.94% |
| i500_6 | 40 | 0.52 | 3.5 | 0.537 | 0.238 | 1.38% | -12.03% | 0.53 | 3.2 | 0.565 | 0.249 | 2.79% | -14.32% |
| i500_6 | 40 | 0.52 | 7.0 | 0.544 | 0.227 | 2.65% | -15.96% | 0.53 | 6.5 | 0.568 | 0.244 | 3.30% | -16.27% |
| i500_6 | 40 | 0.52 | 14.0 | 0.553 | 0.219 | 4.33% | -18.86% | 0.53 | 13.0 | 0.571 | 0.241 | 3.75% | -17.14% |
| i500_6 | 40 | 0.52 | 21.0 | 0.555 | 0.217 | 4.75% | -19.57% | 0.53 | 20.0 | 0.571 | 0.241 | 3.84% | -17.14% |
| i500_6 | 40 | 0.52 | 28.0 | 0.555 | 0.217 | 4.75% | -19.57% | 0.53 | 26.0 | 0.571 | 0.241 | 3.84% | -17.14% |

Table 3 shows the solution results from the five instances with different values of the parameter $\beta$. In the table, columns 'Mean' and 'SD' show the mean distance and the standard deviation, respectively; columns 'dM' and 'dSD' indicate the change of the mean and the standard deviation compared with the PMP (CPMP) solution. There are two finding from Table 3. First, the spatial equality of all the solutions is improved substantially. Second, for $\beta \geq 0.5\beta_0$, the mean instance remains same or increases slightly, and the standard deviation remains same or decreases slightly. The solution changes with the parameter $\beta$ can be illustrated in Figure S1 and Figure S2 in the Supplementary Material. Table 3, Figure S1 and Figure S2 indicate that the MDELP and CMDELP solutions are not sensitive to the parameter $\beta$ when $\beta \approx [0.5\beta_0, 2.0\beta_0]$.

Table 4: Solutions from selected instances with different values of the parameter $d^*$

| | | MDELP | | | | | | CMDELP | | | | | |
|---|---|---|---|---|---|---|---|---|---|---|---|---|---|
| Inst | P | $d^*$ | $\beta$ | Mean | SD | dM | dSD | $d^*$ | $\beta$ | Mean | SD | dM | dSD |
| GY | 24 | 1.40 | 2.5 | 1.904 | 1.162 | 4.49% | -14.22% | 1.50 | 2.5 | 1.937 | 1.360 | 3.33% | -8.14% |
| GY | 24 | 1.60 | 2.5 | 1.922 | 1.159 | 5.48% | -14.44% | 1.70 | 2.5 | 1.952 | 1.340 | 4.10% | -9.44% |
| GY | 24 | 1.80 | 2.5 | 1.922 | 1.159 | 5.48% | -14.44% | 1.90 | 2.5 | 1.995 | 1.294 | 6.38% | -12.56% |
| GY | 24 | 2.00 | 2.5 | 1.938 | 1.150 | 6.37% | -15.10% | 2.10 | 2.5 | 1.974 | 1.317 | 5.25% | -11.04% |
| GY | 24 | 2.20 | 2.5 | 1.938 | 1.150 | 6.37% | -15.10% | 2.30 | 2.5 | 1.982 | 1.316 | 5.70% | -11.09% |
| ZY | 12 | 0.36 | 20.0 | 0.413 | 0.174 | 3.77% | -15.83% | 0.36 | 20.0 | 0.414 | 0.175 | 4.00% | -15.70% |
| ZY | 12 | 0.38 | 20.0 | 0.413 | 0.174 | 3.77% | -15.83% | 0.38 | 20.0 | 0.414 | 0.175 | 4.00% | -15.70% |
| ZY | 12 | 0.40 | 20.0 | 0.413 | 0.174 | 3.77% | -15.83% | 0.40 | 20.0 | 0.417 | 0.181 | 4.85% | -13.18% |
| ZY | 12 | 0.42 | 20.0 | 0.413 | 0.174 | 3.77% | -15.83% | 0.42 | 20.0 | 0.417 | 0.181 | 4.85% | -13.18% |
| ZY | 12 | 0.44 | 20.0 | 0.412 | 0.177 | 3.54% | -14.26% | 0.44 | 20.0 | 0.417 | 0.181 | 4.87% | -13.18% |
| KF | 20 | 0.63 | 10.0 | 0.807 | 0.391 | 2.54% | -9.82% | 0.66 | 10.0 | 0.809 | 0.389 | 2.77% | -10.28% |
| KF | 20 | 0.71 | 10.0 | 0.818 | 0.382 | 3.96% | -12.06% | 0.74 | 10.0 | 0.822 | 0.381 | 4.39% | -12.21% |
| KF | 20 | 0.79 | 10.0 | 0.821 | 0.381 | 4.37% | -12.21% | 0.82 | 10.0 | 0.809 | 0.395 | 2.78% | -9.09% |
| KF | 20 | 0.87 | 10.0 | 0.822 | 0.381 | 4.41% | -12.19% | 0.90 | 10.0 | 0.822 | 0.381 | 4.43% | -12.19% |
| KF | 20 | 0.95 | 10.0 | 0.822 | 0.381 | 4.41% | -12.19% | 0.98 | 10.0 | 0.822 | 0.381 | 4.43% | -12.19% |
| i300_6 | 30 | 0.46 | 13.0 | 0.593 | 0.252 | 2.38% | -13.79% | 0.48 | 13.0 | 0.619 | 0.262 | 1.79% | -14.72% |
| i300_6 | 30 | 0.51 | 13.0 | 0.601 | 0.243 | 3.80% | -16.63% | 0.54 | 13.0 | 0.621 | 0.262 | 2.06% | -14.68% |
| i300_6 | 30 | 0.57 | 13.0 | 0.601 | 0.243 | 3.80% | -16.63% | 0.60 | 13.0 | 0.618 | 0.267 | 1.64% | -13.03% |
| i300_6 | 30 | 0.63 | 13.0 | 0.601 | 0.243 | 3.80% | -16.63% | 0.66 | 13.0 | 0.618 | 0.267 | 1.64% | -13.03% |
| i300_6 | 30 | 0.68 | 13.0 | 0.601 | 0.243 | 3.80% | -16.63% | 0.72 | 13.0 | 0.625 | 0.259 | 2.72% | -15.50% |
| i500_6 | 40 | 0.42 | 13.0 | 0.546 | 0.223 | 3.11% | -17.58% | 0.42 | 13.0 | 0.570 | 0.241 | 3.64% | -17.07% |
| i500_6 | 40 | 0.47 | 13.0 | 0.546 | 0.223 | 3.11% | -17.58% | 0.48 | 13.0 | 0.570 | 0.241 | 3.64% | -17.07% |
| i500_6 | 40 | 0.52 | 13.0 | 0.553 | 0.219 | 4.33% | -18.86% | 0.53 | 13.0 | 0.571 | 0.241 | 3.75% | -17.14% |
| i500_6 | 40 | 0.57 | 13.0 | 0.553 | 0.219 | 4.33% | -18.86% | 0.58 | 13.0 | 0.571 | 0.241 | 3.84% | -17.21% |
| i500_6 | 40 | 0.62 | 13.0 | 0.548 | 0.229 | 3.39% | -15.15% | 0.64 | 13.0 | 0.570 | 0.244 | 3.68% | -16.29% |

Table 4 shows the solution results from the five instances with different values of the parameter $d^*$. In the table, columns 'Mean' and 'SD' show the mean distance and the standard deviation, respectively; columns 'dM' and 'dSD' indicate the changes of the mean distance and the standard deviation compared with the PMP (CPMP) solution. Table 4 show that the spatial equality of all the solutions is improved substantially. At the same time, for $d^* \approx 0.8d_0{\sim}1.2d_0$, the mean distance remains same or changes slightly, and the standard deviation remains same or changes slightly. This indicates that problem solution is not sensitive to the parameter $d^*$ when it is estimated with an error less than 20%. Figure S3 and Figure S4 in the Supplementary Material show the changes of the mean distance and the standard deviation along with the parameter $d^*$. The figures also illustrate that the problem solutions are not sensitive to the parameter $d^*$ when $d^* \approx 0.8d_0 \sim 1.2d_0$.

## 4.4 Pareto optimality analysis

The Pareto optimality of MELP and MDELP solution is investigated by case studies on 13 instances: ZY (P=10, 12, and 14), Cap121 (P=4, 6, 8, 10, and 12) and Pmed1~Pmed5 (https:// people.brunel.ac.uk/ ~mastjjb/jeb/orlib/pmedinfo.html). A mean-variance bi-objective location problem is formulated in Appendix 3 in the Supplementary Material, and solved by the ϵ-constrained method with constraint on mean distance: $d' - \epsilon \leq \sum_{i \in I} \sum_{j \in J} w_j d_{ij} x_{ij} / \sum_{j \in J} w_j \leq d' + \epsilon$. For the real-world instance ZY with $|I|$=105, $|J|$=324, and $I \subset J$, a set of Pareto optimal solutions can be obtained in about 4 hours; for the instance Cap121 with $|I|$=50, $|J|$=50, and $I \neq J$, a set of Pareto optimal solutions can be obtained in about 5~7 minutes; and for instances Pmed1~Pmed5 with $|I|$=100, $|J|$=100, $I = J$, and $w_j$=1, a set of Pareto optimal solutions can be obtained in about 1~2 hours.

Pareto frontiers, MDELP and MELP solutions from the 13 instances are shown in Figure 3. The figure shows that almost all the MDELP and MELP solutions are Pareto optimal. All the MELP solutions, except the Cap121 (P=4), and all the MDELP solutions, except the Cap121 (P=4 and 5), are located at middle points of their Pareto frontiers, where the mean travel distance and the standard deviation of distances are well balanced.

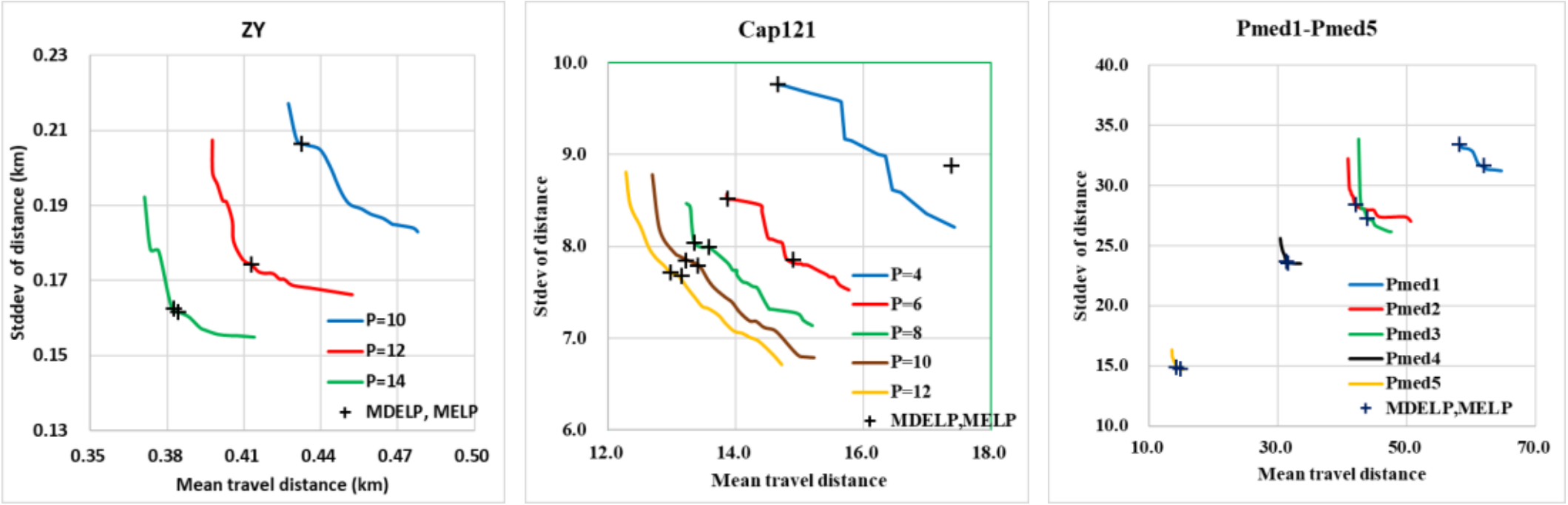

Figure 3: Pareto frontiers, and MDELP/MELP solutions from instances ZY, Cap121, and Pmed1~5

## 4.5 Service supply, spatial access and spatial equality

In public service planning, whether the service efficiency and equality can both be achieved? Can the spatial equality of service be improved by the increase in service supply? What are the relationships between service supply, spatial access and spatial equality? Answers on these questions are critical for the general public and the decision makers.

First of all, when the service supply is given in terms of the number of facilities, the service efficiency and equality can be balanced by the objective function of weighed distance and envy. Model properties illustrate that the distance and spatial envy can be balanced by the weight parameter in the objective function. Table 2 shows that the standard deviation obtained by the MDELP is averagely decreased by 12.44% compared with the PMP solutions, while the mean distance is averagely increased by 2.66%. For the capacitated problem, CMDELP, the standard deviation is averagely decreased by 11.49% compared with the CPMP solutions, while the mean distance is increased by 2.76%.

Second, solution results from three sets of instances also indicate that when the service supply is increased in terms of the number of facilities, both the service efficiency and equality can be improved. Figure S5 and Figure S6 in the Supplementary Material show the relationships between the mean distance, the standard deviation, and the number of facilities from six instances: GY, ZY, KF, ZZ, i300_1, and i700_1. For both the MDELP solutions and the CMDELP solutions, a downtrend of the mean distance and the standard deviation occurs when the supply is increased in terms of the number of facilities. This trend is also validated by the Pareto frontiers from instances ZY and Cap121, as illustrated in Figure 3.

## 4.6 Comparison with other approaches

The PCP, which minimizes the worst-off travel cost, has been widely considered in public sector applications. Therefore, the MDELP solutions from selected instances were compared with the PCP solutions. Table 5 shows all the MDELP and PCP solutions in terms of maximum (Max), mean (Mean), standard deviation (SD), the Gini coefficient (Gini) of the travel distances, and computational time in seconds. The solutions from selected instances indicate that the maximum travel distances can be significantly improved by the PCP, but at the same time, the mean travel distances are increased substantially, especially for the instances i500_6, GY, and ZY. Compared with the PMP solutions, the MDELP can robustly balance the service efficiency and equality, while the PCP can only minimize the worst distance without considering the average travel distance and the service equality. Table 5 also indicates that the solution time of the PCP is longer than that of the MDELP.

Table 5: Comparison of MDELP solutions with PCP solutions

| Inst. | P | Model | Max | Mean | SD | Time | Model | Max | Mean | SD | Time |
|---|---|---|---|---|---|---|---|---|---|---|---|
| Capa1 | 10 | MDELP | 286.0 | 139.7 | 54.51 | 31.4 | PCP | 266.3 | 145.1 | 58.39 | 44.5 |
| Capb1 | 12 | MDELP | 271.3 | 125.3 | 52.27 | 55.9 | PCP | 239.5 | 129.5 | 52.67 | 67.1 |
| i300_1 | 40 | MDELP | 1.315 | 0.542 | 0.213 | 14.1 | PCP | 0.922 | 0.551 | 0.231 | 14.8 |
| i500_6 | 50 | MDELP | 1.200 | 0.480 | 0.204 | 25.5 | PCP | 0.854 | 0.517 | 0.206 | 81.1 |
| GY | 22 | MDELP | 6.372 | 1.999 | 1.210 | 22.9 | PCP | 4.879 | 2.549 | 1.194 | 25.5 |
| ZY | 10 | MDELP | 1.053 | 0.432 | 0.206 | 17.6 | PCP | 0.879 | 0.469 | 0.199 | 19.7 |

The advantages of MDELP and MELP are further examined by comparing the MDELP/MELP with a number of most commonly used location models. First of all, four groups of location problems were selected: (1) the PMP; (2) four variants of the ordered median problem, such as the p-center, k-centrum, $\alpha$-centdian, and k-centdian problems; (3) the minimum inequity location problems (MinIneq) that minimize of an inequity indicator, such as the standard deviation (MinSD), the mean absolute deviation (MinMAD), the mean absolute difference (MinAD), the coefficient of variance (MinCV), the Schutz indicator (MinSI), or the Gini coefficient (MinGC); and (4) the mean-variance bi-objective location problem. All the problems are mathematically defined in Appendix 3 in the Supplementary Material. Second, three instances with different sizes of demand points were chosen for model comparative analysis: Cap121 ($|I|$=50, $|J|$=50, $I \neq J$ and $P$=8), Pmed2 ($|I|$=100, $|J|$=100, $I = J$, $w_j$=1, and $P$ =10), and ZY ($|I|$=105, $|J|$=324, $I \subset J$ and $P$=12). Third, each instance was solved using different location models. Finally, the location models were evaluated in terms of computational time and Pareto optimality.

Various model solutions from instances ZY, Cap121, and Pmed2 are shown in Table S18, Table S19 and Table S20 in the Supplementary Material, respectively. The solution indicators such as the maximum distance (Max), mean distance (Mean), standard deviation (SD), mean absolute deviation (MAD), mean absolute deviation (AD), Schutz's index (SI), coefficient of variance (CV), Gini coefficient (GC), and solution time (Time), are shown in these table. The empty values mean that the model cannot be optimally or near-optimally solved by Gurobi Optimizer 9 in 4 hours. In addition, Table S21 shows the mean-variance Pareto optimal solutions from the three instances.

There are three findings from the solution results. First of all, among the selected location problems, the PMP, the MDLP, the MDELP and the PCP are relatively easy to solve. Of course, this does not mean that the computational complexity of these problems is low. The solution times of the k-centrum, $\alpha$-centdian, and k-centdian problems vary substantially, depending on the parameters $k$ and $\alpha$. Meanwhile, the minimum inequity location problems such as MinMAD, MinSI, MinAD, MinGC, MinSD, and MinCV are extremely difficult to solve, especially for the instances with more than 100 demand points.

Second, the optimal solutions of selected location problems vary substantially in terms of the Mean, Max, SD, MAD, AD, CV, SI, and GC. For example, the mean of distance of Cap121 changes from 13.225 to 35.103; the standard deviation of distance changes from 3.820 to 10.994; and the Gini coefficient changes from 0.071 to 0.347. The minimum values of Max, Mean, SD, MAD, AD, SI, CV, and GC can be obtained by location models PCP, PMP, MinSD, MinMAD, MinAD, MinSI, MinCV, and MinGC, respectively. The minimum inequity problems, especially the MinSI, MinCV, and MinGC, always seriously distort efficiency in terms of the mean distance.

Third, the chosen problems exhibit significant variation in Pareto optimality, shown in Figure 4. The PMP solution is positioned at the starting point of the Pareto frontier, while the MinSD solution is at the endpoint. The MDELP and MELP solutions are situated at turning points of the Pareto frontier. Unlike the MDELP and MELP solutions, most PCP, k-centrum, $\alpha$-centdian, and k-centdian solutions are not Pareto optimal or Pareto near-optimal. It is also observed that in the starting part of the Pareto frontier of instance Pmed and ZY, the travel distance is increased slightly while the standard deviation is decreased significantly. This means that more equitable solutions can be achieved with little sacrifice in efficiency.

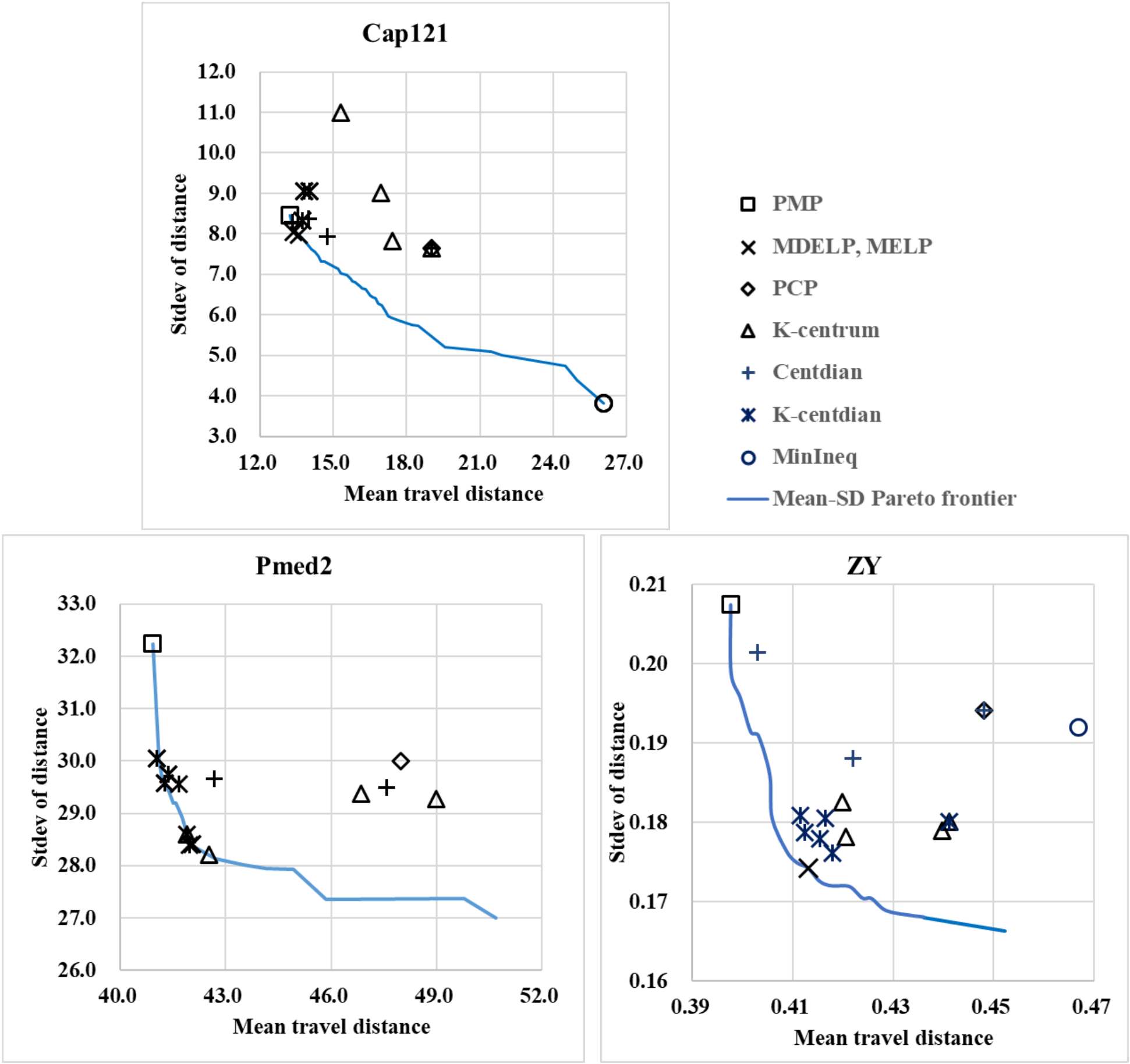

Figure 3: The position of various model solutions on mean-standard deviation space

Some special cases of the ordered median problem have been used to obtain equitable service solutions. Filippi et al. (2021a) formulated a fair facility location problem (FFLP), in which the service demand is set to 1 at each demand location, and the capacity of each facility is assumed to be unlimited. An average time of 6,463 seconds is needed to optimally or near-optimally solve one of the instances with $|\mathrm{I}| = |\mathrm{J}| = 100 \sim 900$ and $\mathrm{P} = 5 \sim 140$ using CPLEX 12.8. Considering the computational complexity of the ordered weighted objective function, it is impossible to solve large instances with varied demands and capacitated facilities. The $k$-centdian problem proposed in this article, a generalized version of the FFLP, performs differently on different instances: low-quality solutions for the instance Cap121, modest-quality solutions for the instance ZY, and Pareto optimal solutions for the instance Pmed2. Obviously, the MDELP and DELP outperform the k-centdian problem in terms of solution quality and computation time.

# 5 Discussion

In this article, the concept of envy is quantified in a manner distinct from that in existing literature. The envy functions in Espejo et al. (2009), Chanta et al. (2011), and Chanta et al. (2014) are defined based on the comparison of any two customers. Like many other equity problems, the envy function has the disadvantage that absolute values must be calculated. Consequently, the computational complexity of these location problems is very high. In addition, such envy functions do not satisfy the property of monotonicity and thus are not inequity-averse. We simplify the concept of envy by comparing an individual's travel distance to a service facility against a threshold distance, such as the mean distance or a preferred distance. As a result, our envy function is not only inequity-averse but also relatively easy to solve. For example, the minimum-envy solution times range from 1568 to 2653 seconds for the benchmark instances with $|I| = |J| = 40$ and $P = 8$ (Espejo et al., 2009). While for the instance i300 with $|I| = |J| = 300$, and $P = 10 \sim 50$, the MELP and MDELP solution times range between 8.9 and 38.1 seconds. Obviously, the new definition of envy is simple and easy to solve.

The advantages of the proposed problems, such as their simplicity, effectiveness, and ease of solution, can be demonstrated through mathematical and comparative analysis. The PMP provide the most efficient but unequal solutions. The PCP can minimize the maximum travel distance, but without considering service efficiency and equality. All the objective functions of MinSD, MinMAD, MinAD, MinCV, MinSI and MinGC strictly or weakly satisfy the Pigou-Dalton principle of transfers, but does not satisfy the property of monotonicity. Consequently, the inequality measures can be minimized while the mean travel distance is usually seriously distorted. The objective functions of the $k$-centrum, $\alpha$-centdian, and $k$-centdian problems strictly or weakly satisfy the properties of monotonicity and Pigou-Dalton principle of transfers, and thus are considered as the most fair location models. However, the case studies indicate that the quality of solutions obtained from these problems depends on their problem parameters, and varies substantially in terms of Pareto optimality. In addition, these problems are hard to solve since the travel distances need to be ordered for calculating their objective values.

The proposed problems have potential in real-world applications. Firstly, it is observed that large instances and solved efficiently. For example, the MELP, MDELP, CMELP, and CMDELP on instances GY ($|I| = 135$, $|J| = 1276$ and $P = 30$) can be optimally solved by Gurobi Optimize in 24.1, 32.6, 3543.2 and 367.3 seconds, respectively. Using heuristic algorithms, the large instances can be solved much more efficiently with high-quality solutions. Second, experiments show that the MDELP solutions are effective: the service equality in terms of the standard deviation can be significantly improved by slightly increasing travel distance. The MELP solutions are more equitable. The CDMELP and CMELP can be used in the applications with constraints on facility capacities. Third, new users might be puzzled about setting model parameters. Theoretical analysis and experiments show that $d^* \approx \bar{d}$ and $\beta \approx 2d^*/\sigma^2$ are the best parameters. Therefore, it is suggested to solve the PMP (or CPMP) first and set the model parameters using the mean distance and its standard deviation from the PMP (or CPMP) solution. Sensitivity analysis indicates that the problem solutions are not very sensitive to the parameters, and thus it is relatively easy to set the model parameters. Considering the discussions above, we believe that the MELP, MDELP, CMELP, and CMDELP are useful in real-world public services.

The envy function (12) can be generalized as function (17), where $p > 0$.

$$f_{envy} = \sum_{i \in I} \sum_{j \in J} w_j \left(d_{ij} - d^*\right)_+^p x_{ij} \tag{17}$$

When $d^* = \bar{d}$ and $p = 1$, $f_{envy}$ is the sum of absolute upper semi-deviation. While $d^* = \bar{d}$ and $p = 2$, $f_{envy}$ is the sum of squared upper semi-variance. The $f_{envy}$ can be aggregated with the total travel distance, resulting in an inequity-averse function (18).

$$\text{Minimize:} \sum_{i \in I} \sum_{j \in J} w_j d_{ij} x_{ij} + \beta \sum_{i \in I} \sum_{j \in J} w_j \left(d_{ij} - d^*\right)_+^p x_{ij} \tag{18}$$

When $p = 1$, it is suggested $d^* \approx \bar{d}$ and $\beta \approx 2d^*/\sigma$ which will equally optimize two parts of the objective.

It's interesting to note that while most solutions are Pareto optimal, there are exceptions like the instance Cap121. As shown in Figure 3, the MELP solution from the instance Cap121 (P=4) is non-optimal, and the MDELP solution is the same as the PMP solution. This observation indeed raises an important question about the variability in performance across different instances. It might be useful to further investigate the characteristics of these instances to understand why certain solutions are non-optimal. Additionally, exploring the underlying factors that contribute to the performance differences could provide valuable insights for improving the robustness of these models. Further research on this topic could investigate the problem performance on different sets of benchmark instances. The design of the benchmark datasets should consider the real-world geographic environment, such as space (Euclidean or network), demand distribution (regular, random, clustered, and core-periphery), and demand variation (homogeneous or heterogeneous).

Furthermore, our problems are proposed based on classical deterministic location models, and thus do not consider uncertainties such as demand uncertainty and travel time uncertainty in service planning. Further research should investigate the possibility of adapting these deterministic problems to stochastic location problems.

# 6 Conclusion

In this article, four new location problems, DELP, MELP, CMDELP, and CMELP, are proposed for balancing service efficiency and equality. These problems are formulated based on the inequity-averse spatial envy function and offer several practical and theoretical advantages. First, experiments on three set of instances show that the four problems are effective to balance service efficiency and spatial equality. Second, the model solution is not sensitive to its parameters. Even if the parameters are estimated with considerable bias, service equality can be significantly improved. Third, according to the model properties, there exist the upper and lower bounds on the travel distance and the spatial envy. The bounds can be obtained by solving the PMP (CPMP) and the MELP (CMELP). Forth, the MDELP and the CMDELP are formulated as mixed integer linear programming models without any additional decision variables and constraints. Therefore, the MDELP instances with up to 1000 facilities and 1000 customers can be optimally solved by the state-of-the-art MIP optimizer; and the CMDELP instances with up to 500 facilities and 500 customers can be optimally solved. Finally, comprehensive comparison of solutions from selected instances using four groups of location problems suggests that the proposed problems outperform the existing problems in terms of service equality.

There are two important findings on the complicated relationships between the service supply, spatial access and spatial equality. First, when the service supply is given in terms of the number of facilities, the service equality in terms of the standard deviation can be significantly improved by slightly increasing the travel distance. Second, when the service supply is increased in terms of the number of facilities, both the service efficiency and spatial equality can be significantly improved. This means that, in public service systems, the spatial equality of service can be improved by spending more money on service supply.

# Disclosure statement

The authors declared no potential conflicts of interest with respect to the research, authorship, and/or publication of this article.

# Code availability

The code is available to download at https://github.com/yfkong/PMPequality.

# Acknowledgments

Research partially supported by the National Natural Science Foundation of China (No. 41871307).

# Supplementary material

# for

# Extended p-median problems for balancing service efficiency and equality

## Appendix 1: Proof of the inequity-averse property of MDELP

Let $I = \{1,2,\dots,n\}$ be a set of $n$ candidate facility locations, $J = \{1,2,\dots,m\}$ be a set of $m$ demand locations, each location $j$ $(j \in J)$ has service demand $w_j$. Variable $d_{ij}$ is the distance between locations $i$ and $j$. The travel distance vector of the demand locations is denoted $D = (d_1, d_2, \dots, d_m)$, where $d_j = \sum_{i \in I} d_{ij} x_{ij}$. The objective function (13) can be equivalently transformed into:

$$f(D) = \sum_{j \in J} w_j d_j + \beta \sum_{j \in J} w_j (d_j - d^*)_+^2$$

where weight parameter $\beta > 0$, $(d_j - d^*)_+ = max(0, d_j - d^*)$.

The objective function $f(D)$ satisfies the following conditions of inequity averse function:

***Condition* 1** (Anonymity): $f(D) = f(\prod(D))$, where$\prod(D)$ is an arbitrary permutation of the $D$ vector.

***Condition* 2** (Monotonicity): $f(y_1, y_2, \dots, y_n) < f(y_1, y_2, \dots y_i + \varepsilon, \dots, y_n)$, where $\varepsilon > 0$.

***Condition* 3** (Pigou–Dalton Principle of Transfers): $f(d_1, d_2, \dots, d_n) \geq f\big(d_1, d_2, \dots d_i + \varepsilon/w_i, \dots, d_j - \varepsilon/w_j, \dots d_n\big)$, where $0 < \varepsilon < \big(d_j - d_i\big) * \min(w_i, w_j)$.

### Appendix 1.1: Proof of Condition 1: Anonymity

The objective function $f(D)$ does not take into account special groups or individuals, so the ***Condition* 1** (anonymity) is naturally satisfied.

### Appendix 1.2 Proof of Condition 2: Monotonicity

Assuming $d_j \geq d^*$, it can be established that the inequality $f\big(d_1, \dots, d_j, \dots, d_m\big) < f\big(d_1, \dots, d_j + \varepsilon, \dots, d_m\big)$ is true.

Assuming $d_j < d^*$, we consider the following two scenarios:

If $d_j + \varepsilon \leq d^*$, the first term in function $f(D)$ experiences an increment while the second term remains unchanged. Consequently, the overall function value increases, and the aforementioned inequality remains valid.

If $d_j + \varepsilon > d^*$, both terms in the function $f(D)$ increase, thereby satisfying the inequality.

In summary, the function $f(D)$ satisfies the property of monotonicity.

**Appendix 1.3 Proof of Condition 3: Pigou–Dalton Principle of Transfers**

Let $D' = (d_1, d_2, \ldots d_i + \varepsilon/w_i, \ldots, d_j - \varepsilon/w_j, \ldots d_m)$, where $0 < \varepsilon < (d_j - d_i) * \min(w_i, w_j)$, it can be established that the inequalities $d_i + \varepsilon/w_i < d_j$ and $d_j - \varepsilon/w_j > d_i$. Given that the transfer from $d_j$ to $d_i$ do not affect the total travel distance, the first part of function $f(D)$, $\sum_{j\in J} w_j d_j$, can be neglected. As a result, the weight $\beta$ in the second section can be ignored in the proofs.

Based on the initial position relationship of $d_i$, $d_j$ and $d^*$, we consider six cases:

***Case1***: $d_i < d_j \le d^*$;

***Case2***: $d^* \le d_i < d_j$;

***Case3a***: $d_i < d^* < d_j$, $d_i + \varepsilon/w_i < d^*$ and $d_j - \varepsilon/w_j < d^*$;

***Case3b***: $d_i < d^* < d_j$, $d_i + \varepsilon/w_i < d^*$ and $d_j - \varepsilon/w_j \ge d^*$;

***Case3c***: $d_i < d^* < d_j$, $d_i + \varepsilon/w_i \ge d^*$ and $d_j - \varepsilon/w_j < d^*$; and

***Case3d***: $d_i < d^* < d_j$, $d_i + \varepsilon/w_i \ge d^*$ and $d_j - \varepsilon/w_j \ge d^*$.

For ***Case1***, since $d_i + \varepsilon/w_i < d_j \le d^*$ and $d_j - \varepsilon/w_j < d_j \le d^*$, the equality $f(D') = f(D)$ holds.

For ***Case2***,
$$\begin{aligned} f(D') - f(D) &= w_i(d_i + \varepsilon/w_i - d^*)^2 + w_j(d_j - \varepsilon/w_j - d^*)^2 \\ &\quad - w_i(d_i - d^*)^2 - w_j(d_j - d^*)^2 \\ &= \varepsilon^2/w_i + 2d_i\varepsilon + \varepsilon^2/w_j - 2d_j\varepsilon \\ &= \varepsilon(\varepsilon/w_i + 2d_i + \varepsilon/w_j - 2d_j) \\ &= \varepsilon(d_i + \varepsilon/w_i - d_j) + \varepsilon(d_i + \varepsilon/w_j - d_j) \\ &< 0. \end{aligned}$$

For ***Case3a***, $f(D') - f(D) = -w_j(d_j - d^*)^2 < 0$.

For ***Case3b***, due to $0 \le d_j - \varepsilon/w_j - d^* < d_j - d^*$, we have:
$$f(D') - f(D) = w_j(d_j - \varepsilon/w_j - d^*)^2 - w_j(d_j - d^*)^2 < 0.$$

For ***Case3c***, depending on the relationship between $w_i$ and $w_j$, ***Case3c*** is divided into the following two scenarios:

If $w_i \ge w_j$: Since $0 < \varepsilon < (d_j - d_i)w_j$, $d_j > \varepsilon/w_j + d_i \ge d^*$, and the inequality $(d_j - d^*)^2 > (d_i + \varepsilon/w_j - d^*)^2$ holds. Therefore,
$$\begin{aligned} f(D') - f(D) &= w_i(d_i + \varepsilon/w_i - d^*)^2 - w_j(d_j - d^*)^2 \\ &< w_i(d_i + \varepsilon/w_i - d^*)^2 - w_j(d_i + \varepsilon/w_j - d^*)^2 \\ &= (w_i - w_j)((d^* - d_i)^2 - \varepsilon^2/w_i w_j). \end{aligned}$$
Given $\varepsilon/w_i \ge d^* - d_i > 0$ and $\varepsilon/w_j \ge \varepsilon/w_i \ge d^* - d_i > 0$, it follows that $\varepsilon^2/w_i w_j \ge (d^* - d_i)^2$, hence $f(D') - f(D) < 0$.

If $w_i < w_j$: Since $0 < \varepsilon < (d_j - d_i)w_i$, $d_j > d_i + \varepsilon/w_i \ge d^*$, and $(d_j - d^*)^2 >$

$(d_i + ε/w_i - d^*)^2$ , we have:

$$f(D') - f(D) = w_i(d_i + ε/w_i - d^*)^2 - w_j(d_j - d^*)^2$$
$$< w_i(d_i + ε/w_i - d^*)^2 - w_j(d_i + ε/w_i - d^*)^2$$
$$< 0.$$

For ***Case3d***, under the condition that $0 < d^* - d_i \le ε/w_i$, we can derive

$$w_i(d^* - d_i)^2/ε \le w_i(d^* - d_i)\, ε/w_i\, /ε = d^* - d_i.$$

Additionally, given that $0 < ε < (d_j - d_i) * \min(w_i, w_j)$ , we proceed to analyze the difference in function values scaled by ε:

$$f(D')/ε - f(D)/ε = w_i(d^* - d_i)^2/ε + ε/w_i + ε/w_j + 2d_i - 2d_j$$
$$\le (d^* + ε/w_j - d_j) + (ε/w_i + d_i - d_j).$$

Relying on the fact that $d^* \le d_j - ε/w_j$ and $d_i + ε/w_i < d_j$, we conclude

$$f(D')/ε - f(D)/ε < 0.$$

Hence $f(D') < f(D)$.

In essence, for all considered cases, the function $f(D)$ satisfies the Pigou–Dalton principle of transfers in a weak sense, as it either remains constant or decreases in value when subject to the specified type of transfer.

# Appendix 2: Proof of the analytical properties of MDELP

There are four analytical properties to understand the relationships between parameter $\beta$ and model performance.

Let $s_m$ be the optimal solution of the PMP. Let $\mu_m$ and $\gamma_m$ be the total travel distance and the total spatial envy of solution $s_m$, respectively.

Let $s_e$ be the optimal solution of the MELP. Let $\mu_e$ and $\gamma_e$ be the total travel distance and the total spatial envy of solution $s_e$, respectively. the total spatial envy.

Let $s_1$ be the optimal solution of the MDELP with parameters $\beta_1$ and $d_1^*$. Let $f_1 = \mu_1 + \beta_1\gamma_1$ be the optimal objective value, where $\mu_1$ is total travel distance, and $\gamma_1$ is the total spatial envy.

Let $s_2$ be the optimal solution of the MDELP with parameters $\beta_2$ and $d_2^*$. Let $f_2 = \mu_2 + \beta_2\gamma_2$ be the optimal objective value, where $\mu_2$ is total travel distance, and $\gamma_2$ is the total spatial envy.

***Property 1: If*** $\boldsymbol{\beta_1 \ge 0}$ ***and*** $\boldsymbol{d_1^* \ge 0}$***, then*** $\boldsymbol{\mu_1 \ge \mu_m}$ ***and*** $\boldsymbol{\gamma_1 \ge \gamma_e}$***.***

If $s_m \equiv s_1$, then $\mu_m = \mu_1$. If $s_m \ne s_1$, then $\mu_m \le \mu_1$, since $s_m$ is the optimal solution of the PMP.

If $s_e \equiv s_1$, then $\gamma_e = \gamma_1$. If $s_e \ne s_1$, then $\gamma_e \le \gamma_1$, since $s_e$ is the optimal solution of the MELP.

*Property* 1 indicates that, for the optimal solution of MDELP with any parameters $\beta$ and $d^*$, $\mu_m$ is the lower bound of total travel distance, and $\gamma_e$ is the lower bound of total spatial envy.

***Property 2: If*** $\mathbf{0 \leq \beta_1 < \beta_2}$ ***and*** $\mathbf{d_1^* = d_2^*}$***, then*** $\mathbf{f_1 \leq f_2}$***.***

If $s_1 \equiv s_2$, obviously, $\mu_1 = \mu_2$, $\gamma_1 = \gamma_2$, then $\mu_1 + \beta_1\gamma_1 = \mu_2 + \beta_2\gamma_2$.

If $s_1 \neq s_2$, then $f_1 = \mu_1 + \beta_1\gamma_1 \leq \mu_2 + \beta_1\gamma_2$, since solution $s_1$ is an optimal solution of the MDELP with parameters $\beta_1$ and $d^*$. As a result, $f_1 = \mu_1 + \beta_1\gamma_1 \leq \mu_2 + \beta_1\gamma_2 \leq \mu_2 + \beta_2\gamma_2 = f_2$.

*Property* 2 indicates that keeping the value of $d^*$ the same, and increasing the weight of spatial envy, $\beta$, the optimal objective of the MDELP will be increased or remained the same.

***Property 3: If*** $\mathbf{0 \leq \beta_1 < \beta_2}$ ***and*** $\mathbf{d_1^* = d_2^*}$***, then*** $\mathbf{\mu_1 \leq \mu_2}$ ***and*** $\mathbf{\gamma_1 \geq \gamma_2}$***.***

Since solution $s_1$ is an optimal solution of the MDELP with parameters $\beta_1$ and $d_1^*$, $\mu_1 + \beta_1\gamma_1 \leq \mu_2 + \beta_1\gamma_2$. Similarly, $\mu_2 + \beta_2\gamma_2 \leq \mu_1 + \beta_2\gamma_1$.

A new inequality is yielded by adding the two inequalities: $\mu_1 + \beta_1\gamma_1 + \mu_2 + \beta_2\gamma_2 \leq \mu_2 + \beta_1\gamma_2 + \mu_1 + \beta_2\gamma_1$. It can be rewritten as $\beta_1\gamma_1 + \beta_2\gamma_2 \leq \beta_1\gamma_2 + \beta_2\gamma_1$, and $(\beta_2 - \beta_1)\gamma_2 \leq (\beta_2 - \beta_1)\gamma_1$. Since $\beta_2 - \beta_1 > 0$, then $\gamma_2 \leq \gamma_1$.

Rewrite the inequality $\mu_1 + \beta_1\gamma_1 \leq \mu_2 + \beta_1\gamma_2$ as $\mu_1 \leq \mu_2 - \beta_1(\gamma_1 - \gamma_2)$. Since $\gamma_1 - \gamma_2 \geq 0$, then $\mu_1 \leq \mu_2$.

*Property* 3 shows that keeping the value of parameter $d^*$ the same, and increasing the weight of spatial envy, $\beta$, the optimal travel distance of the MDELP will be increased or remained the same, and the optimal spatial envy of the MDELP will be reduced or remained the same.

***Property 4: If*** $\mathbf{\beta_1 \geq 0}$ ***and*** $\mathbf{d_1^* \geq 0}$***, then*** $\mathbf{\mu_1 \leq \mu_e}$ ***and*** $\mathbf{\gamma_1 \leq \gamma_m}$***.***

For $\beta_1 = \infty$, the optimal solution is dominated by the envy, then $s_1 \equiv s_e$, and $\mu_1 = \mu_e$.

For $\beta_1 \to 0$, the optimal solution is dominated by the distance, then $s_1 \equiv s_m$, and $\gamma_1 = \gamma_m$.

According to ***Property*** 3, $\mu_1 \leq \mu_e$ and $\gamma_1 \leq \gamma_m$.

*Properties* 1, 3 and 4 show that $\mu_m \leq \mu_1 \leq \mu_e$ and $\gamma_e \leq \gamma_1 \leq \gamma_m$. This indicates that the optimal travel distance of the MDELP located in the interval $[\mu_m, \mu_e]$, and the optimal spatial envy of the MDELP located in the interval $[\gamma_e, \gamma_m]$.

# Appendix 3: Location models used in Section 4.4 and 4.5

Let $I = \{1,2 \cdots n\}$ be a set of *n* candidate facility locations. Let $J = \{1,2 \cdots m\}$ be a set of *m* demand locations, each location $j$ $(j \in J)$ has service demand $w_j$. Variable $d_{ij}$ is the distance between locations $i$ and $j$. Let $y_i$ $(i \in I)$ be a binary variable indicating whether a facility is opened at location $i$. Let $x_{ij}$ $(i \in I, j \in J)$ be a binary variable denoting whether customer $j$ is served by facility at location $i$. The well-known PMP is formulated as follows.

$$\text{Minimize: } \sum_{i \in I} \sum_{j \in J} w_j d_{ij} x_{ij} \tag{S1}$$

$$\text{Subject to: } \sum_{i \in I} x_{ij} = 1, \forall j \in J \tag{S2}$$

$$x_{ij} \le y_i, \forall i \in I, j \in J \quad \text{(S3)}$$

$$\sum_{i \in I} y_i = P \quad \text{(S4)}$$

$$x_{ij}, y_i \in \{0,1\}, \forall i \in I, j \in J \quad \text{(S5)}$$

In public services, spatial equity is one of the most important criteria for service planning and delivering. Let $d_j = \sum_{j \in J} w_j d_{ij} x_{ij}$ be the travel distance from demand location $j$ to it its nearest opened facility, and $\overline{d}$ be the mean travel distance. Spatial inequity indicators, the standard deviation (SD), the mean absolute deviation (MAD), the sum of absolute difference (AD), the coefficient of variance (CV), the Schutz indicator (SI), the Gini coefficient (GC), can be defined as follows.

$$SD = (\sum_{j \in J} w_j (d_j - \overline{d})^2 / \sum_{j \in J} w_j)^{0.5} \quad \text{(S6)}$$

$$MAD = \sum_{j \in J} w_i |d_i - \overline{d}| / \sum_{j \in J} w_j \quad \text{(S7)}$$

$$AD = \sum_{i \in J} \sum_{j \in J} w_i \, w_j \, |d_i - d_j| / \sum_{j \in J} w_j \, / \sum_{j \in J} w_j \quad \text{(S8)}$$

$$VC = SD / \overline{d} \quad \text{(S9)}$$

$$SI = MAD / 2\overline{d} \quad \text{(S10)}$$

$$GC = AD / 2\overline{d} \quad \text{(S11)}$$

Note that the inequity indicators defined above are different from others in two aspects. First, the demand weight is considered in all indicators. Second, the MAD and AD are averaged, and thus they can be compared with the mean distance and the SD.

The objective function of the PMP can be directly replaced by minimizing an inequity indictor. Using indictors (S6)~(S11), six minimum inequity location problems can be formulated, denoted as MinSD, MinMAD, MinAD, MinCV, MinSI and MinGC, respectively. Note that, different from the location models in Barbati and Bruno (2018), the demand weights are considered in the objective functions. Usually, constraints (S12) are used to ensure that each demand location is assigned to the nearest facility (Barbati and Bruno, 2018), where $M$ is a constant, $M > \max(d_{ij})$.

$$d_{ij} x_{ij} + (M - d_{ij}) y_i \le M, \forall i \in I, j \in J \quad \text{(S12)}$$

The objective functions (6) ~ (11) can be formulated as mixed integer linear programs (MILP) or mixed integer quadratic programs (MIQP) by introducing additional variables and constraints. The MinSD is formulated as follows.

$$\text{Minimize:} \sum_{j \in J} w_j (v_j - t)^2 \quad \text{(S13)}$$

$$\text{Subject to: (S2), (S3), (S4), (S5), (S12), and}$$

$$v_j = \sum_{i \in I} d_{ij} x_{ij}, \forall j \in J \quad \text{(S14)}$$

$$\sum_{j \in J} w_j \, t = \sum_{i \in I} \sum_{j \in J} w_j d_{ij} x_{ij} \quad \text{(S15)}$$

$$v_j, t \ge 0, \forall j \in J \quad \text{(S16)}$$

The objective function (S13) minimizes the variance of travel distances, in which, the travel distances ($v_j$) and the mean travel distance ($t$) are defined by equalities (S14) and (S15), respectively.

The MinCV problem is formulated as follows. Since the coefficient of variance is scale-invariant, the CV function can be rewritten as a quadratic function. In objective (S17), the variables $v_j$ are scaled distances $(d_j/\overline{d})$ defined by constraints (S18)~(S21).

$$\text{Minimize:} \sum_{j\in J} w_j (v_j - 1)^2 \tag{S17}$$

$$\text{Subject to: (S2), (S3), (S4), (S5), (S12), and}$$

$$u_{ij} \le M x_{ij}, \forall i \in I, j \in J \tag{S18}$$

$$M(x_{ij} - 1) \le u_{ij} - d_{ij} t \le M(1 - x_{ij}), \forall i \in I, j \in J \tag{S19}$$

$$v_j = \sum_{i\in I} u_{ij}, \forall j \in J \tag{S20}$$

$$\sum_{j\in J} v_j = \sum_{j\in J} w_j \tag{S21}$$

$$u_{ij}, v_j, t \ge 0, \forall i \in I, j \in J \tag{S22}$$

The MinMAD problem is formulated as follows. In objective (S23), variables $v_j$ are absolution deviations $(|d_i - \overline{d}|)$ defined by constraints (S24) and (S25).

$$\text{Minimize:} \sum_{j\in J} w_j v_j \tag{S23}$$

$$\text{Subject to: (S2), (S3), (S4), (S5), (S12), and}$$

$$\sum_{j\in J} w_j t = \sum_{i\in I} \sum_{j\in J} w_j d_{ij} x_{ij} \tag{S24}$$

$$-v_j \le \sum_{i\in I} d_{ij} x_{ij} - t \le v_j, \forall j \in J \tag{S25}$$

$$v_j, t \ge 0, \forall i \in I, j \in J \tag{S26}$$

The MinSI problem is formulated as follows. In objective (S27), variables $v_j$ are scaled absolution deviations $(|d_i - \overline{d}|/\overline{d})$ defined by constraints (S28)~(S31).

$$\text{Minimize:} \sum_{j\in J} w_j v_j \tag{S27}$$

$$\text{Subject to: (S2), (S3), (S4), (S5), (S12), } and$$

$$u_{ij} \le M x_{ij}, \forall i \in I, j \in J \tag{S28}$$

$$\sum_{i\in I} \sum_{j\in J} w_j u_{ij} = \sum_{j\in J} w_j \tag{S29}$$

$$M(x_{ij} - 1) \le u_{ij} - d_{ij} t \le M(1 - x_{ij}), \forall i \in I, j \in J \tag{S30}$$

$$-v_j \le \sum_{i\in I} u_{ij} - 1 \le v_j, \forall j \in J \tag{S31}$$

$$u_{ij}, v_j, t \ge 0, \forall i \in I, j \in J \tag{S32}$$

The MinAD problem is formulated as follows. In objective (S33), variables $z_{ij}$ are absolute differences $(|d_i - d_j|)$ defined by constraints (S34) and (S35).

$$\text{Minimize:} \sum_{i\in J} \sum_{j\in J} w_j w_j z_{ij} \tag{S33}$$

$$\text{Subject to: (S2), (S3), (S4), (S5), (S12), and}$$

$$v_j = \sum_{i\in I} d_{ij} x_{ij}, \forall j \in J \tag{S34}$$

$$-z_{ij} \le v_i - v_j \le z_{ij}, \forall i, j \in J \tag{S35}$$

$$z_{ij} \ge 0, \forall i, j \in J \tag{S36}$$

$$v_j \ge 0, \forall i \in I, j \in J \tag{S37}$$

The MinGC problem is formulated as follows. In objective (S38), variables $z_{ij}$ are scaled absolution difference $(|d_i - d_j|/\overline{d})$ , defined by constraints (S39)~(S42).

$$\text{Minimize:} \sum_{i\in J} \sum_{j\in J} w_j w_j z_{ij} \tag{S38}$$

$$\text{Subject to: (S2), (S3), (S4), (S5), (S12), and}$$

$$u_{ij} \leq My_i, \forall i \in I, j \in J \quad \text{(S39)}$$

$$\sum_{i\in I} \sum_{j\in J} w_j u_{ij} = \sum_{j\in J} w_j \quad \text{(S40)}$$

$$M(x_{ij} - 1) \leq u_{ij} - d_{ij}t \leq M(1 - x_{ij}), \forall i \in I, j \in J \quad \text{(S41)}$$

$$-z_{ij} \leq \sum_{k\in I} u_{ki} - \sum_{k\in I} u_{kj} \leq z_{ij}, \forall i, j \in J \quad \text{(S42)}$$

$$z_{ij} \geq 0, \forall i, j \in J \quad \text{(S43)}$$

$$u_{ij}, t \geq 0, \forall i \in I, j \in J \quad \text{(S44)}$$

The above location problems use particular functions in order to capture equity concerns. The inequality indicators can be combined with the efficiency objective (1), resulting in mean-inequity bi-objective location problems (Berman, 1990). The mean-variance bi-objective optimization technique has been widely used in economy, finance and public service. In this paper, a mean-variance bi-objective location problem (Mean-SD) can be defined by minimizing objectives (S1) and (S13) subject to constraints (S2)~(S5), (S12) and (S14)~(S16). Similarly, a mean-deviation bi-objective location problem (Mean-MAD) can be defined by minimizing objectives (S1) and (S23) subject to constraints (S2)~(S5), (12) and (S24)~(S26).

The ordered median problem (OMP) (Nickel & Puerto, 2005; Puerto & Rodríguez-Chía 2019) is considered by some studies as the "most equitable" solution (Karsu and Morton, 2015). It provides a common framework for several classical location problems such as the median, the center, the k-centrum (Slater 1978; Ogryczak and Zawadzki 2002), and $\alpha$-centdian (Halpern 1976) problems. The trade-off between efficiency and equality can be implementing by assigning appropriate weights to the ordered travel distances. Filippi et al. (2021a) proposed a fair facility location problem (FFLP) which minimizes the average of worst distances traveled by the $\beta$% of customers. It is a special case of OMP with a weighting vector $(0.01, 0.01, \ldots, 0.01, 0.01 + 0.99/k_\beta, 0.01 + 0.99/k_\beta, \ldots, 0.01 + 0.99/k_\beta)$, where $k_\beta = \lceil \beta\% n \rceil$. Inspired by Filippi et al. (2021a), a $k$-centdian problem is formulated in this paper as follows.

$$\text{Minimize: } \alpha \sum_{i\in I} \sum_{j\in J} w_j d_{ij} x_{ij} / \sum_{j\in J} w_j + (1 - \alpha)(ku + \sum_{j\in J} v_j) \quad \text{(S45)}$$

$$\text{Subject to: (S2), (S3), (S4), (S5)}$$

$$ku + kv_j \geq \sum_{i\in I} x_{ij}, \forall j \in J \quad \text{(S46)}$$

$$u, v_j \geq 0, \forall j \in J \quad \text{(S47)}$$

In objective (45), the first part denotes the mean travel distance, and the second part denotes the mean of $k$ worst distances. The two parts are summed with weights $\alpha$ and $1 - \alpha$ ($0 \leq \alpha \leq 1$), respectively. It is different from the FFLP in that the demand weights $w_j$ are considered in the first part of the objective function. Noticeably, some location problems are special cases of the problem: the PMP ($\alpha = 1$), the PCP ($\alpha = 0$ and $k = 1$), the $\alpha$-centdian problem ($\alpha > 0$ and $k = 1$), and the $k$-centrum problem ($\alpha = 0$).

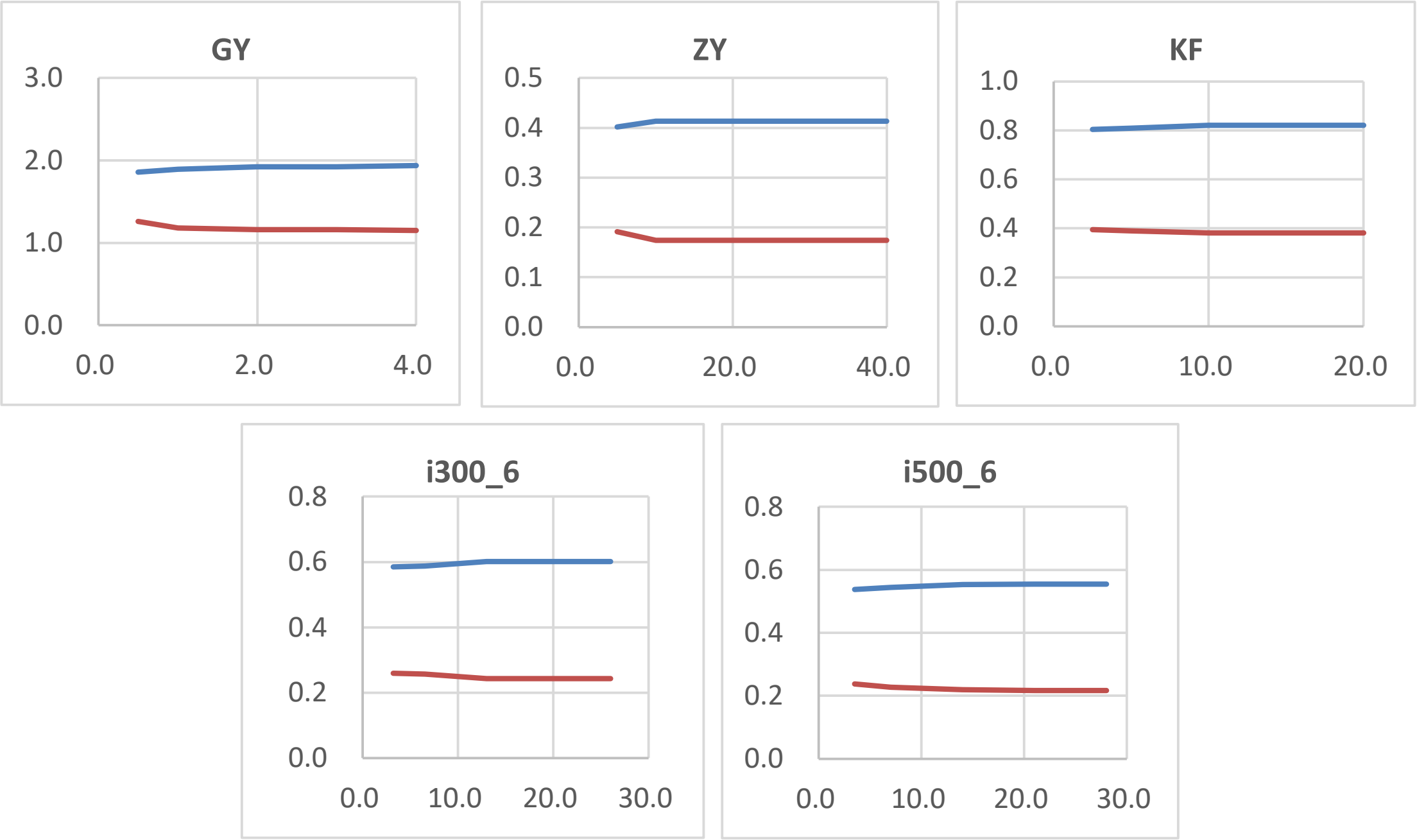

**Figure S1: The mean distance and the standard deviation from the MDELP solutions with different parameter $\beta$. The horizontal axis shows the value of parameter $\beta$; the vertical axis shows the mean distance (upper line) and the standard deviation (lower line).**

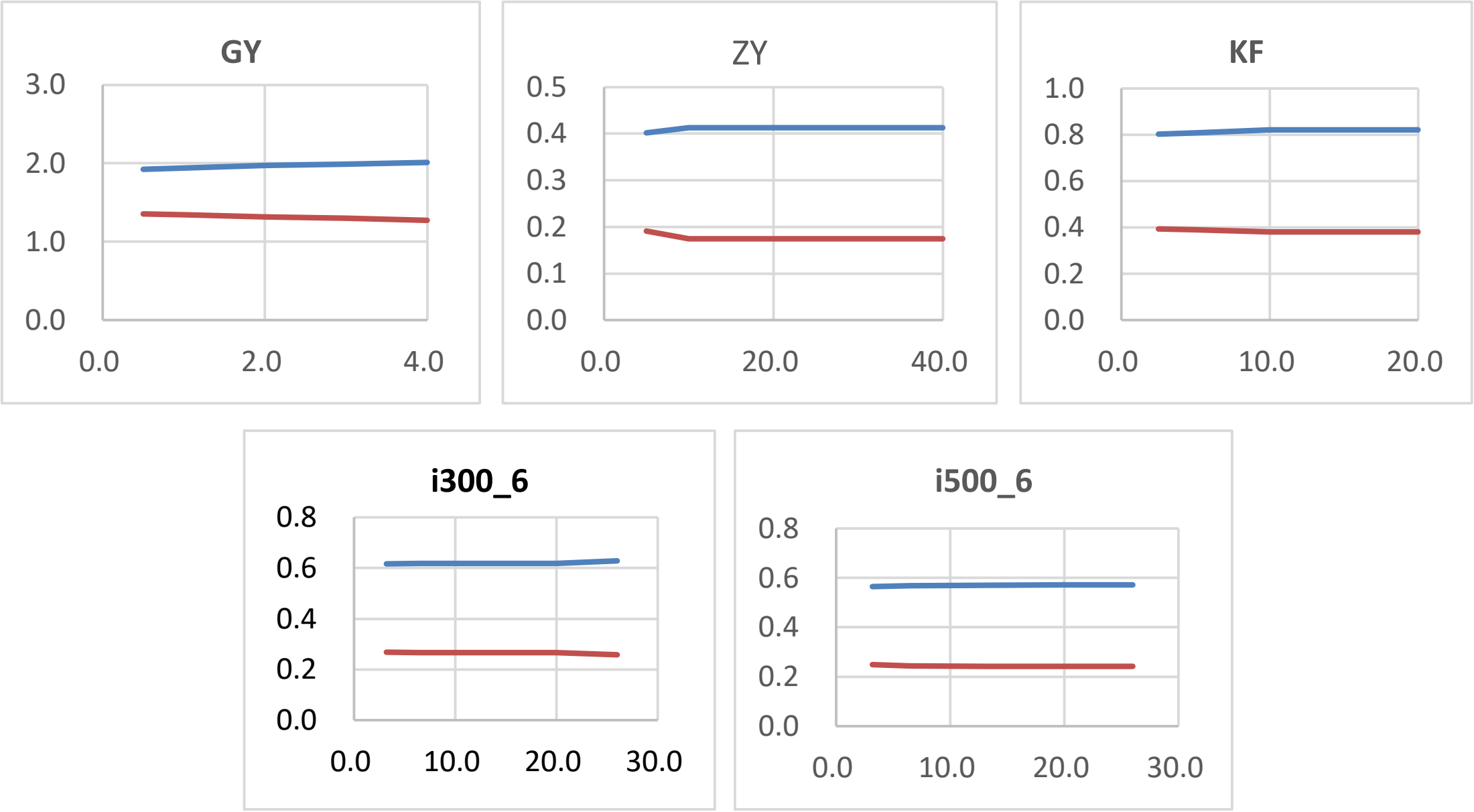

**Figure S2: The mean distance and the standard deviation from the CMDELP solutions with different parameter $\beta$. The horizontal axis shows the value of parameter $\beta$; the vertical axis shows the mean distance (upper line) and the standard deviation (lower line).**

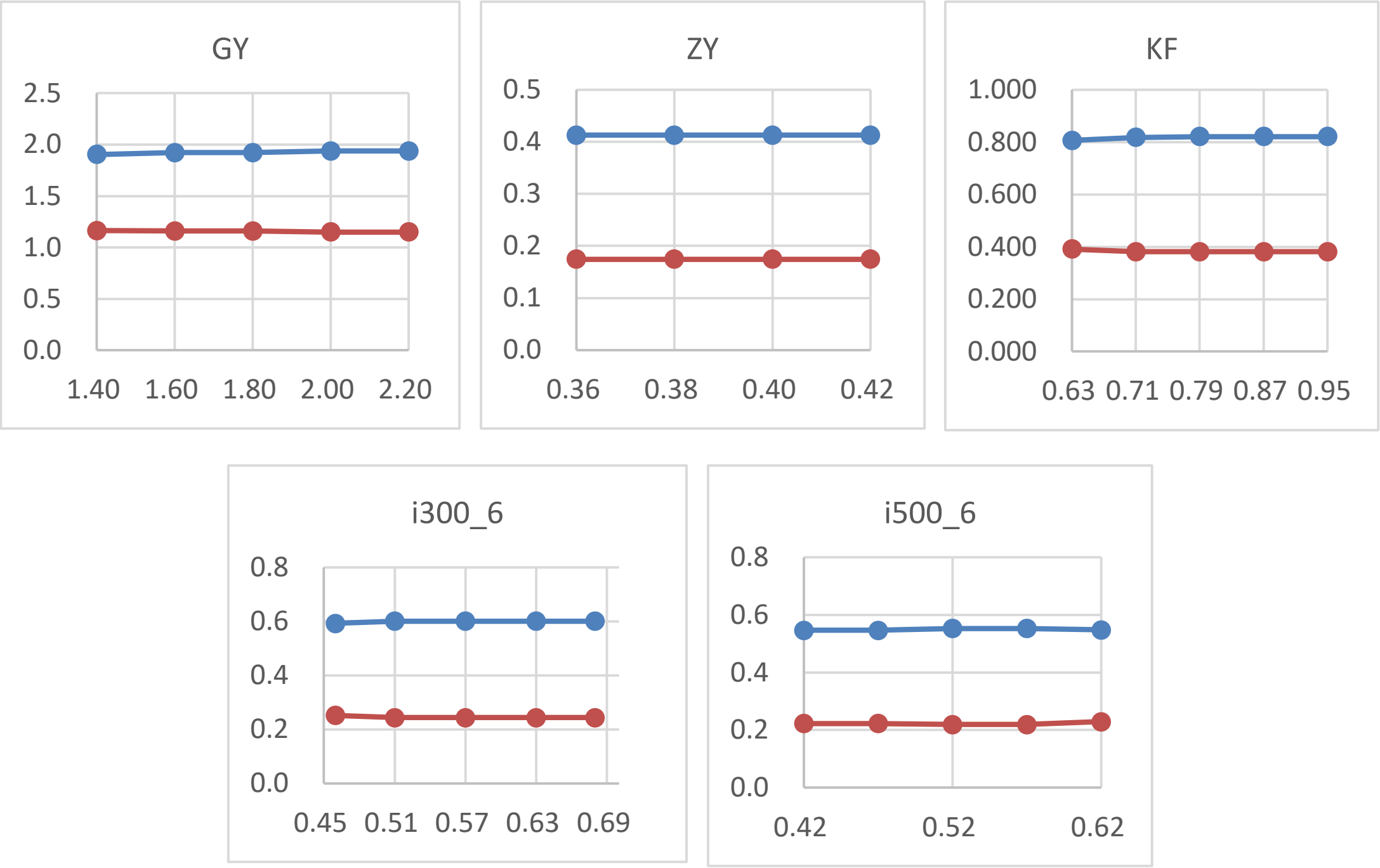

**Figure S3: The mean distance and the standard deviation from the MDELP solutions with different parameter $d^*$. The horizontal axis shows the value of parameter $d^*$; the vertical axis shows the mean distance (upper line) and the standard deviation (lower line).**

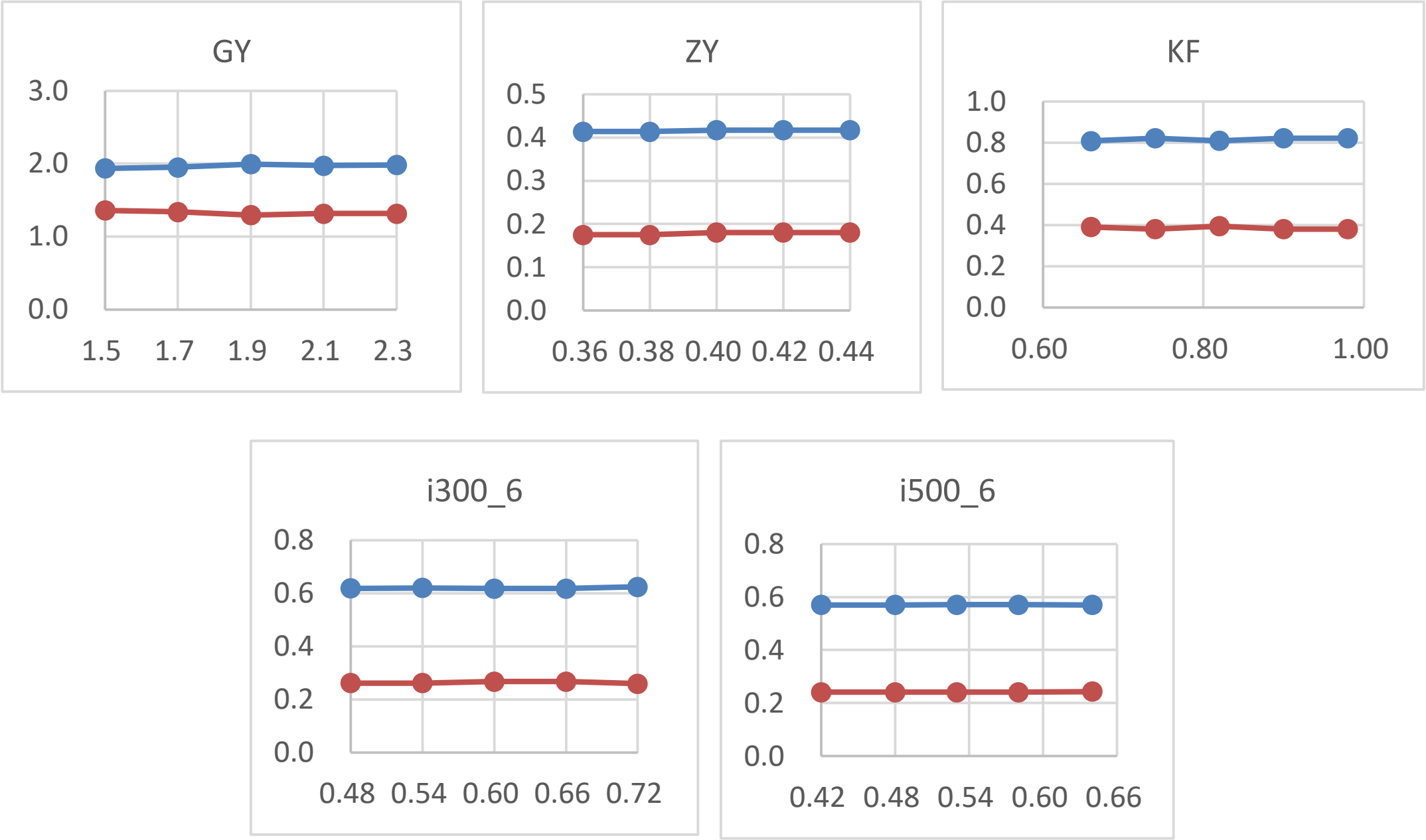

**Figure S4: The mean distance and the standard deviation from the CMDELP solutions with different parameter $d^*$. The horizontal axis shows the value of parameter $d^*$; the vertical axis shows the mean distance (upper line) and the standard deviation (lower line).**

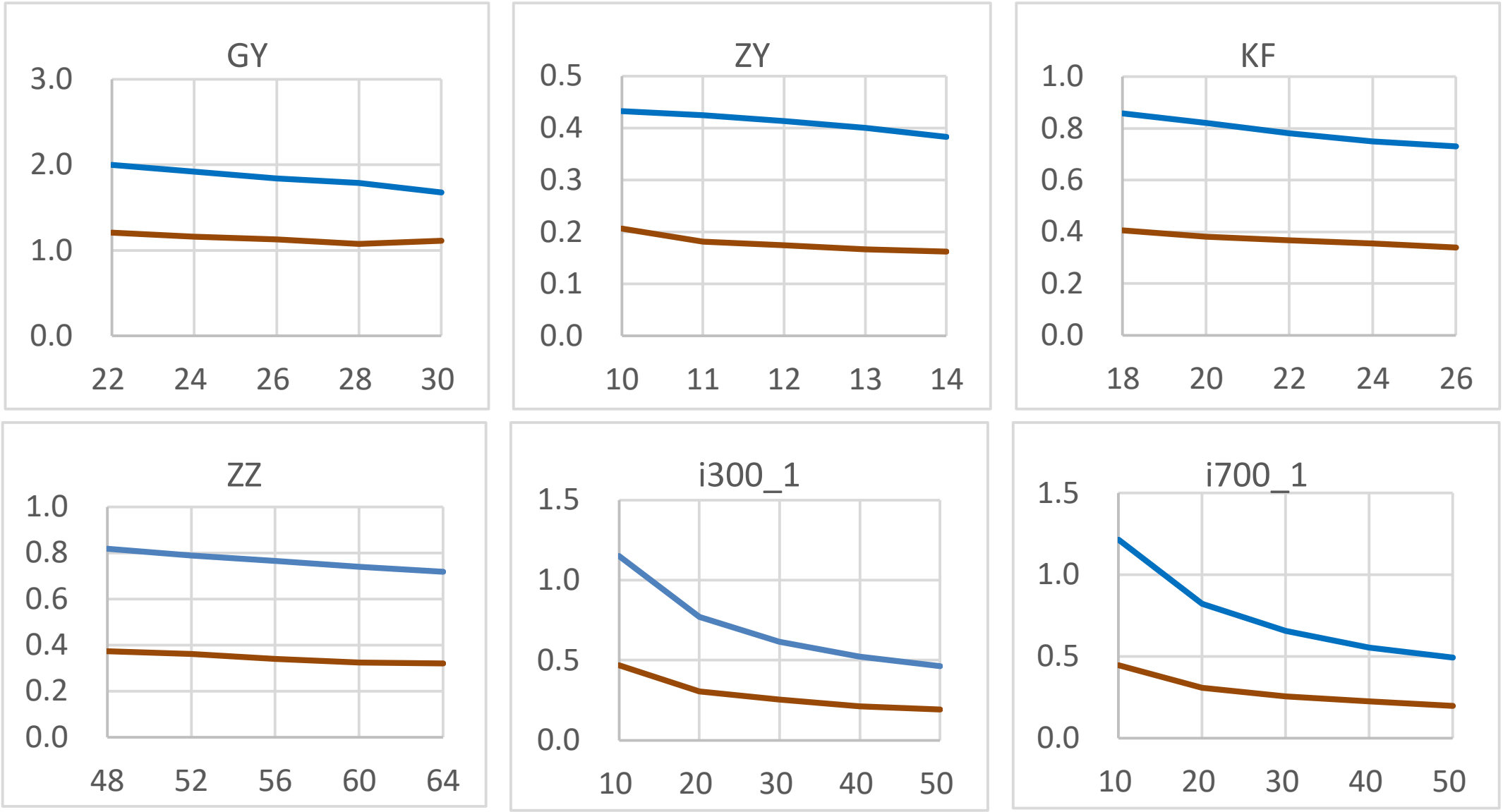

**Figure S5: The mean distance and the standard deviation versus the number of facilities. The horizontal axis shows the number of facilities for the MDELP; the vertical axis shows the mean distance (upper line) and the standard deviation (lower line).**

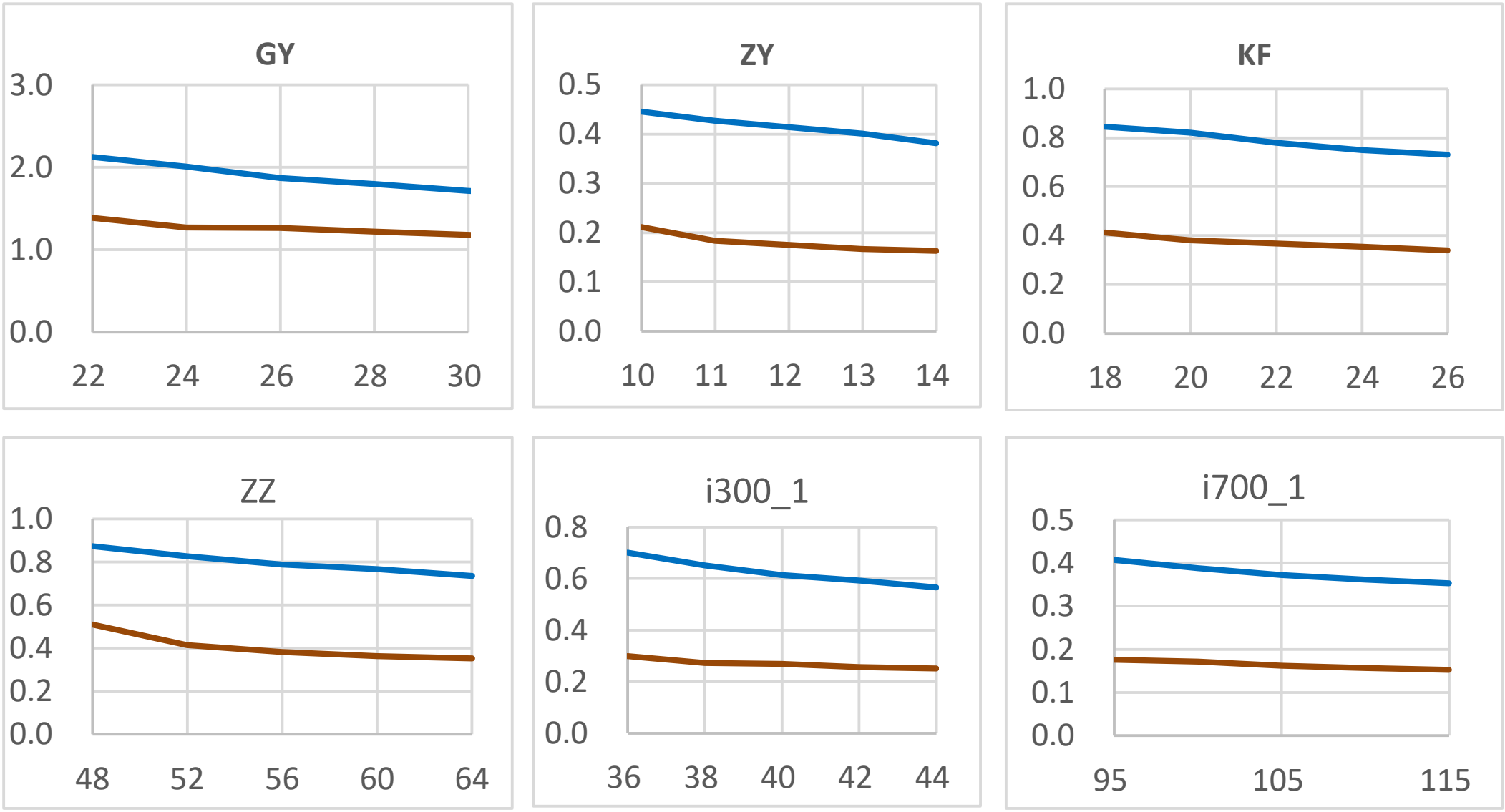

**Figure S6: The mean distance and the standard deviation versus the number of facilities. The horizontal axis shows the number of facilities for the CMDELP; the vertical axis shows the mean distance (upper line) and the standard deviation (lower line).**

## Table S1: Solutions from the instance Capa1

| Problem | Inst. | P | $d^*$ | $\beta$ | Objective | Gap | Time | Mean | SD | Mad | Gini |
|---|---|---|---|---|---|---|---|---|---|---|---|
| PMP | capa1 | 10 | ~ | ~ | 7070447.2 | opt | 51.9 | 138.9 | 59.9 | 49.4 | 0.246 |
| PMP | capa1 | 11 | ~ | ~ | 6726618.9 | opt | 166.9 | 132.2 | 56.3 | 46.2 | 0.243 |
| PMP | capa1 | 12 | ~ | ~ | 6413109.1 | opt | 41.0 | 126.0 | 53.5 | 43.9 | 0.242 |
| PMP | capa1 | 13 | ~ | ~ | 6139107.8 | opt | 34.7 | 120.6 | 51.0 | 41.6 | 0.239 |
| PMP | capa1 | 14 | ~ | ~ | 5870738.4 | opt | 12.8 | 115.4 | 48.0 | 39.2 | 0.237 |
| MDELP | capa1 | 10 | 138.9 | 0.077 | 12905305.5 | opt | 31.4 | 139.7 | 54.5 | 44.7 | 0.223 |
| MDELP | capa1 | 11 | 132.2 | 0.083 | 12683936.2 | opt | 51.5 | 133.1 | 52.2 | 42.9 | 0.224 |
| MDELP | capa1 | 12 | 126.0 | 0.088 | 12080844.5 | opt | 35.5 | 127.3 | 49.3 | 40.9 | 0.222 |
| MDELP | capa1 | 13 | 120.6 | 0.093 | 11910391.6 | opt | 74.1 | 122.1 | 48.3 | 39.9 | 0.226 |
| MDELP | capa1 | 14 | 115.4 | 0.100 | 11707370.8 | opt | 38.2 | 116.5 | 47.2 | 39.1 | 0.231 |
| MELP | capa1 | 10 | 138.9 | ~ | 75232954.6 | opt | 39.5 | 140.1 | 54.2 | 44.3 | 0.221 |
| MELP | capa1 | 11 | 132.2 | ~ | 71110635.9 | opt | 72.2 | 134.3 | 52.3 | 43.8 | 0.223 |
| MELP | capa1 | 12 | 126.0 | ~ | 63689667.4 | opt | 32.6 | 127.3 | 49.3 | 40.9 | 0.222 |
| MELP | capa1 | 13 | 120.6 | ~ | 61077315.6 | opt | 52.2 | 122.7 | 48.0 | 39.6 | 0.223 |
| MELP | capa1 | 14 | 115.4 | ~ | 57796316.2 | opt | 55.1 | 116.5 | 47.2 | 39.1 | 0.231 |
| CPMP | capa1 | 10 | ~ | ~ | 7070447.2 | opt | 1008.3 | 138.9 | 59.9 | 49.4 | 0.246 |
| CPMP | capa1 | 11 | ~ | ~ | 6726618.9 | opt | 1131.4 | 132.2 | 56.3 | 46.2 | 0.243 |
| CPMP | capa1 | 12 | ~ | ~ | 6413109.1 | opt | 842.2 | 126.0 | 53.5 | 43.9 | 0.242 |
| CPMP | capa1 | 13 | ~ | ~ | 6139107.8 | opt | 1066.2 | 120.6 | 51.0 | 41.6 | 0.239 |
| CPMP | capa1 | 14 | ~ | ~ | 5870738.4 | opt | 405.2 | 115.4 | 48.0 | 39.2 | 0.237 |
| CMDELP | capa1 | 10 | 138.9 | 0.077 | 12905305.5 | opt | 1176.2 | 139.7 | 54.5 | 44.7 | 0.223 |
| CMDELP | capa1 | 11 | 132.2 | 0.083 | 12683936.2 | opt | 898.8 | 133.1 | 52.2 | 42.9 | 0.224 |
| CMDELP | capa1 | 12 | 126.0 | 0.088 | 12080844.5 | opt | 878.7 | 127.3 | 49.3 | 40.9 | 0.222 |
| CMDELP | capa1 | 13 | 120.6 | 0.093 | 11910391.6 | opt | 711.3 | 122.1 | 48.3 | 39.9 | 0.226 |
| CMDELP | capa1 | 14 | 115.4 | 0.100 | 11707370.8 | opt | 748.5 | 116.5 | 47.2 | 39.1 | 0.231 |
| MECLP | capa1 | 10 | 138.9 | ~ | 75232954.6 | opt | 1051.2 | 140.1 | 54.2 | 44.3 | 0.221 |
| MECLP | capa1 | 11 | 132.2 | ~ | 71110635.9 | opt | 1091.1 | 134.3 | 52.3 | 43.8 | 0.223 |
| MECLP | capa1 | 12 | 126.0 | ~ | 63689667.4 | opt | 729.5 | 127.3 | 49.3 | 40.9 | 0.222 |
| MECLP | capa1 | 13 | 120.6 | ~ | 61077315.6 | opt | 808.0 | 122.7 | 48.0 | 39.6 | 0.223 |
| MECLP | capa1 | 14 | 115.4 | ~ | 57796316.2 | opt | 538.5 | 116.5 | 47.2 | 39.1 | 0.231 |

## Table S2: Solutions from the instance Capb1

| Problem | Inst. | P | $d^*$ | $\beta$ | Objective | Gap | Time | Mean | SD | Mad | Gini |
|---|---|---|---|---|---|---|---|---|---|---|---|
| PMP | capb1 | 12 | ~ | ~ | 6436697.6 | opt | 14.6 | 125.2 | 52.5 | 43.0 | 0.239 |
| PMP | capb1 | 13 | ~ | ~ | 6191851.2 | opt | 25.2 | 120.4 | 53.8 | 44.5 | 0.256 |
| PMP | capb1 | 14 | ~ | ~ | 5964313.9 | opt | 29.3 | 116.0 | 51.9 | 42.6 | 0.256 |
| PMP | capb1 | 15 | ~ | ~ | 5749904.7 | opt | 23.2 | 111.9 | 50.7 | 41.4 | 0.258 |
| PMP | capb1 | 16 | ~ | ~ | 5555952.6 | opt | 17.8 | 108.1 | 49.5 | 40.2 | 0.260 |
| MDELP | capb1 | 12 | 125.2 | 0.091 | 12656400.5 | opt | 55.9 | 125.3 | 52.3 | 42.9 | 0.238 |
| MDELP | capb1 | 13 | 120.4 | 0.083 | 11778734.6 | opt | 73.5 | 122.0 | 50.0 | 40.8 | 0.233 |
| MDELP | capb1 | 14 | 116.0 | 0.086 | 11452663.1 | opt | 73.7 | 118.3 | 48.0 | 39.4 | 0.231 |
| MDELP | capb1 | 15 | 111.9 | 0.087 | 11011107.6 | opt | 159.4 | 113.4 | 46.0 | 37.1 | 0.230 |
| MDELP | capb1 | 16 | 108.1 | 0.088 | 10459755.8 | opt | 68.7 | 108.8 | 45.1 | 36.2 | 0.235 |
| MELP | capb1 | 12 | 125.2 | ~ | 68043641.8 | opt | 93.5 | 127.2 | 50.2 | 41.1 | 0.225 |
| MELP | capb1 | 13 | 120.4 | ~ | 65622728.7 | opt | 159.5 | 123.6 | 49.1 | 40.7 | 0.227 |
| MELP | capb1 | 14 | 116.0 | ~ | 62164188.8 | opt | 214.6 | 120.1 | 46.7 | 38.6 | 0.222 |
| MELP | capb1 | 15 | 111.9 | ~ | 59480107.2 | opt | 147.2 | 115.5 | 45.8 | 37.5 | 0.226 |
| MELP | capb1 | 16 | 108.1 | ~ | 55031390.2 | opt | 143.7 | 110.3 | 44.8 | 36.6 | 0.232 |
| CPMP | capb1 | 12 | ~ | ~ | 6445140.0 | opt | 537.6 | 125.4 | 52.8 | 43.2 | 0.240 |
| CPMP | capb1 | 13 | ~ | ~ | 6196126.1 | opt | 704.1 | 120.5 | 53.6 | 44.2 | 0.254 |
| CPMP | capb1 | 14 | ~ | ~ | 5967623.3 | opt | 856.5 | 116.1 | 52.0 | 42.7 | 0.256 |
| CPMP | capb1 | 15 | ~ | ~ | 5751877.3 | opt | 747.3 | 111.9 | 50.8 | 41.4 | 0.259 |
| CPMP | capb1 | 16 | ~ | ~ | 5555952.6 | opt | 718.7 | 108.1 | 49.5 | 40.2 | 0.260 |
| CMDELP | capb1 | 12 | 125.4 | 0.090 | 12652978.8 | opt | 1267.0 | 125.5 | 52.5 | 43.1 | 0.239 |
| CMDELP | capb1 | 13 | 120.5 | 0.084 | 11832650.7 | opt | 1384.0 | 123.1 | 49.0 | 40.3 | 0.227 |
| CMDELP | capb1 | 14 | 116.1 | 0.086 | 13987375.0 | opt | 1026.6 | 118.3 | 48.0 | 39.4 | 0.231 |
| CMDELP | capb1 | 15 | 111.9 | 0.087 | 11011107.6 | opt | 958.7 | 113.4 | 46.0 | 37.1 | 0.230 |
| CMDELP | capb1 | 16 | 108.1 | 0.088 | 10459755.8 | opt | 744.6 | 108.8 | 45.1 | 36.2 | 0.235 |
| MECLP | capb1 | 12 | 125.4 | ~ | 68358631.6 | opt | 1418.4 | 127.9 | 50.3 | 41.6 | 0.225 |
| MECLP | capb1 | 13 | 120.5 | ~ | 65481024.2 | opt | 1097.1 | 123.9 | 48.9 | 40.4 | 0.226 |
| MECLP | capb1 | 14 | 116.1 | ~ | 62050757.7 | opt | 1220.5 | 120.2 | 46.7 | 38.6 | 0.222 |
| MECLP | capb1 | 15 | 111.9 | ~ | 59480107.2 | opt | 1154.7 | 115.5 | 45.8 | 37.5 | 0.226 |
| MECLP | capb1 | 16 | 108.1 | ~ | 55031390.2 | opt | 841.6 | 110.3 | 44.8 | 36.5 | 0.231 |

## Table S3: Solutions from the instance Capc1

| Problem | Inst. | P | $d^*$ | $\beta$ | Objective | Gap | Time | Mean | SD | Mad | Gini |
|---|---|---|---|---|---|---|---|---|---|---|---|
| PMP | capc1 | 12 | ~ | ~ | 6250683.7 | opt | 15.5 | 122.7 | 53.8 | 44.6 | 0.250 |
| PMP | capc1 | 13 | ~ | ~ | 6003669.9 | opt | 12.9 | 117.9 | 53.4 | 44.6 | 0.259 |
| PMP | capc1 | 14 | ~ | ~ | 5801256.9 | opt | 13.1 | 113.9 | 52.7 | 43.9 | 0.265 |
| PMP | capc1 | 15 | ~ | ~ | 5627996.4 | opt | 25.7 | 110.5 | 51.8 | 43.0 | 0.268 |
| PMP | capc1 | 16 | ~ | ~ | 5455463.4 | opt | 16.6 | 107.1 | 49.8 | 41.3 | 0.265 |
| MDELP | capc1 | 12 | 122.7 | 0.085 | 12100632.3 | opt | 31.3 | 123.5 | 52.4 | 43.0 | 0.243 |
| MDELP | capc1 | 13 | 117.9 | 0.083 | 11380168.9 | opt | 63.6 | 119.6 | 49.8 | 40.8 | 0.238 |
| MDELP | capc1 | 14 | 113.9 | 0.082 | 10714661.3 | opt | 24.1 | 115.4 | 48.0 | 39.4 | 0.237 |
| MDELP | capc1 | 15 | 110.5 | 0.082 | 10357256.7 | opt | 54.7 | 112.3 | 46.0 | 37.2 | 0.233 |
| MDELP | capc1 | 16 | 107.1 | 0.086 | 10098363.2 | opt | 49.4 | 108.7 | 44.8 | 36.6 | 0.235 |
| MELP | capc1 | 12 | 122.7 | ~ | 68373444.9 | opt | 55.6 | 123.5 | 52.4 | 43.0 | 0.243 |
| MELP | capc1 | 13 | 117.9 | ~ | 63727381.1 | opt | 45.8 | 119.6 | 49.8 | 40.8 | 0.238 |
| MELP | capc1 | 14 | 113.9 | ~ | 58975212.1 | opt | 30.0 | 115.5 | 48.4 | 40.0 | 0.239 |
| MELP | capc1 | 15 | 110.5 | ~ | 56508158.6 | opt | 75.6 | 112.6 | 46.0 | 37.8 | 0.233 |
| MELP | capc1 | 16 | 107.1 | ~ | 52915440.9 | opt | 40.5 | 109.8 | 43.7 | 35.6 | 0.226 |
| CPMP | capc1 | 12 | ~ | ~ | 6258882.1 | opt | 731.8 | 122.9 | 54.1 | 44.7 | 0.251 |
| CPMP | capc1 | 13 | ~ | ~ | 6003669.9 | opt | 610.7 | 117.9 | 53.4 | 44.6 | 0.259 |
| CPMP | capc1 | 14 | ~ | ~ | 5801256.9 | opt | 459.0 | 113.9 | 52.7 | 43.9 | 0.265 |
| CPMP | capc1 | 15 | ~ | ~ | 5627996.4 | opt | 682.7 | 110.5 | 51.8 | 43.0 | 0.268 |
| CPMP | capc1 | 16 | ~ | ~ | 5455463.4 | opt | 527.8 | 107.1 | 49.8 | 41.3 | 0.265 |
| CMDELP | capc1 | 12 | 122.9 | 0.084 | 12010424.8 | opt | 853.1 | 123.3 | 52.6 | 43.2 | 0.243 |
| CMDELP | capc1 | 13 | 117.9 | 0.083 | 11316441.5 | opt | 695.8 | 119.6 | 49.8 | 40.8 | 0.238 |
| CMDELP | capc1 | 14 | 113.9 | 0.082 | 10773649.5 | opt | 557.7 | 115.4 | 48.0 | 39.4 | 0.237 |
| CMDELP | capc1 | 15 | 110.5 | 0.082 | 10357256.7 | opt | 720.0 | 112.3 | 46.0 | 37.2 | 0.233 |
| CMDELP | capc1 | 16 | 107.1 | 0.086 | 10098363.2 | opt | 677.5 | 108.7 | 44.8 | 36.6 | 0.235 |
| MECLP | capc1 | 12 | 122.9 | ~ | 68115396.8 | opt | 1173.0 | 123.5 | 52.4 | 43.0 | 0.243 |
| MECLP | capc1 | 13 | 117.9 | ~ | 63727381.1 | opt | 1121.3 | 119.6 | 49.8 | 40.8 | 0.238 |
| MECLP | capc1 | 14 | 113.9 | ~ | 58975212.1 | opt | 574.8 | 115.5 | 48.4 | 40.0 | 0.239 |
| MECLP | capc1 | 15 | 110.5 | ~ | 56508158.6 | opt | 663.3 | 112.6 | 46.0 | 37.8 | 0.233 |
| MECLP | capc1 | 16 | 107.1 | ~ | 52915440.9 | opt | 510.7 | 109.8 | 43.7 | 35.6 | 0.226 |

## Table S4: Solutions from the instance i300_1

| Problem | Inst. | P | $d^*$ | $\beta$ | Objective | Gap | Time | Mean | SD | Mad | Gini |
|---|---|---|---|---|---|---|---|---|---|---|---|
| PMP | i300_1 | 10 | ~ | ~ | 6596.6 | opt | 8.7 | 1.152 | 0.559 | 0.449 | 0.273 |
| PMP | i300_1 | 20 | ~ | ~ | 4413.7 | opt | 9.7 | 0.771 | 0.368 | 0.302 | 0.271 |
| PMP | i300_1 | 30 | ~ | ~ | 3525.1 | opt | 7.5 | 0.616 | 0.319 | 0.254 | 0.291 |
| PMP | i300_1 | 40 | ~ | ~ | 2983.0 | opt | 9.7 | 0.521 | 0.254 | 0.209 | 0.277 |
| PMP | i300_1 | 50 | ~ | ~ | 2643.0 | opt | 6.7 | 0.462 | 0.228 | 0.190 | 0.280 |
| MDELP | i300_1 | 10 | 1.15 | 7.3 | 11599.8 | opt | 19.5 | 1.182 | 0.469 | 0.388 | 0.225 |
| MDELP | i300_1 | 20 | 0.77 | 11.3 | 7885.8 | opt | 12.0 | 0.797 | 0.304 | 0.247 | 0.216 |
| MDELP | i300_1 | 30 | 0.61 | 12.1 | 6069.8 | opt | 8.9 | 0.629 | 0.254 | 0.206 | 0.230 |
| MDELP | i300_1 | 40 | 0.52 | 16.1 | 5460.3 | opt | 14.1 | 0.542 | 0.213 | 0.173 | 0.223 |
| MDELP | i300_1 | 50 | 0.46 | 17.7 | 4701.0 | opt | 7.0 | 0.474 | 0.193 | 0.160 | 0.231 |
| MELP | i300_1 | 10 | 1.15 | ~ | 661.7 | opt | 38.1 | 1.182 | 0.469 | 0.388 | 0.225 |
| MELP | i300_1 | 20 | 0.77 | ~ | 290.0 | opt | 14.8 | 0.805 | 0.292 | 0.234 | 0.205 |
| MELP | i300_1 | 30 | 0.61 | ~ | 199.7 | opt | 30.3 | 0.657 | 0.235 | 0.187 | 0.201 |
| MELP | i300_1 | 40 | 0.52 | ~ | 144.2 | opt | 14.1 | 0.552 | 0.203 | 0.164 | 0.209 |
| MELP | i300_1 | 50 | 0.46 | ~ | 111.6 | opt | 20.7 | 0.479 | 0.189 | 0.156 | 0.224 |
| CPMP | i300_1 | 36 | ~ | ~ | 3909.7 | 0.94% | 7201.0 | 0.683 | 0.340 | 0.272 | 0.281 |
| CPMP | i300_1 | 38 | ~ | ~ | 3624.4 | opt | 5467.6 | 0.633 | 0.319 | 0.256 | 0.284 |
| CPMP | i300_1 | 40 | ~ | ~ | 3416.2 | opt | 150.8 | 0.597 | 0.308 | 0.247 | 0.291 |
| CPMP | i300_1 | 42 | ~ | ~ | 3278.7 | opt | 53.6 | 0.573 | 0.298 | 0.241 | 0.294 |
| CPMP | i300_1 | 44 | ~ | ~ | 3166.9 | opt | 45.9 | 0.553 | 0.285 | 0.232 | 0.292 |
| CMDELP | i300_1 | 36 | 0.68 | 11.8 | 7565.6 | 2.27% | 7201.0 | 0.702 | 0.299 | 0.234 | 0.239 |
| CMDELP | i300_1 | 38 | 0.63 | 12.4 | 6702.5 | opt | 439.4 | 0.651 | 0.272 | 0.218 | 0.237 |
| CMDELP | i300_1 | 40 | 0.59 | 12.5 | 6383.6 | opt | 208.3 | 0.615 | 0.269 | 0.222 | 0.249 |
| CMDELP | i300_1 | 42 | 0.57 | 12.9 | 6048.6 | opt | 249.0 | 0.593 | 0.256 | 0.213 | 0.247 |
| CMDELP | i300_1 | 44 | 0.55 | 13.6 | 5881.5 | opt | 107.8 | 0.566 | 0.251 | 0.207 | 0.253 |
| CMELP | i300_1 | 36 | 0.68 | ~ | 299.0 | 5.54% | 7211.4 | 0.706 | 0.297 | 0.238 | 0.239 |
| CMELP | i300_1 | 38 | 0.63 | ~ | 238.7 | opt | 431.1 | 0.655 | 0.270 | 0.217 | 0.234 |
| CMELP | i300_1 | 40 | 0.59 | ~ | 229.0 | opt | 428.6 | 0.616 | 0.269 | 0.221 | 0.249 |
| CMELP | i300_1 | 42 | 0.57 | ~ | 202.5 | opt | 414.6 | 0.608 | 0.245 | 0.202 | 0.230 |
| CMELP | i300_1 | 44 | 0.55 | ~ | 187.6 | opt | 146.4 | 0.587 | 0.234 | 0.191 | 0.226 |

## Table S5: Solutions from the instance i300_6

| Problem | Inst. | P | $d^*$ | $\beta$ | Objective | Gap | Time | Mean | SD | Mad | Gini |
|---|---|---|---|---|---|---|---|---|---|---|---|
| PMP | i300_6 | 10 | ~ | ~ | 6684.2 | opt | 8.5 | 1.107 | 0.497 | 0.407 | 0.256 |
| PMP | i300_6 | 20 | ~ | ~ | 4490.5 | opt | 7.7 | 0.744 | 0.353 | 0.296 | 0.270 |
| PMP | i300_6 | 30 | ~ | ~ | 3493.0 | opt | 6.8 | 0.579 | 0.292 | 0.240 | 0.286 |
| PMP | i300_6 | 40 | ~ | ~ | 2975.2 | opt | 7.1 | 0.493 | 0.261 | 0.213 | 0.297 |
| PMP | i300_6 | 50 | ~ | ~ | 2615.3 | opt | 6.7 | 0.433 | 0.221 | 0.178 | 0.285 |
| MDELP | i300_6 | 10 | 1.10 | 8.9 | 12556.4 | opt | 10.2 | 1.132 | 0.435 | 0.353 | 0.219 |
| MDELP | i300_6 | 20 | 0.74 | 11.9 | 8665.6 | opt | 8.0 | 0.761 | 0.325 | 0.275 | 0.245 |
| MDELP | i300_6 | 30 | 0.57 | 13.6 | 6651.6 | opt | 8.6 | 0.601 | 0.243 | 0.198 | 0.231 |
| MDELP | i300_6 | 40 | 0.49 | 14.5 | 5533.6 | opt | 7.9 | 0.522 | 0.210 | 0.173 | 0.228 |
| MDELP | i300_6 | 50 | 0.43 | 17.6 | 5015.4 | opt | 9.5 | 0.447 | 0.189 | 0.156 | 0.241 |
| MELP | i300_6 | 10 | 1.1 | ~ | 642.7 | opt | 36.1 | 1.134 | 0.435 | 0.352 | 0.218 |
| MELP | i300_6 | 20 | 0.74 | ~ | 335.6 | opt | 30.2 | 0.778 | 0.310 | 0.259 | 0.228 |
| MELP | i300_6 | 30 | 0.57 | ~ | 221.4 | opt | 11.0 | 0.613 | 0.233 | 0.186 | 0.215 |
| MELP | i300_6 | 40 | 0.49 | ~ | 164.1 | opt | 26.4 | 0.522 | 0.210 | 0.173 | 0.228 |
| MELP | i300_6 | 50 | 0.43 | ~ | 127.8 | opt | 10.7 | 0.462 | 0.178 | 0.147 | 0.219 |
| CPMP | i300_6 | 22 | ~ | ~ | 4491.6 | opt | 47.3 | 0.744 | 0.354 | 0.291 | 0.270 |
| CPMP | i300_6 | 24 | ~ | ~ | 4232.6 | opt | 30.4 | 0.701 | 0.341 | 0.279 | 0.275 |
| CPMP | i300_6 | 26 | ~ | ~ | 4002.5 | opt | 52.5 | 0.663 | 0.325 | 0.267 | 0.279 |
| CPMP | i300_6 | 28 | ~ | ~ | 3821.3 | opt | 77.3 | 0.633 | 0.324 | 0.263 | 0.289 |
| CPMP | i300_6 | 30 | ~ | ~ | 3671.5 | opt | 41.6 | 0.608 | 0.307 | 0.250 | 0.285 |
| CMDELP | i300_6 | 22 | 0.74 | 11.8 | 8992.6 | opt | 104.1 | 0.763 | 0.329 | 0.275 | 0.247 |
| CMDELP | i300_6 | 24 | 0.70 | 12.0 | 8357.9 | opt | 137.0 | 0.717 | 0.313 | 0.259 | 0.249 |
| CMDELP | i300_6 | 26 | 0.66 | 12.5 | 7845.3 | opt | 75.8 | 0.675 | 0.300 | 0.244 | 0.253 |
| CMDELP | i300_6 | 28 | 0.63 | 12.0 | 7138.4 | opt | 53.9 | 0.644 | 0.284 | 0.231 | 0.251 |
| CMDELP | i300_6 | 30 | 0.60 | 12.9 | 6858.4 | opt | 49.8 | 0.618 | 0.267 | 0.219 | 0.247 |
| CMELP | i300_6 | 22 | 0.74 | ~ | 369.2 | opt | 160.6 | 0.773 | 0.320 | 0.267 | 0.237 |
| CMELP | i300_6 | 24 | 0.70 | ~ | 333.7 | opt | 152.9 | 0.724 | 0.307 | 0.253 | 0.242 |
| CMELP | i300_6 | 26 | 0.66 | ~ | 299.3 | opt | 146.2 | 0.686 | 0.292 | 0.235 | 0.241 |
| CMELP | i300_6 | 28 | 0.63 | ~ | 266.5 | opt | 118.4 | 0.668 | 0.268 | 0.215 | 0.227 |
| CMELP | i300_6 | 30 | 0.60 | ~ | 239.7 | opt | 72.5 | 0.629 | 0.258 | 0.207 | 0.233 |

**Table S6: Solutions from the instance i3001500_1**

| Problem | Inst. | P | $d^*$ | $\beta$ | Objective | Gap | Time | Mean | SD | Mad | Gini |
|---|---|---|---|---|---|---|---|---|---|---|---|
| PMP | i3001500_1 | 10 | ~ | ~ | 347899.8 | opt | 76.2 | 12.163 | 5.224 | 4.206 | 0.243 |
| PMP | i3001500_1 | 20 | ~ | ~ | 238254.3 | opt | 52.1 | 8.330 | 3.513 | 2.900 | 0.241 |
| PMP | i3001500_1 | 30 | ~ | ~ | 194664.3 | opt | 76.4 | 6.806 | 2.860 | 2.307 | 0.239 |
| PMP | i3001500_1 | 40 | ~ | ~ | 169842.5 | opt | 108.4 | 5.938 | 2.541 | 2.067 | 0.243 |
| PMP | i3001500_1 | 50 | ~ | ~ | 151867.8 | opt | 73.5 | 5.310 | 2.328 | 1.901 | 0.250 |
| MDELP | i3001500_1 | 10 | 12.16 | 0.8 | 596242.9 | opt | 270.8 | 12.416 | 4.570 | 3.744 | 0.210 |
| MDELP | i3001500_1 | 20 | 8.32 | 1.3 | 451113.5 | opt | 1447.5 | 8.482 | 3.333 | 2.742 | 0.224 |
| MDELP | i3001500_1 | 30 | 6.80 | 1.6 | 359226.6 | opt | 422.6 | 6.864 | 2.658 | 2.165 | 0.220 |
| MDELP | i3001500_1 | 40 | 5.93 | 1.8 | 308053.3 | opt | 221.4 | 6.001 | 2.313 | 1.904 | 0.220 |
| MDELP | i3001500_1 | 50 | 5.30 | 1.9 | 276205.4 | opt | 332.0 | 5.414 | 2.067 | 1.687 | 0.217 |
| MELP | i3001500_1 | 10 | 12.16 | ~ | 301407.0 | opt | 208.6 | 12.416 | 4.570 | 3.744 | 0.210 |
| MELP | i3001500_1 | 20 | 8.32 | ~ | 160020.6 | opt | 1238.1 | 8.516 | 3.292 | 2.699 | 0.220 |
| MELP | i3001500_1 | 30 | 6.8 | ~ | 100921.3 | opt | 506.8 | 6.934 | 2.645 | 2.145 | 0.217 |
| MELP | i3001500_1 | 40 | 5.93 | ~ | 75610.4 | opt | 475.9 | 6.018 | 2.299 | 1.882 | 0.218 |
| MELP | i3001500_1 | 50 | 5.3 | ~ | 63503.3 | opt | 195.5 | 5.468 | 2.024 | 1.652 | 0.211 |
| CPMP | i3001500_1 | 100 | ~ | ~ | 111856.8 | opt | 228.0 | 3.911 | 1.799 | 1.463 | 0.261 |
| CPMP | i3001500_1 | 105 | ~ | ~ | 109576.7 | opt | 176.5 | 3.831 | 1.776 | 1.445 | 0.263 |
| CPMP | i3001500_1 | 110 | ~ | ~ | 107492.9 | opt | 133.2 | 3.758 | 1.757 | 1.430 | 0.265 |
| CPMP | i3001500_1 | 115 | ~ | ~ | 105550.9 | opt | 120.8 | 3.690 | 1.729 | 1.400 | 0.265 |
| CPMP | i3001500_1 | 120 | ~ | ~ | 103849.7 | opt | 95.3 | 3.631 | 1.693 | 1.367 | 0.264 |
| CMDELP | i3001500_1 | 100 | 3.91 | 2.4 | 217689.3 | opt | 176.7 | 3.989 | 1.670 | 1.358 | 0.238 |
| CMDELP | i3001500_1 | 105 | 3.83 | 2.4 | 212084.5 | opt | 152.3 | 3.891 | 1.654 | 1.344 | 0.241 |
| CMDELP | i3001500_1 | 110 | 3.75 | 2.4 | 207429.7 | opt | 103.4 | 3.807 | 1.633 | 1.318 | 0.243 |
| CMDELP | i3001500_1 | 115 | 3.69 | 2.4 | 201307.3 | opt | 90.2 | 3.736 | 1.606 | 1.296 | 0.243 |
| CMDELP | i3001500_1 | 120 | 3.63 | 2.5 | 201697.4 | opt | 76.8 | 3.690 | 1.575 | 1.269 | 0.241 |
| CMELP | i3001500_1 | 100 | 3.91 | ~ | 43102.5 | opt | 236.1 | 4.005 | 1.656 | 1.346 | 0.235 |
| CMELP | i3001500_1 | 105 | 3.83 | ~ | 41907.6 | opt | 130.6 | 3.910 | 1.637 | 1.327 | 0.237 |
| CMELP | i3001500_1 | 110 | 3.75 | ~ | 40896.4 | opt | 205.8 | 3.857 | 1.597 | 1.285 | 0.234 |
| CMELP | i3001500_1 | 115 | 3.69 | ~ | 39141.8 | opt | 118.7 | 3.774 | 1.574 | 1.268 | 0.236 |
| CMELP | i3001500_1 | 120 | 3.63 | ~ | 38403.6 | opt | 105.8 | 3.721 | 1.551 | 1.242 | 0.235 |

**Table S7: Solutions from the instance i3001500_6**

| Problem | Inst. | P | $d^*$ | $\beta$ | Objective | Gap | Time | Mean | SD | Mad | Gini |
|---|---|---|---|---|---|---|---|---|---|---|---|
| PMP | i3001500_6 | 10 | ~ | ~ | 358628.0 | opt | 78.0 | 12.134 | 4.823 | 3.945 | 0.227 |
| PMP | i3001500_6 | 20 | ~ | ~ | 248351.0 | opt | 78.8 | 8.403 | 3.571 | 2.946 | 0.243 |
| PMP | i3001500_6 | 30 | ~ | ~ | 201462.3 | opt | 60.4 | 6.817 | 3.029 | 2.474 | 0.253 |
| PMP | i3001500_6 | 40 | ~ | ~ | 175256.3 | opt | 60.6 | 5.930 | 2.540 | 2.093 | 0.245 |
| PMP | i3001500_6 | 50 | ~ | ~ | 157304.0 | opt | 69.3 | 5.322 | 2.346 | 1.924 | 0.251 |
| MDELP | i3001500_6 | 10 | 12.13 | 1.0 | 670089.4 | opt | 817.1 | 12.156 | 4.720 | 3.883 | 0.222 |
| MDELP | i3001500_6 | 20 | 8.40 | 1.3 | 471953.7 | opt | 837.3 | 8.555 | 3.373 | 2.786 | 0.225 |
| MDELP | i3001500_6 | 30 | 6.81 | 1.4 | 369852.1 | opt | 591.4 | 6.916 | 2.816 | 2.333 | 0.233 |
| MDELP | i3001500_6 | 40 | 5.92 | 1.8 | 330593.1 | opt | 561.8 | 6.014 | 2.374 | 1.953 | 0.225 |
| MDELP | i3001500_6 | 50 | 5.32 | 1.9 | 293936.5 | opt | 326.0 | 5.425 | 2.142 | 1.752 | 0.225 |
| MELP | i3001500_6 | 10 | 12.13 | ~ | 308755.2 | opt | 717.3 | 12.233 | 4.608 | 3.786 | 0.215 |
| MELP | i3001500_6 | 20 | 8.4 | ~ | 168537.6 | opt | 705.9 | 8.555 | 3.373 | 2.786 | 0.225 |
| MELP | i3001500_6 | 30 | 6.81 | ~ | 116556.2 | opt | 409.3 | 7.022 | 2.735 | 2.242 | 0.222 |
| MELP | i3001500_6 | 40 | 5.92 | ~ | 84911.7 | opt | 464.5 | 6.014 | 2.374 | 1.953 | 0.225 |
| MELP | i3001500_6 | 50 | 5.32 | ~ | 70010.4 | opt | 403.3 | 5.456 | 2.105 | 1.716 | 0.220 |
| CPMP | i3001500_6 | 50 | ~ | ~ | 159125.0 | opt | 3934.4 | 5.384 | 2.440 | 1.992 | 0.258 |
| CPMP | i3001500_6 | 53 | ~ | ~ | 154635.2 | opt | 2110.9 | 5.232 | 2.330 | 1.907 | 0.254 |
| CPMP | i3001500_6 | 56 | ~ | ~ | 150513.7 | opt | 3220.1 | 5.093 | 2.279 | 1.865 | 0.255 |
| CPMP | i3001500_6 | 59 | ~ | ~ | 146532.5 | opt | 1893.0 | 4.958 | 2.245 | 1.827 | 0.257 |
| CPMP | i3001500_6 | 62 | ~ | ~ | 143012.1 | opt | 2535.0 | 4.839 | 2.192 | 1.787 | 0.258 |
| CMDELP | i3001500_6 | 50 | 5.38 | 1.8 | 299391.7 | opt | 1611.4 | 5.512 | 2.208 | 1.358 | 0.229 |
| CMDELP | i3001500_6 | 53 | 5.23 | 1.9 | 290629.7 | opt | 1903.7 | 5.330 | 2.178 | 1.347 | 0.234 |
| CMDELP | i3001500_6 | 56 | 5.09 | 2.0 | 291828.3 | opt | 3565.9 | 5.200 | 2.117 | 1.343 | 0.233 |
| CMDELP | i3001500_6 | 59 | 4.96 | 2.0 | 281882.6 | opt | 2234.3 | 5.043 | 2.065 | 1.342 | 0.234 |
| CMDELP | i3001500_6 | 62 | 4.84 | 2.0 | 270590.3 | opt | 2777.0 | 4.910 | 2.018 | 1.332 | 0.235 |
| CMELP | i3001500_6 | 50 | 5.38 | ~ | 75780.8 | opt | 2726.3 | 5.563 | 2.166 | 1.781 | 0.222 |
| CMELP | i3001500_6 | 53 | 5.23 | ~ | 72654.4 | opt | 4729.5 | 5.383 | 2.148 | 1.771 | 0.228 |
| CMELP | i3001500_6 | 56 | 5.09 | ~ | 69072.6 | opt | 2976.3 | 5.204 | 2.114 | 1.745 | 0.232 |
| CMELP | i3001500_6 | 59 | 4.96 | ~ | 65354.4 | opt | 1241.6 | 5.047 | 2.063 | 1.694 | 0.233 |
| CMELP | i3001500_6 | 62 | 4.84 | ~ | 62512.9 | opt | 1066.8 | 4.931 | 2.006 | 1.637 | 0.232 |

**Table S8: Solutions from the instance i500_1**

| Problem | Inst. | P | $d^*$ | $\beta$ | Objective | Gap | Time | Mean | SD | Mad | Gini |
|---|---|---|---|---|---|---|---|---|---|---|---|
| PMP | i500_1 | 10 | ~ | ~ | 11029.3 | opt | 25.8 | 1.138 | 0.513 | 0.416 | 0.257 |
| PMP | i500_1 | 20 | ~ | ~ | 7614.5 | opt | 27.8 | 0.785 | 0.394 | 0.324 | 0.287 |
| PMP | i500_1 | 30 | ~ | ~ | 5937.0 | opt | 21.1 | 0.612 | 0.312 | 0.256 | 0.288 |
| PMP | i500_1 | 40 | ~ | ~ | 5029.1 | opt | 21.2 | 0.519 | 0.273 | 0.222 | 0.297 |
| PMP | i500_1 | 50 | ~ | ~ | 4393.8 | opt | 20.3 | 0.453 | 0.244 | 0.195 | 0.301 |
| MDELP | i500_1 | 10 | 1.13 | 8.6 | 21102.2 | opt | 37.1 | 1.149 | 0.488 | 0.406 | 0.243 |
| MDELP | i500_1 | 20 | 0.78 | 10.1 | 13562.7 | opt | 64.2 | 0.815 | 0.320 | 0.260 | 0.223 |
| MDELP | i500_1 | 30 | 0.61 | 12.6 | 10957.2 | opt | 39.4 | 0.633 | 0.269 | 0.225 | 0.243 |
| MDELP | i500_1 | 40 | 0.51 | 13.8 | 9397.3 | opt | 38.1 | 0.542 | 0.221 | 0.183 | 0.233 |
| MDELP | i500_1 | 50 | 0.45 | 15.2 | 8013.6 | opt | 25.5 | 0.469 | 0.198 | 0.161 | 0.239 |
| MELP | i500_1 | 10 | 1.13 | ~ | 1158.0 | opt | 67.0 | 1.149 | 0.488 | 0.406 | 0.243 |
| MELP | i500_1 | 20 | 0.78 | ~ | 558.8 | opt | 66.9 | 0.823 | 0.317 | 0.260 | 0.218 |
| MELP | i500_1 | 30 | 0.61 | ~ | 374.1 | opt | 45.7 | 0.650 | 0.250 | 0.206 | 0.219 |
| MELP | i500_1 | 40 | 0.51 | ~ | 299.6 | opt | 61.7 | 0.543 | 0.221 | 0.182 | 0.232 |
| MELP | i500_1 | 50 | 0.45 | ~ | 226.6 | opt | 35.8 | 0.476 | 0.195 | 0.158 | 0.232 |
| CPMP | i500_1 | 70 | ~ | ~ | 4357.7 | 1.04% | 7200.9 | 0.449 | 0.247 | 0.200 | 0.305 |
| CPMP | i500_1 | 73 | ~ | ~ | 4190.1 | 0.40% | 7200.8 | 0.432 | 0.238 | 0.189 | 0.305 |
| CPMP | i500_1 | 76 | ~ | ~ | 4047.6 | opt | 5772.4 | 0.417 | 0.224 | 0.178 | 0.296 |
| CPMP | i500_1 | 79 | ~ | ~ | 3921.4 | opt | 1415.7 | 0.404 | 0.222 | 0.178 | 0.303 |
| CPMP | i500_1 | 82 | ~ | ~ | 3811.1 | opt | 426.9 | 0.393 | 0.217 | 0.175 | 0.306 |
| CMDELP | i500_1 | 70 | 0.44 | 14.7 | 8496.7 | 5.83% | 7201.9 | 0.471 | 0.204 | 0.169 | 0.246 |
| CMDELP | i500_1 | 73 | 0.43 | 15.2 | 7854.2 | 3.25% | 7201.1 | 0.452 | 0.196 | 0.162 | 0.247 |
| CMDELP | i500_1 | 76 | 0.41 | 16.5 | 7863.4 | 2.37% | 7201.3 | 0.434 | 0.189 | 0.156 | 0.247 |
| CMDELP | i500_1 | 79 | 0.40 | 16.4 | 7412.3 | 1.45% | 7201.2 | 0.423 | 0.182 | 0.150 | 0.245 |
| CMDELP | i500_1 | 82 | 0.39 | 16.6 | 7147.2 | 0.70% | 7200.7 | 0.412 | 0.177 | 0.145 | 0.244 |
| CMELP | i500_1 | 70 | 0.44 | ~ | 264.7 | 11.01% | 7219.7 | 0.476 | 0.203 | 0.166 | 0.242 |
| CMELP | i500_1 | 73 | 0.43 | ~ | 224.2 | 5.33% | 7225.8 | 0.456 | 0.191 | 0.159 | 0.238 |
| CMELP | i500_1 | 76 | 0.41 | ~ | 217.4 | 3.02% | 7223.5 | 0.440 | 0.185 | 0.152 | 0.240 |
| CMELP | i500_1 | 79 | 0.40 | ~ | 199.6 | 2.20% | 7222.1 | 0.429 | 0.178 | 0.146 | 0.236 |
| CMELP | i500_1 | 82 | 0.39 | ~ | 188.1 | 1.52% | 7220.2 | 0.417 | 0.173 | 0.143 | 0.237 |

## Table S9: Solutions from the instance i500_6

| Problem | Inst. | P | $d^*$ | $\beta$ | Objective | Gap | Time | Mean | SD | Mad | Gini |
|---|---|---|---|---|---|---|---|---|---|---|---|
| PMP | i500_6 | 10 | ~ | ~ | 11936.5 | opt | 45.5 | 1.166 | 0.521 | 0.442 | 0.257 |
| PMP | i500_6 | 20 | ~ | ~ | 8189.5 | opt | 39.4 | 0.800 | 0.386 | 0.311 | 0.273 |
| PMP | i500_6 | 30 | ~ | ~ | 6445.8 | opt | 32.6 | 0.630 | 0.304 | 0.249 | 0.274 |
| PMP | i500_6 | 40 | ~ | ~ | 5421.8 | opt | 21.0 | 0.530 | 0.270 | 0.223 | 0.290 |
| PMP | i500_6 | 50 | ~ | ~ | 4755.3 | opt | 21.1 | 0.465 | 0.234 | 0.189 | 0.285 |
| MDELP | i500_6 | 10 | 1.16 | 8.5 | 21242.3 | opt | 34.6 | 1.203 | 0.423 | 0.346 | 0.201 |
| MDELP | i500_6 | 20 | 0.80 | 10.7 | 14065.2 | opt | 60.8 | 0.822 | 0.312 | 0.256 | 0.216 |
| MDELP | i500_6 | 30 | 0.62 | 13.6 | 11976.2 | opt | 192.7 | 0.641 | 0.265 | 0.219 | 0.236 |
| MDELP | i500_6 | 40 | 0.52 | 14.4 | 10061.7 | opt | 55.9 | 0.553 | 0.219 | 0.180 | 0.225 |
| MDELP | i500_6 | 50 | 0.46 | 16.9 | 9051.5 | opt | 58.5 | 0.480 | 0.204 | 0.167 | 0.242 |
| MELP | i500_6 | 10 | 1.16 | ~ | 1051.0 | opt | 51.7 | 1.203 | 0.423 | 0.346 | 0.201 |
| MELP | i500_6 | 20 | 0.8 | ~ | 527.8 | opt | 45.0 | 0.822 | 0.312 | 0.256 | 0.216 |
| MELP | i500_6 | 30 | 0.62 | ~ | 396.2 | opt | 45.9 | 0.645 | 0.263 | 0.216 | 0.233 |
| MELP | i500_6 | 40 | 0.52 | ~ | 304.4 | opt | 67.8 | 0.555 | 0.217 | 0.177 | 0.222 |
| MELP | i500_6 | 50 | 0.46 | ~ | 237.6 | opt | 43.0 | 0.493 | 0.189 | 0.151 | 0.216 |
| CPMP | i500_6 | 36 | ~ | ~ | 6055.5 | opt | 697.0 | 0.592 | 0.307 | 0.256 | 0.296 |
| CPMP | i500_6 | 38 | ~ | ~ | 5827.5 | opt | 441.0 | 0.569 | 0.298 | 0.249 | 0.298 |
| CPMP | i500_6 | 40 | ~ | ~ | 5625.9 | opt | 217.3 | 0.550 | 0.291 | 0.242 | 0.301 |
| CPMP | i500_6 | 42 | ~ | ~ | 5448.8 | opt | 117.0 | 0.532 | 0.281 | 0.234 | 0.300 |
| CPMP | i500_6 | 44 | ~ | ~ | 5285.1 | opt | 192.0 | 0.516 | 0.270 | 0.222 | 0.297 |
| CMDELP | i500_6 | 36 | 0.59 | 12.5 | 11050.8 | opt | 816.9 | 0.607 | 0.262 | 0.217 | 0.247 |
| CMDELP | i500_6 | 38 | 0.56 | 12.8 | 10901.0 | opt | 911.1 | 0.589 | 0.250 | 0.206 | 0.242 |
| CMDELP | i500_6 | 40 | 0.54 | 12.9 | 10472.9 | opt | 830.7 | 0.571 | 0.241 | 0.200 | 0.241 |
| CMDELP | i500_6 | 42 | 0.53 | 13.4 | 9993.8 | opt | 513.4 | 0.556 | 0.231 | 0.190 | 0.236 |
| CMDELP | i500_6 | 44 | 0.51 | 14.1 | 9902.7 | opt | 260.2 | 0.532 | 0.229 | 0.187 | 0.245 |
| CMELP | i500_6 | 36 | 0.59 | ~ | 381.7 | opt | 984.9 | 0.614 | 0.254 | 0.210 | 0.237 |
| CMELP | i500_6 | 38 | 0.56 | ~ | 380.6 | opt | 1601.3 | 0.589 | 0.250 | 0.206 | 0.242 |
| CMELP | i500_6 | 40 | 0.54 | ~ | 358.7 | opt | 680.7 | 0.571 | 0.241 | 0.200 | 0.241 |
| CMELP | i500_6 | 42 | 0.53 | ~ | 320.6 | opt | 632.3 | 0.557 | 0.231 | 0.190 | 0.237 |
| CMELP | i500_6 | 44 | 0.51 | ~ | 311.1 | opt | 254.7 | 0.542 | 0.223 | 0.183 | 0.234 |

## Table S10: Solutions from the instance i700_1

| Problem | Inst. | P | $d^*$ | $\beta$ | Objective | Gap | Time | Mean | SD | Mad | Gini |
|---|---|---|---|---|---|---|---|---|---|---|---|
| PMP | i700_1 | 10 | ~ | ~ | 16017.3 | opt | 130.3 | 1.184 | 0.515 | 0.417 | 0.247 |
| PMP | i700_1 | 20 | ~ | ~ | 10778.7 | opt | 56.2 | 0.797 | 0.368 | 0.305 | 0.265 |
| PMP | i700_1 | 30 | ~ | ~ | 8597.8 | opt | 62.7 | 0.636 | 0.316 | 0.260 | 0.283 |
| PMP | i700_1 | 40 | ~ | ~ | 7295.3 | opt | 43.2 | 0.539 | 0.274 | 0.222 | 0.288 |
| PMP | i700_1 | 50 | ~ | ~ | 6423.5 | opt | 55.3 | 0.475 | 0.238 | 0.193 | 0.283 |
| MDELP | i700_1 | 10 | 1.18 | 8.9 | 29148.3 | opt | 385.2 | 1.215 | 0.447 | 0.373 | 0.210 |
| MDELP | i700_1 | 20 | 0.79 | 11.7 | 19556.2 | opt | 507.8 | 0.823 | 0.308 | 0.248 | 0.212 |
| MDELP | i700_1 | 30 | 0.63 | 12.7 | 15342.1 | opt | 255.7 | 0.657 | 0.256 | 0.209 | 0.221 |
| MDELP | i700_1 | 40 | 0.53 | 14.3 | 13184.0 | opt | 217.8 | 0.554 | 0.225 | 0.185 | 0.232 |
| MDELP | i700_1 | 50 | 0.47 | 16.7 | 11731.7 | opt | 144.8 | 0.492 | 0.197 | 0.161 | 0.228 |
| MELP | i700_1 | 10 | 1.18 | ~ | 1428.9 | opt | 91.9 | 1.215 | 0.447 | 0.373 | 0.210 |
| MELP | i700_1 | 20 | 0.79 | ~ | 719.1 | opt | 130.3 | 0.829 | 0.302 | 0.244 | 0.207 |
| MELP | i700_1 | 30 | 0.63 | ~ | 500.6 | opt | 298.2 | 0.671 | 0.246 | 0.198 | 0.207 |
| MELP | i700_1 | 40 | 0.53 | ~ | 389.4 | opt | 166.6 | 0.564 | 0.217 | 0.178 | 0.219 |
| MELP | i700_1 | 50 | 0.47 | ~ | 303.2 | opt | 202.5 | 0.494 | 0.195 | 0.160 | 0.225 |
| CPMP | i700_1 | 95 | ~ | ~ | 5266.9 | 1.15% | 7201.7 | 0.389 | 0.209 | 0.173 | 0.303 |
| CPMP | i700_1 | 100 | ~ | ~ | 5038.2 | 0.45% | 7201.2 | 0.373 | 0.196 | 0.162 | 0.297 |
| CPMP | i700_1 | 105 | ~ | ~ | 4850.4 | opt | 1831.3 | 0.359 | 0.186 | 0.156 | 0.296 |
| CPMP | i700_1 | 110 | ~ | ~ | 4695.5 | opt | 1288.6 | 0.347 | 0.183 | 0.154 | 0.300 |
| CPMP | i700_1 | 115 | ~ | ~ | 4562.0 | opt | 1103.4 | 0.337 | 0.175 | 0.145 | 0.295 |
| CMDELP | i700_1 | 95 | 0.38 | 17.8 | 10223.1 | 5.51% | 7201.1 | 0.408 | 0.176 | 0.146 | 0.246 |
| CMDELP | i700_1 | 100 | 0.37 | 19.4 | 9712.8 | 4.16% | 7201.5 | 0.388 | 0.171 | 0.143 | 0.252 |
| CMDELP | i700_1 | 105 | 0.35 | 20.6 | 9522.6 | 0.49% | 7201.9 | 0.372 | 0.162 | 0.134 | 0.248 |
| CMDELP | i700_1 | 110 | 0.34 | 20.7 | 9071.1 | 0.59% | 7200.9 | 0.361 | 0.157 | 0.129 | 0.247 |
| CMDELP | i700_1 | 115 | 0.33 | 21.9 | 8954.1 | 0.76% | 7201.6 | 0.353 | 0.152 | 0.126 | 0.245 |
| CMELP | i700_1 | 95 | 0.38 | ~ | 260.3 | 9.92% | 7228.4 | 0.409 | 0.174 | 0.144 | 0.243 |
| CMELP | i700_1 | 100 | 0.37 | ~ | 224.3 | 6.73% | 7237.8 | 0.392 | 0.166 | 0.138 | 0.242 |
| CMELP | i700_1 | 105 | 0.35 | ~ | 217.5 | 3.07% | 7264.8 | 0.374 | 0.159 | 0.132 | 0.243 |
| CMELP | i700_1 | 110 | 0.34 | ~ | 200.3 | 1.70% | 7234.3 | 0.366 | 0.152 | 0.125 | 0.236 |
| CMELP | i700_1 | 115 | 0.33 | ~ | 189.7 | 2.07% | 7242.9 | 0.358 | 0.147 | 0.121 | 0.233 |

## Table S11: Solutions from the instance i700_6

| Problem | Inst. | P | $d^*$ | $\beta$ | Objective | Gap | Time | Mean | SD | Mad | Gini |
|---|---|---|---|---|---|---|---|---|---|---|---|
| PMP | i700_6 | 10 | ~ | ~ | 15872.9 | opt | 131.0 | 1.169 | 0.484 | 0.400 | 0.237 |
| PMP | i700_6 | 20 | ~ | ~ | 10977.6 | opt | 89.6 | 0.808 | 0.380 | 0.314 | 0.268 |
| PMP | i700_6 | 30 | ~ | ~ | 8669.8 | opt | 46.9 | 0.639 | 0.295 | 0.243 | 0.264 |
| PMP | i700_6 | 40 | ~ | ~ | 7393.7 | opt | 46.1 | 0.545 | 0.269 | 0.223 | 0.282 |
| PMP | i700_6 | 50 | ~ | ~ | 6497.1 | opt | 49.2 | 0.478 | 0.246 | 0.199 | 0.290 |
| MDELP | i700_6 | 10 | 1.16 | 9.9 | 29997.4 | opt | 132.2 | 1.178 | 0.457 | 0.379 | 0.221 |
| MDELP | i700_6 | 20 | 0.80 | 11.2 | 19642.2 | opt | 248.9 | 0.829 | 0.323 | 0.263 | 0.221 |
| MDELP | i700_6 | 30 | 0.63 | 14.6 | 16132.7 | opt | 74.4 | 0.656 | 0.259 | 0.212 | 0.225 |
| MDELP | i700_6 | 40 | 0.54 | 15.0 | 13226.7 | opt | 162.3 | 0.562 | 0.222 | 0.182 | 0.224 |
| MDELP | i700_6 | 50 | 0.47 | 15.8 | 11674.7 | opt | 76.6 | 0.495 | 0.196 | 0.158 | 0.225 |
| MELP | i700_6 | 10 | 1.16 | ~ | 1414.0 | opt | 122.6 | 1.178 | 0.457 | 0.379 | 0.221 |
| MELP | i700_6 | 20 | 0.80 | ~ | 745.4 | opt | 651.9 | 0.833 | 0.321 | 0.261 | 0.219 |
| MELP | i700_6 | 30 | 0.63 | ~ | 494.7 | opt | 195.8 | 0.657 | 0.258 | 0.211 | 0.223 |
| MELP | i700_6 | 40 | 0.54 | ~ | 369.8 | opt | 184.4 | 0.568 | 0.217 | 0.176 | 0.215 |
| MELP | i700_6 | 50 | 0.47 | ~ | 312.6 | opt | 140.3 | 0.502 | 0.193 | 0.157 | 0.218 |
| CPMP | i700_6 | 50 | ~ | ~ | 6955.9 | 0.43% | 7201.2 | 0.512 | 0.242 | 0.198 | 0.269 |
| CPMP | i700_6 | 53 | ~ | ~ | 6676.4 | opt | 5255.0 | 0.492 | 0.234 | 0.191 | 0.271 |
| CPMP | i700_6 | 56 | ~ | ~ | 6448.1 | opt | 2506.8 | 0.475 | 0.229 | 0.187 | 0.273 |
| CPMP | i700_6 | 59 | ~ | ~ | 6248.4 | opt | 2539.3 | 0.460 | 0.226 | 0.184 | 0.278 |
| CPMP | i700_6 | 62 | ~ | ~ | 6076.0 | opt | 1295.5 | 0.447 | 0.223 | 0.181 | 0.282 |
| CMDELP | i700_6 | 50 | 0.51 | 17.5 | 13447.7 | 0.54% | 7201.0 | 0.531 | 0.215 | 0.176 | 0.231 |
| CMDELP | i700_6 | 53 | 0.49 | 17.8 | 12910.8 | 0.65% | 7201.1 | 0.506 | 0.210 | 0.172 | 0.237 |
| CMDELP | i700_6 | 56 | 0.47 | 18.1 | 12312.8 | opt | 2017.4 | 0.488 | 0.201 | 0.164 | 0.234 |
| CMDELP | i700_6 | 59 | 0.46 | 18.0 | 11560.8 | opt | 1760.9 | 0.478 | 0.192 | 0.158 | 0.229 |
| CMDELP | i700_6 | 62 | 0.44 | 18.0 | 11412.2 | opt | 3578.1 | 0.462 | 0.188 | 0.155 | 0.232 |
| CMELP | i700_6 | 50 | 0.51 | ~ | 356.5 | 1.13% | 7271.6 | 0.532 | 0.214 | 0.175 | 0.229 |
| CMELP | i700_6 | 53 | 0.49 | ~ | 336.2 | 3.19% | 7231.7 | 0.517 | 0.204 | 0.167 | 0.224 |
| CMELP | i700_6 | 56 | 0.47 | ~ | 313.7 | opt | 3147.8 | 0.491 | 0.198 | 0.163 | 0.230 |
| CMELP | i700_6 | 59 | 0.46 | ~ | 281.8 | opt | 2973.5 | 0.480 | 0.191 | 0.157 | 0.227 |
| CMELP | i700_6 | 62 | 0.44 | ~ | 284.0 | 0.94% | 7228.7 | 0.467 | 0.184 | 0.153 | 0.225 |

## Table S12: Solutions from the instance i1000_1

| Problem | Inst. | P | $d^*$ | $\beta$ | Objective | Gap | Time | Mean | SD | Mad | Gini |
|---|---|---|---|---|---|---|---|---|---|---|---|
| PMP | i1000_1 | 10 | ~ | ~ | 22574.9 | opt | 759.8 | 1.195 | 0.478 | 0.391 | 0.228 |
| PMP | i1000_1 | 20 | ~ | ~ | 15449.4 | opt | 237.1 | 0.818 | 0.368 | 0.298 | 0.256 |
| PMP | i1000_1 | 30 | ~ | ~ | 12237.4 | opt | 161.2 | 0.648 | 0.291 | 0.237 | 0.255 |
| PMP | i1000_1 | 40 | ~ | ~ | 10462.2 | opt | 214.8 | 0.554 | 0.256 | 0.210 | 0.263 |
| PMP | i1000_1 | 50 | ~ | ~ | 9212.9 | opt | 262.4 | 0.488 | 0.231 | 0.190 | 0.269 |
| MDELP | i1000_1 | 10 | 1.19 | 10.4 | 42263.6 | opt | 313.7 | 1.207 | 0.449 | 0.370 | 0.212 |
| MDELP | i1000_1 | 20 | 0.81 | 12.0 | 28129.4 | opt | 3607.7 | 0.844 | 0.309 | 0.258 | 0.208 |
| MDELP | i1000_1 | 30 | 0.64 | 15.3 | 23042.7 | opt | 957.6 | 0.671 | 0.248 | 0.203 | 0.210 |
| MDELP | i1000_1 | 40 | 0.55 | 16.9 | 18671.4 | opt | 688.4 | 0.571 | 0.215 | 0.176 | 0.213 |
| MDELP | i1000_1 | 50 | 0.48 | 18.3 | 17004.5 | opt | 1958.2 | 0.505 | 0.191 | 0.156 | 0.215 |
| MELP | i1000_1 | 10 | 1.19 | ~ | 1870.7 | opt | 236.6 | 1.207 | 0.449 | 0.370 | 0.212 |
| MELP | i1000_1 | 20 | 0.81 | ~ | 1014.4 | opt | 1569.7 | 0.845 | 0.307 | 0.248 | 0.206 |
| MELP | i1000_1 | 30 | 0.64 | ~ | 677.5 | opt | 1091.1 | 0.671 | 0.248 | 0.203 | 0.210 |
| MELP | i1000_1 | 40 | 0.55 | ~ | 464.6 | opt | 501.9 | 0.574 | 0.212 | 0.174 | 0.209 |
| MELP | i1000_1 | 50 | 0.48 | ~ | 407.1 | opt | 1280.4 | 0.508 | 0.189 | 0.153 | 0.211 |
| CPMP | i1000_1 | 140 | ~ | ~ | 5876.1 | 1.16% | 7204.7 | 0.311 | 0.161 | 0.135 | 0.289 |
| CPMP | i1000_1 | 145 | ~ | ~ | 5709.3 | 0.94% | 7200.5 | 0.302 | 0.160 | 0.126 | 0.295 |
| CPMP | i1000_1 | 150 | ~ | ~ | 5557.0 | 0.61% | 7201.7 | 0.294 | 0.157 | 0.122 | 0.297 |
| CPMP | i1000_1 | 155 | ~ | ~ | 5418.3 | 0.09% | 7203.5 | 0.287 | 0.153 | 0.122 | 0.298 |
| CPMP | i1000_1 | 160 | ~ | ~ | 5301.6 | opt | 4337.4 | 0.287 | 0.153 | 0.121 | 0.298 |
| CMDELP | i1000_1 | 140 | 0.31 | 24.1 | 10981.5 | 5.42% | 7203.6 | 0.322 | 0.136 | 0.110 | 0.238 |
| CMDELP | i1000_1 | 145 | 0.30 | 23.6 | 10414.8 | 3.87% | 7207.2 | 0.313 | 0.132 | 0.104 | 0.237 |
| CMDELP | i1000_1 | 150 | 0.29 | 23.9 | 10193.7 | 2.13% | 7202.3 | 0.305 | 0.128 | 0.100 | 0.234 |
| CMDELP | i1000_1 | 155 | 0.28 | 24.4 | 10104.4 | 1.15% | 7201.8 | 0.298 | 0.126 | 0.100 | 0.237 |
| CMDELP | i1000_1 | 160 | 0.28 | 24.4 | 9537.3 | 1.12% | 7201.5 | 0.292 | 0.123 | 0.098 | 0.236 |
| CMELP | i1000_1 | 140 | 0.31 | ~ | 198.8 | 10.03% | 7256.9 | 0.328 | 0.131 | 0.105 | 0.225 |
| CMELP | i1000_1 | 145 | 0.30 | ~ | 193.3 | 10.18% | 7265.5 | 0.321 | 0.126 | 0.099 | 0.219 |
| CMELP | i1000_1 | 150 | 0.29 | ~ | 184.5 | 5.19% | 7271.9 | 0.310 | 0.124 | 0.097 | 0.225 |
| CMELP | i1000_1 | 155 | 0.28 | ~ | 181.9 | 2.69% | 7264.5 | 0.303 | 0.122 | 0.096 | 0.226 |
| CMELP | i1000_1 | 160 | 0.28 | ~ | 163.0 | 3.20% | 7303.5 | 0.298 | 0.118 | 0.093 | 0.223 |

## Table S13: Solutions from the instance i1000_6

| Problem | Inst. | P | $d^*$ | $\beta$ | Objective | Gap | Time | Mean | SD | Mad | Gini |
|---|---|---|---|---|---|---|---|---|---|---|---|
| PMP | i1000_6 | 10 | ~ | ~ | 23651.4 | opt | 263.3 | 1.191 | 0.512 | 0.419 | 0.245 |
| PMP | i1000_6 | 20 | ~ | ~ | 16093.6 | opt | 255.6 | 0.810 | 0.366 | 0.305 | 0.259 |
| PMP | i1000_6 | 30 | ~ | ~ | 12915.4 | opt | 224.4 | 0.650 | 0.304 | 0.245 | 0.264 |
| PMP | i1000_6 | 40 | ~ | ~ | 10877.1 | opt | 262.9 | 0.548 | 0.255 | 0.211 | 0.265 |
| PMP | i1000_6 | 50 | ~ | ~ | 9519.1 | opt | 169.3 | 0.479 | 0.234 | 0.193 | 0.278 |
| MDELP | i1000_6 | 10 | 1.19 | 9.0 | 41581.4 | opt | 455.0 | 1.199 | 0.458 | 0.376 | 0.217 |
| MDELP | i1000_6 | 20 | 0.81 | 12.1 | 29417.7 | opt | 3383.3 | 0.826 | 0.327 | 0.266 | 0.225 |
| MDELP | i1000_6 | 30 | 0.65 | 14.1 | 22496.3 | opt | 1622.3 | 0.666 | 0.257 | 0.211 | 0.219 |
| MDELP | i1000_6 | 40 | 0.54 | 16.8 | 20263.1 | opt | 226.1 | 0.561 | 0.220 | 0.183 | 0.224 |
| MDELP | i1000_6 | 50 | 0.47 | 17.5 | 17768.2 | opt | 325.2 | 0.491 | 0.201 | 0.164 | 0.233 |
| MELP | i1000_6 | 10 | 1.19 | ~ | 1971.9 | opt | 492.4 | 1.207 | 0.449 | 0.368 | 0.212 |
| MELP | i1000_6 | 20 | 0.81 | ~ | 1074.7 | opt | 3690.7 | 0.838 | 0.318 | 0.261 | 0.216 |
| MELP | i1000_6 | 30 | 0.65 | ~ | 655.5 | opt | 839.5 | 0.669 | 0.256 | 0.209 | 0.216 |
| MELP | i1000_6 | 40 | 0.54 | ~ | 540.0 | opt | 474.4 | 0.564 | 0.217 | 0.179 | 0.219 |
| MELP | i1000_6 | 50 | 0.47 | ~ | 454.9 | opt | 511.5 | 0.499 | 0.193 | 0.158 | 0.220 |
| CPMP | i1000_6 | 72 | ~ | ~ | 8203.3 | 0.58% | 7204.4 | 0.413 | 0.208 | 0.166 | 0.283 |
| CPMP | i1000_6 | 76 | ~ | ~ | 7875.4 | 0.20% | 7202.2 | 0.396 | 0.199 | 0.159 | 0.282 |
| CPMP | i1000_6 | 80 | ~ | ~ | 7613.9 | opt | 4898.4 | 0.383 | 0.190 | 0.153 | 0.280 |
| CPMP | i1000_6 | 84 | ~ | ~ | 7386.2 | opt | 2940.0 | 0.372 | 0.184 | 0.148 | 0.279 |
| CPMP | i1000_6 | 88 | ~ | ~ | 7178.1 | opt | 4061.9 | 0.361 | 0.179 | 0.143 | 0.278 |
| CMDELP | i1000_6 | 72 | 0.41 | 19.1 | 15848.7 | 6.14% | 7200.6 | 0.427 | 0.182 | 0.150 | 0.243 |
| CMDELP | i1000_6 | 76 | 0.40 | 20.0 | 14682.9 | 3.89% | 7201.2 | 0.409 | 0.174 | 0.146 | 0.242 |
| CMDELP | i1000_6 | 80 | 0.38 | 21.1 | 14600.6 | 3.45% | 7201.3 | 0.395 | 0.167 | 0.134 | 0.241 |
| CMDELP | i1000_6 | 84 | 0.37 | 21.9 | 13864.0 | 1.32% | 7202.2 | 0.381 | 0.160 | 0.130 | 0.239 |
| CMDELP | i1000_6 | 88 | 0.36 | 22.6 | 13366.6 | opt | 4863.6 | 0.370 | 0.154 | 0.125 | 0.236 |
| CMELP | i1000_6 | 72 | 0.41 | ~ | 421.3 | 20.62% | 7320.7 | 0.444 | 0.180 | 0.148 | 0.231 |
| CMELP | i1000_6 | 76 | 0.40 | ~ | 316.9 | 5.18% | 7270.9 | 0.413 | 0.169 | 0.136 | 0.232 |
| CMELP | i1000_6 | 80 | 0.38 | ~ | 316.8 | 5.44% | 7365.5 | 0.399 | 0.163 | 0.131 | 0.231 |
| CMELP | i1000_6 | 84 | 0.37 | ~ | 286.7 | 3.92% | 7256.7 | 0.388 | 0.155 | 0.125 | 0.227 |
| CMELP | i1000_6 | 88 | 0.36 | ~ | 263.3 | 1.04% | 7411.0 | 0.377 | 0.148 | 0.120 | 0.222 |

## Table S14: Solutions from the instance GY

| Problem | Inst. | P | $d^*$ | $\beta$ | Objective | Gap | Time | Mean | SD | Mad | Gini |
|---|---|---|---|---|---|---|---|---|---|---|---|
| PMP | GY | 22 | ~ | ~ | 1567390.8 | opt | 29.9 | 1.912 | 1.384 | 1.113 | 0.407 |
| PMP | GY | 24 | ~ | ~ | 1493475.9 | opt | 21.2 | 1.822 | 1.355 | 1.088 | 0.417 |
| PMP | GY | 26 | ~ | ~ | 1427280.8 | opt | 34.6 | 1.741 | 1.341 | 1.093 | 0.431 |
| PMP | GY | 28 | ~ | ~ | 1368159.6 | opt | 27.7 | 1.669 | 1.319 | 1.085 | 0.442 |
| PMP | GY | 30 | ~ | ~ | 1315066.7 | opt | 27.8 | 1.604 | 1.250 | 1.045 | 0.441 |
| MDELP | GY | 22 | 1.912 | 2.0 | 3033435.1 | opt | 22.9 | 1.999 | 1.210 | 0.977 | 0.343 |
| MDELP | GY | 24 | 1.822 | 2.0 | 2868516.6 | opt | 23.1 | 1.922 | 1.159 | 0.943 | 0.343 |
| MDELP | GY | 26 | 1.741 | 1.9 | 2675030.6 | opt | 22.4 | 1.839 | 1.127 | 0.916 | 0.348 |
| MDELP | GY | 28 | 1.669 | 1.9 | 2565966.4 | opt | 42.3 | 1.786 | 1.074 | 0.893 | 0.342 |
| MDELP | GY | 30 | 1.604 | 2.1 | 2567714.0 | opt | 32.6 | 1.675 | 1.112 | 0.929 | 0.379 |
| MELP | GY | 22 | 1.912 | ~ | 679029.1 | opt | 6.8 | 2.134 | 1.095 | 0.844 | 0.287 |
| MELP | GY | 24 | 1.822 | ~ | 642119.7 | opt | 55.4 | 1.944 | 1.149 | 0.931 | 0.335 |
| MELP | GY | 26 | 1.741 | ~ | 614147.9 | opt | 62.7 | 1.841 | 1.125 | 0.918 | 0.347 |
| MELP | GY | 28 | 1.669 | ~ | 577288.6 | opt | 45.0 | 1.798 | 1.078 | 0.879 | 0.341 |
| MELP | GY | 30 | 1.604 | ~ | 544852.2 | opt | 24.1 | 1.794 | 0.988 | 0.793 | 0.312 |
| CPMP | GY | 22 | ~ | ~ | 1670208.4 | opt | 2380.0 | 2.037 | 1.615 | 1.302 | 0.439 |
| CPMP | GY | 24 | ~ | ~ | 1537200.6 | opt | 658.0 | 1.875 | 1.480 | 1.228 | 0.445 |
| CPMP | GY | 26 | ~ | ~ | 1459800.1 | opt | 887.0 | 1.781 | 1.441 | 1.174 | 0.454 |
| CPMP | GY | 28 | ~ | ~ | 1389821.2 | opt | 225.4 | 1.695 | 1.382 | 1.133 | 0.456 |
| CPMP | GY | 30 | ~ | ~ | 1333953.7 | opt | 142.7 | 1.627 | 1.337 | 1.100 | 0.459 |
| CMDELP | GY | 22 | 2.037 | 1.6 | 3219160.3 | opt | 5605.4 | 2.111 | 1.402 | 1.132 | 0.375 |
| CMDELP | GY | 24 | 1.875 | 1.7 | 3007883.6 | opt | 2180.8 | 1.973 | 1.317 | 1.082 | 0.380 |
| CMDELP | GY | 26 | 1.781 | 1.8 | 2864941.0 | opt | 638.9 | 1.858 | 1.267 | 1.040 | 0.387 |
| CMDELP | GY | 28 | 1.695 | 1.8 | 2722588.7 | opt | 287.2 | 1.757 | 1.235 | 1.013 | 0.399 |
| CMDELP | GY | 30 | 1.627 | 1.9 | 2652549.6 | opt | 367.3 | 1.703 | 1.188 | 0.982 | 0.396 |
| CMELP | GY | 22 | 2.037 | ~ | 908204.5 | 0.96% | 7201.2 | 2.318 | 1.224 | 0.942 | 0.293 |
| CMELP | GY | 24 | 1.875 | ~ | 805152.0 | 1.59% | 7201.6 | 2.084 | 1.196 | 0.958 | 0.324 |
| CMELP | GY | 26 | 1.781 | ~ | 738429.7 | 1.65% | 7203.3 | 2.016 | 1.119 | 0.888 | 0.312 |
| CMELP | GY | 28 | 1.695 | ~ | 695903.0 | opt | 6541.0 | 1.909 | 1.115 | 0.917 | 0.331 |
| CMELP | GY | 30 | 1.627 | ~ | 650760.6 | opt | 3543.2 | 1.894 | 1.032 | 0.817 | 0.308 |

## Table S15: Solutions from the instance ZY

| Problem | Inst. | P | $d^*$ | $\beta$ | Objective | Gap | Time | Mean | SD | Mad | Gini |
|---|---|---|---|---|---|---|---|---|---|---|---|
| PMP | ZY | 10 | ~ | ~ | 1655.2 | opt | 4.7 | 0.427 | 0.217 | 0.169 | 0.285 |
| PMP | ZY | 11 | ~ | ~ | 1594.5 | opt | 4.8 | 0.412 | 0.217 | 0.167 | 0.294 |
| PMP | ZY | 12 | ~ | ~ | 1540.1 | opt | 7.1 | 0.398 | 0.207 | 0.160 | 0.291 |
| PMP | ZY | 13 | ~ | ~ | 1487.9 | opt | 9.0 | 0.384 | 0.203 | 0.155 | 0.294 |
| PMP | ZY | 14 | ~ | ~ | 1436.9 | opt | 12.0 | 0.371 | 0.192 | 0.147 | 0.287 |
| MDELP | ZY | 10 | 0.427 | 18.1 | 3148.4 | opt | 17.6 | 0.432 | 0.206 | 0.161 | 0.267 |
| MDELP | ZY | 11 | 0.412 | 17.5 | 2825.4 | opt | 5.1 | 0.425 | 0.182 | 0.140 | 0.238 |
| MDELP | ZY | 12 | 0.398 | 18.5 | 2743.1 | opt | 9.5 | 0.413 | 0.174 | 0.132 | 0.232 |
| MDELP | ZY | 13 | 0.384 | 18.7 | 2605.7 | opt | 7.6 | 0.400 | 0.167 | 0.128 | 0.230 |
| MDELP | ZY | 14 | 0.371 | 20.1 | 2536.8 | opt | 6.6 | 0.383 | 0.162 | 0.126 | 0.236 |
| MELP | ZY | 10 | 0.427 | ~ | 81.4 | opt | 9.9 | 0.432 | 0.206 | 0.161 | 0.267 |
| MELP | ZY | 11 | 0.412 | ~ | 67.4 | opt | 5.9 | 0.425 | 0.182 | 0.140 | 0.238 |
| MELP | ZY | 12 | 0.398 | ~ | 61.8 | opt | 8.6 | 0.413 | 0.174 | 0.132 | 0.232 |
| MELP | ZY | 13 | 0.384 | ~ | 56.2 | opt | 5.1 | 0.402 | 0.166 | 0.128 | 0.228 |
| MELP | ZY | 14 | 0.371 | ~ | 52.3 | opt | 8.0 | 0.384 | 0.161 | 0.125 | 0.233 |
| CPMP | ZY | 10 | ~ | ~ | 1686.8 | opt | 97.4 | 0.436 | 0.221 | 0.171 | 0.284 |
| CPMP | ZY | 11 | ~ | ~ | 1598.2 | opt | 59.3 | 0.413 | 0.218 | 0.168 | 0.295 |
| CPMP | ZY | 12 | ~ | ~ | 1541.7 | opt | 48.6 | 0.398 | 0.208 | 0.160 | 0.292 |
| CPMP | ZY | 13 | ~ | ~ | 1487.9 | opt | 46.8 | 0.384 | 0.203 | 0.155 | 0.294 |
| CPMP | ZY | 14 | ~ | ~ | 1436.9 | opt | 34.7 | 0.371 | 0.192 | 0.147 | 0.287 |
| CMDELP | ZY | 10 | 0.436 | 17.8 | 3277.8 | opt | 215.4 | 0.446 | 0.211 | 0.165 | 0.265 |
| CMDELP | ZY | 11 | 0.413 | 17.4 | 2865.5 | opt | 97.3 | 0.427 | 0.184 | 0.142 | 0.240 |
| CMDELP | ZY | 12 | 0.398 | 18.4 | 2770.2 | opt | 83.9 | 0.414 | 0.175 | 0.132 | 0.234 |
| CMDELP | ZY | 13 | 0.384 | 18.7 | 2617.1 | opt | 52.2 | 0.401 | 0.167 | 0.129 | 0.231 |
| CMDELP | ZY | 14 | 0.371 | 20.1 | 2546.3 | opt | 38.8 | 0.382 | 0.163 | 0.122 | 0.233 |
| CMELP | ZY | 10 | 0.436 | ~ | 87.1 | opt | 166.8 | 0.446 | 0.211 | 0.165 | 0.265 |
| CMELP | ZY | 11 | 0.413 | ~ | 69.7 | opt | 69.0 | 0.427 | 0.184 | 0.142 | 0.240 |
| CMELP | ZY | 12 | 0.398 | ~ | 63.4 | opt | 67.5 | 0.414 | 0.175 | 0.133 | 0.234 |
| CMELP | ZY | 13 | 0.384 | ~ | 57.0 | opt | 52.2 | 0.401 | 0.167 | 0.129 | 0.231 |
| CMELP | ZY | 14 | 0.371 | ~ | 52.9 | opt | 43.8 | 0.387 | 0.162 | 0.124 | 0.231 |

## Table S16: Solutions from the instance KF

| Problem | Inst. | P | $d^*$ | $\beta$ | Objective | Gap | Time | Mean | SD | Mad | Gini |
|---|---|---|---|---|---|---|---|---|---|---|---|
| PMP | KF | 18 | ~ | ~ | 589019.6 | opt | 63.9 | 0.824 | 0.461 | 0.361 | 0.308 |
| PMP | KF | 20 | ~ | ~ | 562264.5 | opt | 70.8 | 0.787 | 0.434 | 0.338 | 0.303 |
| PMP | KF | 22 | ~ | ~ | 538545.4 | opt | 104.0 | 0.754 | 0.417 | 0.321 | 0.302 |
| PMP | KF | 24 | ~ | ~ | 517626.7 | opt | 126.1 | 0.725 | 0.403 | 0.310 | 0.304 |
| PMP | KF | 26 | ~ | ~ | 498859.5 | opt | 102.8 | 0.698 | 0.396 | 0.303 | 0.307 |
| MDELP | KF | 18 | 0.824 | 7.8 | 1172066.8 | opt | 351.7 | 0.858 | 0.405 | 0.328 | 0.266 |
| MDELP | KF | 20 | 0.787 | 8.4 | 1126444.3 | opt | 472.7 | 0.821 | 0.381 | 0.304 | 0.260 |
| MDELP | KF | 22 | 0.754 | 8.7 | 1067318.9 | opt | 237.3 | 0.781 | 0.367 | 0.290 | 0.262 |
| MDELP | KF | 24 | 0.725 | 8.9 | 1023839.7 | opt | 493.3 | 0.749 | 0.355 | 0.280 | 0.264 |
| MDELP | KF | 26 | 0.698 | 8.9 | 979802.4 | opt | 1306.6 | 0.731 | 0.340 | 0.271 | 0.261 |
| MELP | KF | 18 | 0.824 | ~ | 71448.7 | opt | 336.1 | 0.871 | 0.398 | 0.322 | 0.258 |
| MELP | KF | 20 | 0.787 | ~ | 64237.1 | opt | 1597.4 | 0.821 | 0.381 | 0.304 | 0.260 |
| MELP | KF | 22 | 0.754 | ~ | 57743.1 | opt | 401.1 | 0.793 | 0.360 | 0.285 | 0.254 |
| MELP | KF | 24 | 0.725 | ~ | 53935.1 | opt | 541.7 | 0.769 | 0.345 | 0.274 | 0.251 |
| MELP | KF | 26 | 0.698 | ~ | 50919.2 | opt | 2042.2 | 0.752 | 0.329 | 0.264 | 0.246 |
| CPMP | KF | 18 | ~ | ~ | 589019.6 | opt | 1743.9 | 0.824 | 0.461 | 0.361 | 0.308 |
| CPMP | KF | 20 | ~ | ~ | 562264.5 | opt | 2343.4 | 0.787 | 0.434 | 0.338 | 0.303 |
| CPMP | KF | 22 | ~ | ~ | 538545.4 | opt | 1815.2 | 0.754 | 0.417 | 0.321 | 0.296 |
| CPMP | KF | 24 | ~ | ~ | 517626.7 | opt | 2187.6 | 0.725 | 0.403 | 0.310 | 0.303 |
| CPMP | KF | 26 | ~ | ~ | 498859.5 | opt | 2911.1 | 0.698 | 0.396 | 0.303 | 0.307 |
| CMDELP | KF | 18 | 0.820 | 10.0 | 1346720.0 | opt | 3185.2 | 0.845 | 0.413 | 0.331 | 0.274 |
| CMDELP | KF | 20 | 0.790 | 10.0 | 1222288.4 | opt | 4622.2 | 0.822 | 0.381 | 0.304 | 0.260 |
| CMDELP | KF | 22 | 0.750 | 10.0 | 1154739.1 | opt | 2750.0 | 0.780 | 0.368 | 0.290 | 0.263 |
| CMDELP | KF | 24 | 0.730 | 10.0 | 1073426.7 | opt | 2565.6 | 0.749 | 0.355 | 0.280 | 0.264 |
| CMDELP | KF | 26 | 0.700 | 10.0 | 1031951.9 | opt | 4913.9 | 0.731 | 0.340 | 0.271 | 0.261 |
| CMELP | KF | 18 | 0.820 | ~ | 73175.4 | opt | 4204.6 | 0.873 | 0.399 | 0.323 | 0.258 |
| CMELP | KF | 20 | 0.790 | ~ | 63531.8 | opt | 4251.3 | 0.822 | 0.381 | 0.304 | 0.247 |
| CMELP | KF | 22 | 0.750 | ~ | 59112.5 | opt | 2989.6 | 0.794 | 0.361 | 0.286 | 0.254 |
| CMELP | KF | 24 | 0.730 | ~ | 53254.9 | opt | 4501.7 | 0.770 | 0.345 | 0.275 | 0.252 |
| CMELP | KF | 26 | 0.700 | ~ | 50459.0 | opt | 2920.3 | 0.752 | 0.329 | 0.264 | 0.246 |

## Table S17: Solutions from the instance ZZ

| Problem | Inst. | P | $d^*$ | $\beta$ | Objective | Gap | Time | Mean | SD | Mad | Gini |
|---|---|---|---|---|---|---|---|---|---|---|---|
| PMP | ZZ | 48 | ~ | ~ | 3457717.6 | ~ | 615.7 | 0.819 | 0.420 | 0.326 | 0.283 |
| PMP | ZZ | 52 | ~ | ~ | 3335783.4 | ~ | 526.6 | 0.790 | 0.415 | 0.318 | 0.284 |
| PMP | ZZ | 56 | ~ | ~ | 3231183.1 | ~ | 474.5 | 0.765 | 0.403 | 0.308 | 0.287 |
| PMP | ZZ | 60 | ~ | ~ | 3124620.2 | ~ | 667.4 | 0.740 | 0.385 | 0.294 | 0.283 |
| PMP | ZZ | 64 | ~ | ~ | 3032341.2 | ~ | 1028.4 | 0.718 | 0.376 | 0.293 | 0.285 |
| MDELP | ZZ | 48 | 0.819 | 9.3 | 6697982.9 | ~ | 1001.3 | 0.839 | 0.373 | 0.302 | 0.251 |
| MDELP | ZZ | 52 | 0.790 | 9.2 | 6381031.2 | ~ | 771.5 | 0.815 | 0.361 | 0.292 | 0.250 |
| MDELP | ZZ | 56 | 0.765 | 9.4 | 5992260.2 | ~ | 754.7 | 0.789 | 0.340 | 0.271 | 0.242 |
| MDELP | ZZ | 60 | 0.740 | 10.0 | 5897516.8 | ~ | 804.7 | 0.772 | 0.325 | 0.260 | 0.237 |
| MDELP | ZZ | 64 | 0.718 | 10.1 | 5650278.0 | ~ | 1136.8 | 0.741 | 0.320 | 0.255 | 0.244 |
| MELP | ZZ | 48 | 0.819 | ~ | 325283.5 | ~ | 588.6 | 0.851 | 0.365 | 0.296 | 0.243 |
| MELP | ZZ | 52 | 0.790 | ~ | 310763.3 | ~ | 514.3 | 0.827 | 0.349 | 0.282 | 0.239 |
| MELP | ZZ | 56 | 0.765 | ~ | 282470.7 | ~ | 933.2 | 0.795 | 0.337 | 0.269 | 0.239 |
| MELP | ZZ | 60 | 0.740 | ~ | 258928.6 | ~ | 1008.1 | 0.769 | 0.325 | 0.260 | 0.238 |
| MELP | ZZ | 64 | 0.718 | ~ | 243616.0 | ~ | 1115.7 | 0.748 | 0.313 | 0.249 | 0.235 |
| CPMP | ZZ | 48 | ~ | ~ | 3640336.9 | ~ | 4617.5 | 0.862 | 0.546 | 0.408 | 0.333 |
| CPMP | ZZ | 52 | ~ | ~ | 3443420.1 | ~ | 1833.9 | 0.815 | 0.452 | 0.355 | 0.305 |
| CPMP | ZZ | 56 | ~ | ~ | 3288316.7 | ~ | 3543.1 | 0.778 | 0.418 | 0.325 | 0.295 |
| CPMP | ZZ | 60 | ~ | ~ | 3153313.4 | ~ | 1515.4 | 0.747 | 0.401 | 0.312 | 0.295 |
| CPMP | ZZ | 64 | ~ | ~ | 3059409.7 | ~ | 1534.6 | 0.724 | 0.388 | 0.303 | 0.295 |
| CMDELP | ZZ | 48 | 0.862 | 5.8 | 7973447.3 | ~ | 7399.1 | 0.873 | 0.510 | 0.379 | 0.309 |
| CMDELP | ZZ | 52 | 0.815 | 8.0 | 7017243.0 | ~ | 4574.8 | 0.828 | 0.415 | 0.329 | 0.279 |
| CMDELP | ZZ | 56 | 0.778 | 8.9 | 6555152.0 | ~ | 6056.7 | 0.789 | 0.382 | 0.302 | 0.271 |
| CMDELP | ZZ | 60 | 0.747 | 9.3 | 6371246.3 | ~ | 3455.4 | 0.767 | 0.365 | 0.289 | 0.266 |
| CMDELP | ZZ | 64 | 0.724 | 9.6 | 6048518.9 | ~ | 3230.8 | 0.737 | 0.353 | 0.281 | 0.269 |
| CMELP | ZZ | 48 | 0.862 | ~ | 702660.8 | ~ | 8295.0 | 0.899 | 0.489 | 0.365 | 0.292 |
| CMELP | ZZ | 52 | 0.815 | ~ | 451594.2 | ~ | 4247.8 | 0.840 | 0.411 | 0.321 | 0.270 |
| CMELP | ZZ | 56 | 0.778 | ~ | 354079.8 | ~ | 4599.6 | 0.805 | 0.369 | 0.293 | 0.257 |
| CMELP | ZZ | 60 | 0.747 | ~ | 332435.7 | ~ | 6241.8 | 0.776 | 0.356 | 0.279 | 0.256 |
| CMELP | ZZ | 64 | 0.724 | ~ | 303376.8 | ~ | 3714.1 | 0.750 | 0.345 | 0.273 | 0.257 |

## Table S18: Various model solutions from the instance Cap121

| Model | p | Parameters | Max | Mean | SD | MAD | AD | CV | SI | Gini | Time |
|---|---|---|---|---|---|---|---|---|---|---|---|
| PMP | 8 | ~ | 38.650 | 13.225 | 8.464 | 6.684 | 8.783 | 0.640 | 0.253 | 0.332 | 0.5 |
| PCP | 8 | ~ | 31.138 | 19.020 | 7.656 | 6.606 | 8.677 | 0.403 | 0.174 | 0.228 | 0.7 |
| MELP | 8 | $d^*$=13.2 | 36.813 | 13.581 | 7.988 | 6.186 | 8.501 | 0.588 | 0.228 | 0.313 | 0.6 |
| MDELP | 8 | $d^*$=13.2,$\beta$=0.37 | 36.813 | 13.352 | 8.033 | 6.145 | 8.410 | 0.602 | 0.230 | 0.315 | 0.6 |
| MinMAD | 8 | ~ | 32.100 | 26.085 | 3.820 | 2.680 | 4.033 | 0.146 | 0.051 | 0.077 | 30.3 |
| MinSI | 8 | ~ | 44.213 | 35.103 | 4.984 | 3.443 | 5.030 | 0.142 | 0.049 | 0.072 | 276.5 |
| MinAD | 8 | ~ | 32.100 | 26.085 | 3.820 | 2.680 | 4.033 | 0.146 | 0.051 | 0.077 | 222.3 |
| MinGC | 8 | ~ | 44.213 | 35.093 | 4.881 | 3.455 | 5.012 | 0.139 | 0.049 | 0.071 | 3737.8 |
| MinSD | 8 | ~ | 32.100 | 26.085 | 3.820 | 2.680 | 4.033 | 0.146 | 0.051 | 0.077 | 67.3 |
| MinCV | 8 | ~ | 44.213 | 35.275 | 4.902 | 3.542 | 5.092 | 0.139 | 0.050 | 0.072 | 1474.0 |
| K-centrum | 8 | K=5 | 31.138 | 19.020 | 7.656 | 6.606 | 8.677 | 0.403 | 0.174 | 0.228 | 1.2 |
| K-centrum | 8 | K=10 | 34.388 | 17.416 | 7.824 | 6.563 | 8.856 | 0.449 | 0.188 | 0.254 | 2.3 |
| K-centrum | 8 | K=15 | 40.150 | 16.935 | 9.002 | 7.083 | 9.966 | 0.532 | 0.209 | 0.294 | 1.0 |
| K-centrum | 8 | K=25 | 46.213 | 15.313 | 10.994 | 7.866 | 10.622 | 0.718 | 0.257 | 0.347 | 0.8 |
| α-centdian | 8 | α=0.1 | 31.138 | 19.020 | 7.656 | 8.677 | 8.677 | 0.403 | 0.174 | 0.228 | 0.7 |
| α-centdian | 8 | α=0.3 | 32.100 | 14.759 | 7.932 | 6.647 | 8.362 | 0.537 | 0.225 | 0.283 | 1.0 |
| α-centdian | 8 | α=0.5 | 32.525 | 13.987 | 8.371 | 6.776 | 8.980 | 0.598 | 0.242 | 0.321 | 0.8 |
| α-centdian | 8 | α=0.7 | 32.525 | 13.987 | 8.371 | 6.776 | 8.980 | 0.598 | 0.242 | 0.321 | 0.7 |
| α-centdian | 8 | α=0.9 | 35.838 | 13.339 | 8.276 | 6.457 | 8.642 | 0.620 | 0.242 | 0.324 | 0.7 |
| K-centdian | 8 | K=5, α=0.3 | 33.613 | 13.737 | 8.327 | 6.641 | 8.854 | 0.606 | 0.242 | 0.322 | 0.9 |
| K-centdian | 8 | K=5, α=0.5 | 33.613 | 13.737 | 8.327 | 6.641 | 8.854 | 0.606 | 0.242 | 0.322 | 0.7 |
| K-centdian | 8 | K=10, α=0.3 | 40.150 | 13.998 | 9.065 | 6.833 | 9.307 | 0.648 | 0.244 | 0.332 | 0.8 |
| K-centdian | 8 | K=10, α=0.5 | 40.150 | 13.998 | 9.065 | 6.833 | 9.307 | 0.648 | 0.244 | 0.332 | 0.7 |
| K-centdian | 8 | K=25, α=0.3 | 40.150 | 13.804 | 9.050 | 6.778 | 9.235 | 0.656 | 0.246 | 0.334 | 0.6 |
| K-centdian | 8 | K=25, α=0.5 | 40.150 | 13.804 | 9.050 | 6.778 | 9.235 | 0.656 | 0.246 | 0.334 | 0.6 |

## Table S19: Various model solutions from the instance Pmed2

| Model | p | Parameters | Max | Mean | SD | MAD | AD | CV | SI | Gini | Time |
|---|---|---|---|---|---|---|---|---|---|---|---|
| PMP | 10 | ~ | 132 | 40.930 | 32.244 | 25.923 | 35.990 | 0.788 | 0.317 | 0.440 | 1.0 |
| PCP | 10 | ~ | 98 | 47.970 | 29.997 | 26.069 | 34.535 | 0.625 | 0.272 | 0.360 | 2.4 |
| MELP | 10 | $d^*$=41.0 | 118 | 42.040 | 28.414 | 23.442 | 32.342 | 0.676 | 0.279 | 0.385 | 1.0 |
| MDELP | 10 | $d^*$=41.0,$\beta$=0.08 | 118 | 42.040 | 28.414 | 23.442 | 32.342 | 0.676 | 0.279 | 0.385 | 1.1 |
| MinMAD | 10 | ~ | - | - | - | - | - | - | - | - | 4h |
| MinSI | 10 | ~ | - | - | - | - | - | - | - | - | 4h |
| MinAD | 10 | ~ | - | - | - | - | - | - | - | - | 4h |
| MinGC | 10 | ~ | - | - | - | - | - | - | - | - | 4h |
| MinSD | 10 | ~ | - | - | - | - | - | - | - | - | 4h |
| MinCV | 10 | ~ | - | - | - | - | - | - | - | - | 4h |
| K-centrum | 10 | K=5 | 102 | 49.000 | 29.266 | 24.760 | 33.571 | 0.597 | 0.253 | 0.343 | 80.1 |
| K-centrum | 10 | K=10 | 108 | 46.860 | 29.367 | 24.943 | 33.705 | 0.627 | 0.266 | 0.360 | 345.5 |
| K-centrum | 10 | K=20 | 118 | 42.530 | 28.213 | 23.350 | 32.110 | 0.663 | 0.275 | 0.377 | 444.7 |
| K-centrum | 10 | K=50 | 118 | 41.890 | 28.588 | 23.028 | 32.421 | 0.682 | 0.275 | 0.387 | 23.0 |
| α-centdian | 10 | α =0.3 | 98 | 47.570 | 29.493 | 25.321 | 33.941 | 0.620 | 0.266 | 0.357 | 25.9 |
| α-centdian | 10 | α=0.4 | 98 | 47.570 | 29.493 | 25.321 | 33.941 | 0.620 | 0.266 | 0.357 | 55.4 |
| α-centdian | 10 | α=0.5 | 102 | 42.690 | 29.657 | 24.824 | 33.948 | 0.695 | 0.291 | 0.398 | 52.7 |
| K-centdian | 10 | K=10, α=0.3 | 118 | 41.680 | 29.564 | 25.075 | 33.766 | 0.709 | 0.301 | 0.405 | 269.8 |
| K-centdian | 10 | K=10, α=0.5 | 118 | 41.370 | 29.744 | 25.362 | 33.986 | 0.719 | 0.307 | 0.411 | 39.6 |
| K-centdian | 10 | K=10, α=0.7 | 118 | 41.370 | 29.744 | 25.362 | 33.986 | 0.719 | 0.307 | 0.411 | 10.0 |
| K-centdian | 10 | K=20, α=0.3 | 118 | 41.980 | 28.398 | 23.040 | 32.221 | 0.676 | 0.274 | 0.384 | 327.8 |
| K-centdian | 10 | K=20, α=0.5 | 118 | 41.280 | 29.578 | 25.159 | 33.763 | 0.717 | 0.305 | 0.409 | 82.8 |
| K-centdian | 10 | K=20, α=0.7 | 118 | 41.280 | 29.578 | 25.159 | 33.763 | 0.717 | 0.305 | 0.409 | 18.5 |
| K-centdian | 10 | K=50, α=0.3 | 118 | 41.890 | 28.588 | 23.028 | 32.421 | 0.682 | 0.275 | 0.387 | 38.7 |
| K-centdian | 10 | K=50, α=0.5 | 120 | 41.050 | 30.043 | 24.956 | 34.135 | 0.732 | 0.304 | 0.416 | 9.2 |
| K-centdian | 10 | K=50, α=0.7 | 120 | 41.050 | 30.043 | 24.956 | 34.135 | 0.732 | 0.304 | 0.416 | 4.2 |

Note 1: The MinMAD, MinSI, MinAD, MinGC, MinSD, and MinCV models cannot be optimally or near-optimally (MIPGap<10%) solved by Gurobi Optimizer 9 in 4 hours.

**Table S20: Various model solutions from the instance ZY**

| Model | p | Parameters | Max | Mean | SD | MAD | AD | CV | SI | Gini | Time |
|---|---|---|---|---|---|---|---|---|---|---|---|
| PMP | 12 | ~ | 1.071 | 0.398 | 0.207 | 0.160 | 0.232 | 0.522 | 0.201 | 0.291 | 7.1 |
| PCP | 12 | ~ | 0.779 | 0.448 | 0.194 | 0.160 | 0.219 | 0.433 | 0.179 | 0.244 | 13.9 |
| MELP | 12 | $d^*$=0.4 | 1.006 | 0.413 | 0.174 | 0.132 | 0.192 | 0.422 | 0.159 | 0.232 | 8.6 |
| MDELP | 12 | $d^*$=0.4,$\beta$=18.5 | 1.006 | 0.413 | 0.174 | 0.132 | 0.192 | 0.422 | 0.159 | 0.232 | 9.5 |
| MinMAD | 12 | ~ | - | - | - | - | - | - | - | - | 4h |
| MinSI | 12 | ~ | - | - | - | - | - | - | - | - | 4h |
| MinAD | 12 | ~ | - | - | - | - | - | - | - | - | 4h |
| MinGC | 12 | ~ | - | - | - | - | - | - | - | - | 4h |
| MinSD | 12 | ~ | - | - | - | - | - | - | - | - | 4h |
| MinCV | 12 | ~ | - | - | - | - | - | - | - | - | 4h |
| K-centrum | 12 | K=16 | 0.874 | 0.440 | 0.179 | 0.141 | 0.200 | 0.407 | 0.160 | 0.227 | 3447.1 |
| K-centrum | 12 | K=32 | 0.874 | 0.441 | 0.180 | 0.145 | 0.202 | 0.408 | 0.164 | 0.229 | 2h |
| K-centrum | 12 | K=96 | 1.006 | 0.421 | 0.178 | 0.137 | 0.198 | 0.424 | 0.163 | 0.235 | 2h |
| K-centrum | 12 | K=160 | 1.051 | 0.420 | 0.183 | 0.140 | 0.202 | 0.435 | 0.167 | 0.241 | 2h |
| α-centdian | 12 | α=0.1 | 0.779 | 0.448 | 0.194 | 0.145 | 0.202 | 0.408 | 0.164 | 0.229 | 43.8 |
| α-centdian | 12 | α=0.3 | 0.779 | 0.448 | 0.194 | 0.160 | 0.219 | 0.433 | 0.179 | 0.244 | 471.3 |
| α-centdian | 12 | α=0.5 | 0.779 | 0.448 | 0.194 | 0.160 | 0.219 | 0.433 | 0.179 | 0.244 | 1541.4 |
| α-centdian | 12 | α=0.7 | 0.832 | 0.422 | 0.188 | 0.160 | 0.219 | 0.433 | 0.179 | 0.244 | 690.2 |
| α-centdian | 12 | α=0.9 | 0.896 | 0.403 | 0.201 | 0.152 | 0.213 | 0.446 | 0.180 | 0.252 | 41.8 |
| K-centdian | 12 | K=32,α=0.3 | 0.874 | 0.441 | 0.180 | 0.160 | 0.227 | 0.500 | 0.199 | 0.282 | 2h |
| K-centdian | 12 | K=32,α=0.5 | 1.006 | 0.418 | 0.176 | 0.138 | 0.196 | 0.421 | 0.165 | 0.234 | 2h |
| K-centdian | 12 | K=96,α=0.3 | 1.006 | 0.415 | 0.178 | 0.135 | 0.197 | 0.428 | 0.163 | 0.237 | 2h |
| K-centdian | 12 | K=96,α=0.5 | 1.006 | 0.412 | 0.179 | 0.137 | 0.198 | 0.434 | 0.166 | 0.240 | 2h |
| K-centdian | 12 | K=160,α=0.3 | 1.006 | 0.416 | 0.181 | 0.140 | 0.201 | 0.433 | 0.168 | 0.242 | 2h |
| K-centdian | 12 | K=160,α=0.5 | 1.006 | 0.411 | 0.181 | 0.139 | 0.201 | 0.440 | 0.169 | 0.245 | 1576.9 |

Note 1: The MinMAD, MinSI, MinAD, MinGC, MinSD, and MinCV models cannot be optimally or near-optimally (MIPGap<10%) solved by Gurobi Optimizer 9 in 4 hours.

Note 2: Most K-centrum and K-centdian model are near-optimally MIPGap<10%) solved in 2 hours.

**Table S21: Mean-variance Pareto optimal solutions from the instances Cap121, Pmde2 and ZY**

| Cap121 (P=8) | | Pmed2 (P=10) | | ZY (P=12) | |
|---|---|---|---|---|---|
| Mean | SD | Mean | SD | Mean | SD |
| 13.225 | 8.464 | 40.93 | 32.24 | 0.398 | 0.207 |
| 13.310 | 8.217 | 41.10 | 30.00 | 0.398 | 0.199 |
| 13.352 | 8.033 | 41.20 | 29.61 | 0.400 | 0.196 |
| 13.581 | 7.988 | 41.28 | 29.58 | 0.402 | 0.191 |
| 13.947 | 7.743 | 41.36 | 29.47 | 0.403 | 0.191 |
| 14.098 | 7.618 | 41.50 | 29.20 | 0.405 | 0.186 |
| 14.248 | 7.567 | 41.60 | 29.19 | 0.406 | 0.181 |
| 14.384 | 7.456 | 41.77 | 28.92 | 0.409 | 0.176 |
| 14.506 | 7.324 | 41.98 | 28.40 | 0.411 | 0.175 |
| 14.649 | 7.309 | 41.98 | 28.40 | 0.413 | 0.174 |
| 15.038 | 7.196 | 42.71 | 28.13 | 0.415 | 0.173 |
| 15.174 | 7.144 | 43.53 | 28.01 | 0.416 | 0.172 |
| 15.285 | 7.036 | 44.14 | 27.94 | 0.418 | 0.172 |
| 15.390 | 7.013 | 44.94 | 27.93 | 0.421 | 0.172 |
| 15.555 | 6.985 | 45.86 | 27.36 | 0.424 | 0.170 |
| 15.699 | 6.906 | 49.78 | 27.38 | 0.426 | 0.170 |
| 15.769 | 6.835 | 50.68 | 27.00 | 0.429 | 0.169 |
| 15.871 | 6.810 | | | 0.434 | 0.168 |
| 16.088 | 6.697 | | | 0.436 | 0.168 |
| 16.191 | 6.667 | | | 0.452 | 0.166 |
| 16.334 | 6.636 | | | | |
| 16.493 | 6.475 | | | | |
| 16.618 | 6.441 | | | | |
| 16.718 | 6.415 | | | | |
| 16.831 | 6.284 | | | | |
| 16.973 | 6.241 | | | | |
| 17.155 | 6.062 | | | | |
| 17.220 | 5.982 | | | | |
| 17.363 | 5.923 | | | | |
| 17.684 | 5.875 | | | | |
| 17.904 | 5.832 | | | | |
| 18.215 | 5.767 | | | | |
| 18.464 | 5.744 | | | | |
| 19.577 | 5.205 | | | | |
| 21.465 | 5.088 | | | | |
| 21.926 | 5.018 | | | | |
| 24.498 | 4.745 | | | | |
| 24.992 | 4.395 | | | | |
| 26.085 | 3.820 | | | | |